\documentclass[epjH]{svjour}
\usepackage{graphicx}

\usepackage{color}
\usepackage{cite}
\usepackage{url}
\usepackage[T1]{fontenc}
\usepackage{yfonts}
\usepackage{graphicx,url}
\usepackage{verbatim}

\begin{document}
\color{black}
\title{Stellar structure and compact objects before 1940: Towards relativistic astrophysics\footnote{Article submitted to the \textit{European Physical Journal H}}}

\author{Luisa Bonolis\inst{1}\fnmsep\thanks{\email{lbonolis@mpiwg-berlin.mpg.de}} }
\institute{Max Planck Institut f\"ur Wissenschaftsgeschichte, Boltzmannstra\ss e 22, \\14195 Berlin, Germany }
\abstract{
Since the mid-1920s, different strands of research used stars as ``physics laboratories'' for investigating the nature of matter under extreme densities and pressures, impossible to realize on Earth.  
To trace this process this paper is following the evolution of the concept of a dense core in stars, which was important both for an understanding of stellar evolution and as a testing ground for the fast-evolving field of nuclear physics. In spite of the divide between physicists and astrophysicists, some key actors working in the cross-fertilized soil of overlapping but different scientific cultures formulated models and tentative theories that gradually evolved into more realistic and structured astrophysical objects. These investigations culminated in the first contact with general relativity in 1939, when J. Robert Oppenheimer and his students George Volkoff and Hartland Snyder systematically applied the theory to the dense core of a collapsing neutron star. This pioneering application of Einstein's theory to an astrophysical compact object can be regarded as a milestone in the path eventually leading to the emergence of relativistic astrophysics in the early 1960s. 
} 
\maketitle

\section{Introduction}

\label{intro}

Despite its enormous influence on scientific thought in its early years, general relativity experienced  a so-called `low-watermark period', going roughly from the mid-1920s to the mid-1950s \cite{Eisenstaedt1986,Eisenstaedt1987,Eisenstaedt2006}, during which it remained cut off  from  the mainstream of physics  and was  perceived  as a  sterile, highly formalistic subject. Accompanied by a series of major astrophysical discoveries, the status of General Relativity definitely changed in the 1960s, when it became an extremely vital research stream of theoretical physics. Quasars, the cosmic microwave background radiation, and pulsars --- soon identified as rotating neutron stars --- led to the recognition that physical processes and astrophysical objects exist in the universe that are understandable only in terms of the general theory of relativity.  In providing definitive proof of the existence of neutron stars, the discovery of pulsars and binary X-ray sources, made even plausible the possibility of black holes, entities that had previously existed only in the minds of a few theorists.  In raising new challenges to the emerging relativity community, these had of course an important role in strengthening the process which turned general relativity into a ``subdiscipline of physics'' \cite{BlumLalliRenn2015,BlumLalliRenn2016}. 

However, the view of a community of relativists magically awakened from its slumber by the new astrophysical discoveries is too one-dimensional. As Alexander Blum, Roberto Lalli, and J\"urgen Renn have outlined in their historiographical framework exploring the main factors underlying the return of general relativity into the mainstream of physics, a complex series of elements underlying such process must be taken into account: intellectual developments, epistemological problems, technological advances, the characteristics of post-World War II and Cold-War science, as well as the newly emerging institutional settings. Starting from the mid 1950s, further implications began to be explored and general relativity gradually came into focus as a physical theory. This framework, in which they propose to speak of a \textit{reinvention of general relativity}, rather than a \textit{renewal}, is leading to an understanding of the reinvention as a result of two main factors: the recognition of the untapped potential of general relativity and an explicit effort at community-building. These two factors allowed this formerly dispersed field to benefit from the postwar changes in the science landscape. 

The dynamics underlined in \cite{BlumLalliRenn2015} is actually independent from --- and prior to --- the major astrophysical discoveries of the 1960s.  Up to that time, the view prevailed that general relativistic effects were significant only for cosmology. However, the violent events that seemed to occur in the core of radio galaxies involving enormous energies corresponding to a rest-mass energy of $10^{6}$ solar masses ($M_\odot$) \cite{Burbidge1959}, the growing field of nuclear astrophysics \cite{B2FH1957,Cameron1958,Burbidge1962}, and the eventful discovery of quasars, had prepared the stage for the emerging awareness at the beginning of the 1960s of physical processes in which general relativistic effects are dominant and that could release much larger fractions of the rest mass as energy than the small fraction provided by the binding energies of nuclei. Such processes that did seem possible in the framework of general relativity suggested the actual existence of astrophysical objects in the universe satisfying requirements that appeared to be beyond the scope of nuclear physics. 

The problem of finding the source of the tremendous energy stored in cosmic rays and magnetic fields of some powerful radio galaxies, led to a theory put forward by William Fowler and Fred Hoyle in January 1963. They suggested that exceedingly massive star-like objects probably could exist with masses up to $10^8$ times that of the sun at the center of those galaxies. The gravitational collapse of such supermassive stars could be the driving force behind  the great amount of energy emitted by those strong radio sources \cite{FowlerHoyle1963a}. Their opinion was that in the process of contraction of a mass of $10^{7}-10^{8} M_{\odot}$ ``general relativity must be used'' in order to obtain the energies of the strongest ``stellar-type'' sources \cite[p. 535]{FowlerHoyle1963b}.

A few months after this proposal, new objects were discovered, having apparently masses of this order of magnitude, dimensions of about a light week, and having a luminosity  two orders of magnitude larger than the luminosity of a large galaxy having dimensions a million times larger and containing something like $10^{11}$ stars. In particular, the crucial identification of the high redshift of the already known radio source 3C273 \cite{HazardMackeySchimmins1963,Schmidt1963,Oke1963} and of the source 3C48 \cite{GreensteinMatthews1963}, made now even more pressing the problem to explain the mechanism whereby these and other sources that were masquerading as a star and were  thus identified as ``quasi-stellar'' objects,  managed to radiate away the energy equivalent of five hundred thousand suns at a very fast rate.

The  ``supermassive stars'' suggested by Fowler and Hoyle immediately became an attractive explanation for these new peculiar astrophysical objects, that appeared to be farther away than most known galaxies but were luminous enough to be observed by optical telescopes. Their enormous luminosity could also sharply change in the course of one week, as analysis of historical plate material of Harvard Observatory showed \cite{SmithHoffleit1963}. As such enormous energies must be emitted by regions less than one light-week across, collapsed objects became candidates for the engine of quasi-stellar radio sources.

The intriguing discovery of quasi-stellar radio sources --- soon renamed \textit{quasars} \cite[p. 21]{Chiu1964} --- with their large red-shifts and corresponding unprecedented-large radio and optical luminosities, opened up the discussion on a series of exciting questions. Among the problems raised were the following:  Were these objects  the debris of a gravitational implosion? By what machinery could gravitational energy be converted into radio waves? Would gravitational collapse lead to indefinite contraction and a singularity in space time? If so, how should  theoretical assumptions be changed to avoid this catastrophe? \cite[Preface]{RobinsonEtAl1965}.

``The topic was just right for reporting and sorting out observations as well as for theoretical analysis'' \cite[p. 51]{Schucking1989}: during the summer 1963, three relativists in Dallas, Ivor Robinson, Alfred Schild, Engelbert Schucking, realized that a  conference bridging the gap  between the still exotic world of  general relativity and the realm of astrophysics, might be well timed, and it would be a perfect occasion to make known the recently created Southwest Center for Advanced Studies. They immediately involved  Peter Bergmann, an influential relativist who had been associated with Einstein at the Institute for Advanced Study in Princeton since 1936, and sent out letters of invitation. Three hundred relativists, optical and radio astronomers, and theoretical astrophysicists attended the \textit{International Symposium on Quasi-Stellar Sources and Gravitational Collapse} \cite{RobinsonEtAl1965}, the first of the long series of Texas Symposia, which set up the stage merging two seemingly distant fields: general relativity and astrophysics, so distant that the organizers had to invent a new label for this brand new field: ``The suspicion existed that quasars might have something to do with relativity and thus might fit into an imaginary discipline combining astronomy with relativity. One of us --- Alfred, Ivor or I? --- invented a catch phrase for this new field of science: \textit{relativistic astrophysics} [emphasis added]'' \cite[p. 50]{Schucking1989}.

 Robert Oppenheimer was asked to chair the first session, a most natural choice, because of his involvement in the first systematic application of Einstein's general theory of relativity to a compact astrophysical object. Oppenheimer's three papers published between 1938 and 1939, each with a different collaborator \cite{OppenheimerSerber1938,OppenheimerVolkoff1939,OppenheimerSnyder1939}, are regarded as a milestone both in his scientific production and in the path eventually leading to the emergence of relativistic astrophysics in the early 1960s. In speaking of observations of ``incredible grandeur'' \cite[p. 50]{Schucking1989}, Oppenheimer officially opened the discussion on topics such as neutron stars or the possibility of gravitational collapse to a singularity in space-time, topics investigated within the context of the considerable revival of interest in the properties of matter at high densities and compact stars  going on since the end of the 1950s \cite{WheelerEtAl1958,Cameron1959,Migdal1959,AmbartsumyanSaakyan1960,Salpeter1960,TauberWeinberg1961,Salpeter1961,HamadaSalpeter1961,Zeldovich1962a,Zeldovich1962b,AmbartsumyanSaakyan1962a,AmbartsumyanSaakyan1962b,Beckedorff1962}.   During the conference Bergmann remarked that in the past ``general relativistic effects had been observed only in the weak field limit. Now new developments of astrophysics have made relativity a more physical theory'' \cite[p. 34]{Chiu1964}. Following the 1963 Texas Symposium \cite{RobinsonEtAl1965,WheelerEtAl1965}, many important advances in understanding black holes developed from new astrophysical observations and theoretical developments.

Cosmology, with its strong connections with general relativity since its early days, provided a continuity through the low-water-mark period, up to the post-war years, even if it was generally considered an ``esoteric'' field without any real connection with physics and having a scant observational basis.  In cosmology, general relativity directed the course of the observational researches in the realm of the galaxies, once the paradigm of an expanding universe became firmly established. However, while in the past it was the geometry, kinematics, and dynamics of the universe which were in the foreground, in the post-war development of cosmology, physical processes in the universe, involving elementary particles, electromagnetic radiation, and nuclear reactions, became a dominant interest, establishing a new and wider interaction with other fields. 

From a different perspective, studies on dense matter and compact astrophysical objects, merging interdisciplinary fields like nuclear physics and astrophysics --- both having many intersecting topics especially with post-war cosmology --- provided since the 1920s another form of continuity, through the 1930s and the 1940s.  During the post- and Cold War period, implosion and explosion problems, related to the design of thermonuclear weapons, brought about renewed interest in investigations on highly dense stellar matter and on the abandoned problem of gravitational collapse within Einstein's theory. New tools, typical of post-war science, were now available: the impressive advances in nuclear science combined with the first powerful computers, designed to perform the complex calculations for thermonuclear weapons, were now used to calculate the equation of state of condensed stellar matter up to the endpoint of thermonuclear evolution. While a new community of researchers in general relativity was achieving novel fundamental theoretical insights into Einstein's equations, the interaction between different scientific communities tackling interconnected astrophysical problems led to a resurgent awareness of processes in which general relativistic effects might play a dominant role. 

A reconstruction of how the emergence of relativistic astrophysics in the early 1960s can be understood as the culmination of a complex process including the longstanding tradition of the astrophysical study of compact objects and its connections with general relativity, will be the subject of a forthcoming article. 

The present contribution is examining how fundamental premises were laid during the period going from the mid 1920s to  the the end of the 1930s by theoretical investigations on such a basic topic, officially inaugurated in 1926 by Ralph Fowler's pioneering paper examining the problem of degenerate dense matter in white dwarf stars \cite{Fowler1926a}. These studies were accomplished at the intersection of different theoretical frameworks involving several disciplines and sub-disciplines and developed into a knowledge network involving some leading actors whose multidisciplinary competences were instrumental in catalyzing the flourishing of this process. Such developments led at the end of the 1930s to Oppenheimer's contributions on the relativistic gravitational collapse of a neutron star. These works were rediscovered after the war and became a starting point for further investigations on the connection between compact objects and general relativity, eventually leading to what Kip Thorne called ``the \textit{golden age} of black hole research''  \cite[pp. 74-80]{Thorne2003}, the decade going from 1964 to the mid-1970s, ``an era that revolutionized our understanding of general relativity's predictions'' \cite[p. 258]{Thorne1994}.

In the renaissance of general relativity and cosmology, central themes have been of course the study of relativistic gravitational collapse, black holes and neutron stars \cite{Miller1993}. Fascinating reconstructions of the evolution of ideas about black holes and other `dark astrophysical objects' were offered by physicists who have been protagonists in the quest to understand Einstein's legacy and its `predictions about the Universe'. I am especially referring to Kip Thorne  \cite{Thorne1994} and Werner Israel \cite{Israel1987}, whose valuable efforts have contributed to outline fundamental steps along the `meandering paths' of this history, providing an important basis for reflecting on the evolution of scientific ideas and the formulation of new concepts, that together with astronomical observations fuelled the actual merging of astrophysics with general relativity. Other excellent essays have addressed the related evolution of the concept of neutron stars  \cite{Baym1983} or more specifically have discussed the contributions to this story by main actors in this narrative, like Georges Lema\^itre \cite{Kragh1987,Eisenstaedt1993}, Robert J. Oppenheimer and Lev Landau \cite{Hufbauer2006,Hufbauer2007,Yakovlev1994}, or Subrahmanyan Chandrasekhar \cite{Wali1990} and George Gamow \cite{Kragh2005,Hufbauer2009}. Other relevant references will be cited in due course. 

The present attempt has instead adopted the perspective to follow the path of the evolving concept of a dense core in stars using it as a guiding key to reconstruct in detail the tapestry of interrelated ideas and changing models related to a series of fundamental questions: on one side  the theoretical problem of the structure and evolution of stars up to their endpoint states, and on the other the role of such core as a virtual laboratory to investigate the behaviour of matter under extreme conditions of densities and pressures prevailing in stars that are impossible to realize on Earth. Its evolution as a challenging physical object --- constantly connected to the problem of the origin of stellar energy --- transformed the core into a testing ground for the emerging field of nuclear physics, also testifying the quickly changing relationship between physics and astrophysics during the 1930s. Such investigations, resulting from the interaction between different material and intellectual cultural practices, provided the multifaceted context and the theoretical framework within which Oppenheimer and his collaborators were able to work out the final fate of a collapsing neutron star at the end of the 1930s, which in retrospect was considered ``the greatest of his discoveries: the black hole'' \cite[p. 50]{Schucking1989}.


\section{Prologue: The `nonsensical message' of white-dwarf stars}
\label{whitedwarf}

In 1925, while Einstein was generalizing Bose's distribution function for the case of a fixed number of particles, Wolfgang Pauli, stimulated by Edmund C. Stoner's analysis  of the quantum states of the electrons in complex atoms \cite{Stoner1924}, proposed the exclusion principle as a general phenomenological rule governing the behavior of electrons in multi-electron systems \cite{Pauli1925}. 

Pauli's proposal triggered the development, independently by Enrico Fermi \cite{Fermi1926a,Fermi1926b} and P. A. M. Dirac \cite{Dirac1926}, of a quantum statistics applicable to a gas of particles that obey the exclusion principle. As a fundamental physical principle rooted in quantum theory, the new quantum statistics provided the tool to treat an assembly of identical particles like a gas of electrons, and in turn, it immediately prompted Pauli's quick reaction \cite{Pauli1927}. In order to prove that the Fermi-Dirac statistics --- and not Bose-Einstein --- was the right statistics to be applied to the degenerate electron gas (``beim materiellen Gas die Fermische und nicht die Einstein-Bosesche Statistik die zutreffende ist''), Pauli derived a physical consequence that could be experimentally verified: he pointed out that the weak temperature-independent paramagnetism of the metals, might be interpreted semi-quantitatively by representing the conduction electrons --- free to move inside the metal --- as a `Fermi gas' of free particles and  demonstrated that the electron gas in a typical metal is highly degenerate.\footnote{Dirac derived the general theory of the behavior of quantum particles including both Fermi's result (which he apparently did not know about) and the Bose-Einstein result as special cases of his general theory. But Fermi's more `physical' approach, discussing the problem of the quantization of a monoatomic ideal gas in a harmonic trap, probably explains why  the expression `Fermi gas' --- and not Fermi-Dirac gas --- is since then generally used in referring to an ensemble of a large number of fermions.} 

In  December of that same 1926, while Pauli was submitting his pivotal contribution applying the Fermi-Dirac statistics to metals, the only form of dense matter known on Earth, Ralph Howard Fowler's paper `On Dense matter', actually the \textit{very first} application of the new statistics, discussing a degenerate gas of electrons in white dwarf stars, had already appeared in the 10 December issue of the \textit{Monthly Notices of the Royal Astronomical Society} \cite{Fowler1926a}. 

Fowler's interest in the quantum theory and in the applications of physical ideas to the theory of valence, made him especially enthusiastic of the new quantum mechanics and its application to various areas of mathematical physics.\footnote{After the great war he had started working on quantum theory, the kinetic theory of gases and in particular statistical mechanics, a field in which he made remarkable contributions. During his collaboration with Charles G. Darwin between 1922 and 1923, Fowler developed new methods in statistical mechanics that were later also applied to deal with the equilibrium states of ionized gases at high temperature.} 
Fowler's early experience in problems of the behavior of solutions of second-order differential equations was at the root of his investigations with Edward A. Milne on stellar structure and the application of kinetic theory and statistical mechanics to stellar atmospheres \cite{FowlerMilne1923}. The statistical-mechanical investigations continued with further papers on the absorption lines in stellar spectra and on the ionization in stellar interiors written alone or in collaboration, like the seminal studies of gases in stars \cite{FowlerGuggenheim1925a,FowlerGuggenheim1925b}.

 By this time, Fowler's studies of gases in stars, matched with his deep knowledge of statistical mechanics \cite{Milne1945}, had fully set the stage for his interest in what appeared to be very peculiar stellar objects that had puzzled astronomers since many years \cite{Holberg2009}.
 
Since 1862, the astronomer Alvan Clark, Jr. had been able to see the companion of Sirius, a very faint star, almost exactly ten thousand times fainter than Sirius itself, whose existence had already been discovered many years earlier, only through its gravitational influence.  By 1910, reliable data showed that the faint companion, Sirius B, had a mass equal to 0.96 of that of the sun ($0.96 M_\odot$), but was 400 times less luminous. ``Nothing unusual thus far, but then came the bombshell'' recalled Willem J. Luyten many years later  \cite[p. 30]{Luyten1960}. From 1921 Luyten began a systematic general survey of the whole sky to search for white dwarfs, and it appears that he was the first to use the term that was subsequently popularized by Arthur Eddington \cite{Holberg2009}.

 In 1915 the American astronomer Walter Adams, an expert in stellar spectra, was able to secure the spectrum of the faint companion of Sirius A: ``The great mass of the star, equal to that of the Sun and about one-half that of Sirius, and its low luminosity,  one-hundredth part of that of the Sun and one ten-thousandth part of that of Sirius, make the character of its spectrum a matter of exceptional interest'' \cite{Adams1915}. The spectrum of Sirius B was quite puzzling: contrary to every expectation, Sirius B was white, in spite of its very low intrinsic brightness. Its spectrum was not very different from that of Sirius A.\footnote{At that time, many dwarf, faint stars were already known, but they were all red.}  

In 1924 Arthur S. Eddington, the most influential astrophysicist of his time, brought these remarkable properties to the attention of the astronomical world. At the time, the conventional wisdom was that  equilibrium against gravitational collapse was  maintained in all stars  by the internal pressure of the matter composing the star which had been heated into a gas, presumably by `subatomic energy', as Eddington pointed out. He had actually been one of the first to put forward such hypothesis.
In discussing the relation between the masses and luminosities of the stars \cite{Eddington1924a}, Eddington dedicated a specific section to white dwarfs, that  ``have long presented a difficult problem''.     Eddington then synthetized his views in an article sent in parallel to \textit{Nature} \cite{Eddington1924b} in which he emphasized the importance of giant stars and white dwarfs as objects apparently escaping the standard laws that at that time allowed the construction of the Hertzsprung-Russell diagram, which gave the relationship between luminosity and surface temperature of a star, and according to which all stars appeared to be arranged in a practically continuous sequence. 
 
An ordinary gas becomes comparatively incompressible at high density because of the finite volume occupied by its atoms or molecules. 
However, argued Eddington, atoms are mainly empty space  \cite[p. 787]{Eddington1924b}: ``at the high temperature within a star these sphere are completely destroyed, and this limit to the compression disappears. The stellar atom is highly ionized, and the peripheral electrons which determine its effective size have been detached [\dots] the ions, or broken atoms, can be packed much more tightly [\dots] There might thus exist stars far more dense than any material yet known to us. This may be the key to a puzzle presented by the companion of Sirius and a few other stars known as `white dwarfs' ''. As he himself mentioned in \cite[p. 10]{Eddington1926}, ``it had been suggested to him independently by Newall, Jeans and Lindemann that in stellar conditions the atoms themselves would break up to a considerable degree, many of the satellite electrons being detached''.\footnote{The problem was thoroughly discussed in \cite[pp. 165-172]{Eddington1926}.} 
The conclusion was that the deduced very high density, according to the views he had presented, should not be accepted ``as absurd''. 
 
 Eddington, who was already well known for his commitment to Einstein's general theory of relativity, immediately added that it seemed unnecessary to debate the proposed alternatives at length, because, as several writers had pointed out, ``the question could probably be settled by measuring the Einstein shift of the spectrum'' for which Eddington proposed a value of about 20 km per second, ``if the high density is correct''. 

 Within a year, Adams, following Eddington's request, had carried out careful spectroscopic observations with the 100-inch telescope and measured the redshift. He found that, after allowance was made for the relative orbital motion of the two stars, the observed displacement was 19 km/s \cite{Adams1925a} \cite[p. 387]{Adams1925b}: ``The results may be considered, therefore, as affording direct evidence from stellar spectra for the validity of the third test of the theory of general relativity, and for the remarkable densities predicted by Eddington for the dwarf stars of early type of spectrum''. 
 Eddington commented \cite[p. 173]{Eddington1926}:  ``This observation is so important that I do not like to accept it too hastily until the spectroscopic experts have had full time to criticize or challenge it; but so far as I know it seems entirely dependable. If so, Prof. Adams has killed two birds with one stone; he has carried out a new test of Einstein's general theory of relativity and he has confirmed our suspicion that matter 2000 times denser than platinum is not only possible, but is actually present in the universe''.  According to the astronomer Henry Norris Russell, this remarkable result was marking ``a very definite advance in our knowledge of both the foundations of science and the constitution of matter'' \cite{Russell1925a},\footnote{See also Russell article `Remarkable new tests favor the Einstein theory' \cite{Russell1925b}.} and  Hans Thirring considered this effect a new, useful tool for the astrophysicist \cite{Thirring1926}.
 
 Adam's measurement, which would be strongly revised at the end of the 1960s \cite{GreensteinEtAl1971,Holberg2010}, at the moment provided an evidence for the extreme compression of stellar matter, as emphasized by Eddington, and made clear that the existence of stars having the extraordinary qualities of what were by 1924 cited in the literature as \textit{white dwarfs}, not only was removing the necessity of confining the Einstein test to the sun, but was establishing for the first time a connection between general relativity and compact objects lying light years away, well beyond the solar system. But this fundamental thread of our story remained suspended and isolated for a long time. 
  
 Eddington's major monograph  \textit{The Internal Constitution of the Stars}, which concluded and summarized the results obtained during the first quarter of our century, was published in 1926 \cite{Eddington1926}. Great progress had been made in the preceding years in the study of stellar interiors. The fundamental equations governing the structure of a star in radiative equilibrium had been established, and the role of ionization in determining the properties of interior stellar matter had been clearly recognized. Eddington's `standard model' of stellar structure  based on stars for which the perfect-gas law held and energy transport via radiation prevailed, yielded information on temperature and density in the interior of main-sequence stars and it was realized that the ideal-gas equation of state was a good approximation for all these stars.
 
 Eddington dedicated a large discussion  to white dwarfs. The extremely high density of the companion of Sirius A had been confirmed by Adams -- but the puzzle remained. He was in fact uneasy as to what would ultimately happen to these superdense stars: ``I do not see how a star which has once got into this compressed condition is ever going to get out of it. So far as we know, the close packing of matter is only possible so long as the temperature is great enough to ionise the material. When the star cools down and regains the normal density ordinarily associated with solids, it must expand and do work against gravity. \textit{The star will need energy in order to cool}''. At zero temperature all random motion should cease, according to ideas generally accepted up to 1926. In a cold star, nothing should prevent electrons and nuclei from recombining. The star would need to expand 10,000-fold to accommodate the volume of its neutral atoms. Where would it find the energy to do this? Its available fuel has been exhausted and it has no other resources: ``We can scarcely credit the star with sufficient foresight to retain more than 90 per cent. in reserve for the difficulty awaiting it. It would seem that the star will be in an awkward predicament when its supply of sub-atomic energy ultimately fails. Imagine a body continually losing heat but with insufficient energy to grow cold!'' concluded Eddington \cite[p. 172]{Eddington1926}.

Eddington had already remarked in 1922, during his Royal Astronomical Society Centenary address, that ``Strange objects which persist in showing a type of spectrum entirely out of keeping with their luminosity, may ultimately teach us more than a host which radiate according to rule''   \cite{Eddington1922}. But what was the meaning of the apparently `nonsensical message' \cite[p. 48]{Eddington1927} coming from the companion of Sirius? Eddington had actually materialized a veil, whose corner was lifted by Ralph Fowler, who promptly responded, taking up the challenge and addressing the brand new astrophysical problem related to the nature of such dense stars.

Eddington's book \textit{The internal constitution of stars} had been written between May 1924 and November 1925. As explained in the Preface dated July 1926, Eddington worked on the proofs up to March, and Fowler himself probably read at least parts of the volume in a preliminary stage. At the end of the preface Eddington is acknowledging him as his ``referee in difficulties over points of theoretical physics'' while Milne had read the proof sheets. Eddington, Milne and Fowler,  were member of a small circle of very influential scientists, with strong common interests, all working in Cambridge. Their scientific discussions stimulated Fowler to apply  quantum mechanics to the white dwarf problem raised by Eddington, whose enigma perfectly matched the advent of quantum mechanics and the Pauli exclusion principle.  

Fowler's own  interests and publications of 1926 are self explanatory in this sense. In April 1926 he had published  `The statistical mechanics of assemblies of ionized atoms and electrons', a detailed theoretical analysis in terms of electrons and positive nuclei that allowed him to tackle the properties of matter at the temperatures and densities occurring in stars \cite{Fowler1926b}. On August 26, 1926, Dirac's paper containing the Fermi-Dirac distribution was communicated by Fowler to the Royal Society \cite{Dirac1926}.\footnote{At that time, P. A. M. Dirac was Fowler's research student. Working under his influence, Dirac wrote papers on quantum theory and they also collaborated writing several articles in the period 1924-1926. Fowler's great commitment to the new quantum mechanics, testified by his scientific production of those years, was instrumental also in the early formation of Robert J. Oppenheimer, who was in Cambridge during the period 1925-1926. See for example his first paper related to his sojourn in Cambridge, entitled `On the Quantum Theory of the Problem of the Two Bodies' presented in July 1926 by Fowler at the Cambridge Philosophical Society \cite{Oppenheimer1926}. Apart from thanking Fowler and Dirac, Oppenheimer collaborated with Fowler in two papers published in that same 1926.} 
On November 3, Fowler presented his own work to the Royal Society in which he systematically worked out the quantum statistics of identical particles, exploring the relationship between statistical mechanics and the new quantum mechanics, especially in connection with the Fermi-Dirac statistics \cite{Fowler1926c}. After having thoroughly delved into the question, Fowler could thus devote his attention to Eddington's paradox, ``\textit{A star will need energy to cool}'' and he tackled the problem in a most general perspective. Fowler reformulated the paradox posed by Eddington in clearer physical terms and resolved it introducing the notion of electron degeneracy for the first time. At the temperatures and densities that may be expected to prevail in the interiors of the white-dwarf stars, the electrons will be highly degenerate and all the available parts of the phase space with momenta less than the Fermi threshold are occupied, consistently with the Pauli exclusion principle. They fill all the energy levels, exactly like the electrons in an atom on the Earth. Therefore, the total kinetic energy evaluated according to such distribution will be about two to four times the negative potential-energy and Eddington's paradox does not arise \cite[p. 115]{Fowler1926a}: ``The apparent difficulty was due to the use of a wrong correlation between energy and temperature, suggested by classical statistical mechanics''.  

In his classical monumental volume \textit{Statistical Mechanics} Fowler well described the ``absolute final state'' --- which he named \textit{black-dwarf stage} --- in which there is only one possible configuration left, when temperature ceases to have any meaning, and the pressure of the fully degenerate electron gas is large enough to balance the weight of the stellar layers attempting to collapse inward due to the gravitational pull \cite[p 552]{Fowler1929}: ``As these stars go on radiating they will if anything condense still further and ultimately may well lose all their superfluous energy and fall to zero temperature. We may perhaps venture to refer to their probable final state as the black dwarf stages [\dots] The black-dwarf material is best likened to a single gigantic molecule in its lowest quantum state. On the Fermi-Dirac statistics, its high density can be achieved in one and only one way, in virtue of a correspondingly great energy content. But this energy can no more be expended in radiation than the energy of a normal atom or molecule. The only difference between black-dwarf matter and a normal molecule is that the molecule can exist in a free state while the black-dwarf matter can only so exist under very high external pressure''. 

 Fowler's 1926 paper constituted a major breakthrough in astrophysical theory and would become one of the great landmark works in the realm of stellar structure.  It was the first demonstration that the new quantum statistics could explain an important property of bulk matter and at the same time, in accounting in a general way for the observed characteristics of white-dwarf stars, it was a clear-cut example of the solution of an astrophysical problem depending upon features in which quantum mechanics differs essentially from any previous theory. 
 
In mentioning white dwarfs and difficulties in the problem of stellar evolution, Eddington concluded his small volume \textit{Stars and Atoms} expressing in the preface the feeling that the whole difficulty seemed to have been removed by R. H. Fowler's investigations, but cautiously adding that ``there is something of fundamental importance that remains undiscovered''. The very last words of the volume were dedicated to what was believed to be the final state of the white dwarf and perhaps therefore of every star: ``If any stars have reached state No. 1 they are invisible; like atoms in the normal (lowest) state they give no light. The binding of the atom which defies the classical conception of forces has extended to cover the star. I little imagined when this survey of Stars and Atoms was begun that it would end with a glimpse of a \textit{Star-Atom} [emphasis added]''.  Eddington could not imagine how prescient he was in saying that ``white dwarfs appeared to be a happy hunting ground for the most revolutionary developments of theoretical physics''  \cite[p. 125]{Eddington1927}.   It was only the beginning.  
Further developments in the study of dense matter, the emergence of modern nuclear age as well as a new generation of scientists, would set the stage for modern challenging theoretical questions stemming from this new state of matter.

\section{Metals and star interiors: Yacov Ilich Frenkel}

 Up to 1924, no one had given serious thought to abnormally dense matter. It was a remarkable coincidence that just at the time when matter of exceedingly great density was discovered in astronomy, physicists were developing the tools to tackle this subject. The idea of electrons free to move, but subject as an ensemble to the laws of quantum statistics,  contained such a basic concept that  was independently applied to electrons in metals.  In seeing the proofs of Pauli's paper in spring 1927 \cite{Pauli1927}, Arnold Sommerfeld, who well knew the problems of the classical electron theory of metals, was so impressed that he said that ``one should make further application to other parts of metal theory''.\footnote{W. Pauli to F. Rasetti, October 30, 1956. The Pauli Letter Collection, CERN Archives. \label{PauliRasettiLetter}
 As recalled in \cite[p. 102]{HoddesonEtAl1992}: ``Sommerfeld liked the electron theory of metals well enough to make it a theme at his institute. In the summer of 1927, he lectured in his special-topics course on the `structure of matter' to a small circle of advanced students, showing basic consequences of the application of Fermi-Dirac statistics to the electron gas [\dots] Electrons in metals were the main concern in Sommerfeld's theoretical research seminar in the winter 1927/1928''. }
 Introducing the idea that the free electrons in a metal constitute a Fermi gas, he was in fact able to explain the heat capacity catastrophe within the framework of Fermi-Dirac statistics. 
 
Sommerfeld presented his theory at the International Volta Congress, held in September 1927 in Como, Italy, \cite{Sommerfeld1928a} \cite{Sommerfeld1928b}. As Pauli later recalled:  ``I met Fermi personally the first time at the Volta-congress in Como, 1927 [\dots] Heisenberg introduced us with the words `May I introduce the applications of the exclusion principle to each other', or with some similar joke''.\footnote{Pauli to Rasetti, see footnote \ref{PauliRasettiLetter}.}
Following his talk Hendrik A. Lorentz, Enrico Fermi, Edwin Hall, and in particular Yacov Ilich Frenkel, participated in the discussion. Having worked since 1924 on the theory of metals, on which Frenkel was considered an authority, he had been invited to the Volta Memorial Conference, where he delivered a paper on the theory of metals in which he first formulated the main premises of quantum theory of electric conductivity \cite{Frenkel1928}.\footnote{Frenkel worked at the Leningrad Physico-Technical Institute on topics connected with problems of the structure of matter --- especially solid and liquid bodies. Between 1922 and 1924 he had published \textit{The Structure of Matter}, a complete theoretical analysis of the field, and in 1924 the \textit{Electron Theory of Solids}, which later served as a basis for further original work. In particular, he also published papers on the theory of electric conductivity of metals and on the electron theory of solids, being considered one of the outstanding physicists of the Soviet Union.  He spent in the period 1925-1926 in Germany, at a time coincident with the foundation of quantum mechanics. During  1927-1929 he published mostly on the theory of metals.
See for example  \cite{Frenkel1928a,Frenkel1928c}.} 
 In March 1928, Frenkel continued by letter the discussion with Sommerfeld begun at the Volta Congress \cite[p. 130]{Frenkel1966} and in mid June he submitted his article in which he applied the `Pauli-Fermi' electron gas theory to the problem of the cohesive forces \cite{Frenkel1928b}.\footnote{The details of this pioneering work, in which Frenkel cites his previous research on metal theory, are discussed in \cite{Yakovlev1994}.}

The atoms in metals lose their last one or two electrons, and these are free to move inside the metal. This problem was quite similar to atoms in white dwarf matter: electrons stripped from atoms were free to move between the compressed nuclei over the entire star. Sommerfeld had now extended the classical electron theory of metals developed by Paul Drude and H. A. Lorentz including quantum statistics.\footnote{In this regard, see Sommerfeld and Bethe's review in the 1933 \textit{Handbuch der Physik} \cite{SommerfeldBethe1933} and \cite{HoddesonBaymEckert1987} for a reconstruction of the development of the quantum-mecanical electron theory of metals during the period 1928-1933.}
The great interest arisen around the quantum properties of the electron theory of metals led in a natural way many people --- some of whom were already working in the field --- to discuss dense matter in white dwarfs, which became a virtual laboratory to test theories on degenerate matter in more extreme realms. Frenkel, described by Peierls as ``a man of great versatility and originality'' \cite[p. 318]{Peierls1996},  wrote in this regard an article on the application of the Pauli-Fermi electron gas theory to the problem of the cohesive forces, whose importance for the theory of white dwarfs remained almost unknown to astrophysicists. The fourth, and last section of the paper is entitled `Superdense stars' \cite[p. 244]{Frenkel1928b}. Frenkel never uses the term `white dwarf', probably because he wants to present the theory in a more general `physical' sense, not necessarily connected with specific astrophysical objects. Moreover, he does not cite \cite{Fowler1926a}. In his third section, specifically dedicated to the Thomas-Fermi atom, Frenkel had attempted to transfer the statistical atom model to the nucleus regarded as a sphere filled with \textit{protons and electrons} (following the spread general view on nuclear models of that time) and reasoned that the electrons inside the nuclear volume were a strongly compressed gas. As a second step Frenkel transferred this analogy to white dwarfs, assuming them to be essentially homogeneous  spheres consisting of a ``mixture of electrons and ions gases'', that he called  ``Kerngas'' [nuclear gas]. It is then remarkable that, basing on  such a ``bad'' model of the nucleus, he was nevertheless able to get a good representation of white dwarf matter calculating the relationship among the mass, radius, and density of the star.
   
  As others would do in the following months, Frenkel remarked that the strongly compressed electron gas becomes relativistic. He stressed how from condensation of a metal vapour, metallic matter is obtained that can be considered as a `single gigantic molecule' in which outer electrons are no more bound to single atoms, but are forming a `gas system'. A further compression would set all electrons free. It is not possible to get such conditions in an Earth laboratory, hence Frenkel is naturally led to reason on dense stars, with density $\rho =10^6 g/cm^{3}$: ``Such pressure can actually exist only inside stars with a sufficiently large mass''. 
  
  Frenkel's investigations not only show how quickly stars were becoming a  testing ground for theoretical physicists, but testify the ongoing transition of interest from the outer atomic layers to the nuclear realm. The method of analogies, as a search for connection among different physical contexts, was typical for Frenkel. He applied these considerations to reason on the superdense matter inside heavy nuclei treating them as solid bodies, a model having in turn a strong analogy with Gamow's  coeval treatment of nuclei as an assembly of $\alpha$-particles ``treated somewhat as a small drops of water in which the particles are held together by surface tension'' \cite[p. 386]{Gamow1929,Gamow1930}.

  Frenkel's remarkable article  also discusses the existence of two types of superdense stars, consisting of non relativistic and ultrarelativistic electron gas and correctly estimated that the mass of a stable star, which is in a relativistic degenerate state, cannot exceed a definite maximum, $M\geq M_\odot$, somewhat larger than the mass of the sun. This really unexpected result  went completely unnoticed at the moment.

  \section{First debate on dense matter in stars}
 
 Fowler's work had already led to a first qualitative understanding of the structure of white dwarfs, but a quantitative theory was still needed. Like Frenkel, others used the Fermi gas model to calculate the relationship among the mass, radius, and density of the white dwarf stars, assuming them to be essentially homogeneous spheres of electron gas.

 The fundamental question of the degeneracy of electrons inside stars was also discussed by the German astrophysicist Wilhelm Anderson, working at Tartu University in Estonia, in an article submitted to the \textit{Zeischrift f\"ur Physik} in July 1928 \cite{Anderson1928a}. In the last part he mentioned an hypothesis put forward by the Australian physicist Kerr Grant, who had quickly reacted to Eddington's discussion on the unusual density of white dwarfs proposing that the mean density at the centre of the star could even be ``fifty' million'' times the mean density, instead of only ``fifty'' times according to Eddington's guess \cite{Grant1926}. Based on the assumption that the properties of stellar material do not vary in a continuous manner from the star's surface to its centre, Grant also suggested a \textit{central core} in which formation of heavy elements could take place with conversion of matter into radiation.

Anderson then submitted a second paper at the end of December \cite{Anderson1928b} following an article by  the Soviet physicist Georgii Pokrowski \cite{Pokrowski1928}, who had put forward a theory according to which ``the mass of a star must have a maximum value'' that would be obtained when the nuclei of completely ionized atoms touched each other and had  estimated this density to be $4\times10^{13\pm 1}g/cm^{3}$. Provided that the nuclei could not be compressed, this should be the maximum density that matter could be in, that is the state of nuclear matter.\footnote{According to Pokrowski, when a star having the limiting mass is also reaching its maximum density  the gravitational potential at  surface must have the critical value  $\phi= c^2$. ``In this case, recalled Anderson, no energy can leave the surface of the star, because  to remove the mass of a quantum of energy $\frac{h\nu}{c^2}$ a work of exactly $\frac{h \nu \phi}{c^2}= \frac{h\nu c^2}{c^2} = h\nu$ would be required \cite[p. 389]{Anderson1929a}.}

In a subsequent article, Anderson continued the analysis of Pokrowski's theory extending these considerations to the whole mass of the star  \cite{Anderson1929a}. His opinion was that the highest possible density, of the order of $10^{13} g/cm^{3}$ could be only reached only under a very high pressure, that is in a central ``core'' of a star where the main part of the mass should be concentrated. This meant that it could happen not only within white dwarfs, but also in giant stars. In this case the idea of gravitational contraction as a source of energy for the star could not be a surprising possibility. 

Anderson did not mention Frenkel's contribution that had already appeared \cite{Frenkel1928b}, but an added note at the end cited an article by  Edmund Clifton Stoner \cite{Stoner1929} discussing the limiting density in white dwarf stars. Anderson correctly remarked that in Stoner's formula from which the maximum possible density in a star could be calculated based on its mass, the latter had ``ignored the variability of the mass of the electron'' attaining velocities of the order of the velocity of light for stars having masses of the order of the sun and announced a new article meant to discuss Stoner's theory.

Stoner, who had been working at the Cavendish Laboratory since the early 1920s, developed theoretical interests  encouraged by Fowler, and in 1924 published the already mentioned work on the distribution of electrons among atomic levels, a problem of topical interest to chemists as well as physicists. It attracted much attention thanks to Sommerfeld, who mentioned it as ``a great advancement''  in the preface of the fourth edition of his classic book, \textit{Atomic Structure and Spectral Lines}.\footnote{See Stoner's biography \cite{Bates1969}, especially p. 214, where it is reported how the scheme came to Stoner's mind and how he wrote a note that he submitted to Rutherford, who in turn passed it to Fowler, with whom Stoner had at that time ``several most helpful discussion on theoretical points''. Fowler was so impressed that he asked him to write a full and detailed paper about it. As already mentioned, Stoner's article contained a scheme for the treatment of electron distribution derived from experimental data, that actually had a strong impact on Pauli, and influenced his formulation of  the explicit statement that later became known as the Pauli exclusion principle.} 
  
Later Stoner developed a strong interest in magnetism and made a relevant contribution by introducing quantum ideas in the elucidation of the magnetic behaviour of matter, and did pioneering work on magnetism and the application of Fermi-Dirac statistics to the theory of para-and ferromagnetic phenomena. But his years in Cambridge --- where eminent astrophysicists like Eddington and Milne worked --- and of course Fowler's relevant paper, sparkled his interest on astrophysical topics, which became his lifelong interest. 

 Stoner wrote a paper to investigate ``the question as to whether there is a limit to electron `congestion' [\dots] under the gravitational conditions in the stars'' \cite[pp. 64-65]{Stoner1929}. He was inspired by Eddington and Fowler, but especially by James Jeans's ideas about the departure from the ideal gas laws in some stellar interiors behaving ``as if in a `quasi-liquid' condition owing to the congestion of the atoms'' \cite{Jeans1927},

He started  from Fowler's idea that  white dwarfs are supported by electron degeneracy pressure but went further, discussing under simplifying assumption whether there might be a limiting density ``due to the `jamming' of the electrons (owing to the exclusion principle which forms the basis of the Fermi statistics)''. He modeled the star as a sphere of uniform density of material composed of completely ionized atoms: ``the density increases as the sphere shrinks, and the limit will be reached when the gravitational energy released just supplies the energy required to squeeze the electrons closer together. The limiting case of high density occurs when the effective temperature is zero''. In these calculations he followed Fowler in neglecting the electrostatic potential energy and considered only the kinetic energy in the degenerate electron gas and the gravitational energy of the star as given by Eddington \cite[p. 87]{Eddington1926}. He also neglected the kinetic energy of the nuclei, which is small, and obtained a limiting density of electrons, $n=9.46\times 10^{29} (M/M_\odot)^{2} cm ^{-3}$, in which M is the mass of the star in question and $M_\odot$ is the mass of the sun. 
He found an expression for the maximum density of a star of mass $M$, consisting of a mixture of fully-ionized atoms, approximately given by $\rho=3.85\times 10^{6} (M/M_\odot)^{2} kg m^{-3}$, giving a value for the mean density of Sirius B in fairly good agreement with the modern value.
The concept of a superdense `core' within stars already materialized by Grant and Anderson, was taken up by Stoner who concluded suggesting that ``white dwarfs contain a core of material approaching the limiting density'' being in an ``almost incompressible or `quasi-liquid' state, due to the `congestion' of the electrons''. 

Such limiting density, observed Y\^usuke Hagihara, a Japanese astronomer who had also studied in Cambridge under Eddington and Henry F. Baker during the 1920s, was considerably lower than a density of the order of $10^{17}$, corresponding to the density reached by a star like the sun, whose volume had been reduced to a radius ``equal to its Schwarzschild singularity $\left(\frac{2GM_\odot}{c^2}\right)$'', that is ``a few kilometers''. ``The most reasonable explanation'' for this reassuring value, confirmed by the observations, would be ``that this is the limit of the relativistically possible density''. Hagihara also cited \cite{Pokrowski1928}, who had given a value of the order of $10^{13}$ \cite[pp. 107]{Hagihara1931}. In his `Theory of the Relativistic Trajectories in a Gravitational Field of Schwarzschild', he thus emphasized that the solutions to the motion inside the circle corresponding to the Schwarzschild radius ``is inadmissible from the principle of relativity'', being ``quite improbable that in any star the distance $r=\alpha$ or $2m$ from the center lies outside its radius'' and that ``the statement that a very massive star can entirely absorb the light emitted from its surface and never be seen from outside, is quite fallacious'' \cite[pp. 173-174]{Hagihara1931}.

 It  seemed for a while that the white-dwarf stage --- or rather the `black-dwarf' stage as Fowler described it --- represented the last stage of stellar evolution for all stars and thus their density appeared both from a theoretical and a physical point of view a limiting density. Moreover, since a finite state seemed possible for any assigned mass, one could rest with the comfortable assurance that all stars would have the `necessary energy to cool', according to Eddington's expression. But this assurance was soon broken when it was realized that the electrons in the centers of degenerate masses begin to have momenta comparable to $m_{e}c$ and the electron gas must thus be treated relativistically.

As already remarked, Anderson immediately reacted to Stoner's paper criticizing calculations in which Stoner had used the rest-mass for the mass of electrons,  and demonstrated that as the density increases the degenerate  electrons in the centers of  white dwarf stars comparable to or higher than the mass of the Sun, begin to attain velocities on the order of the velocity of light and that in this case the variation of the electron mass with velocity must be taken into account by using the equations of special relativity. He thus concluded that Stoner's assumptions led to ``gr\"oblich falschen Resultaten'' [gross false results] in the case of white dwarfs having a mass comparable to the mass of our sun \cite[p. 852]{Anderson1929b}. His attempt to extend the equation of state of a degenerate electron gas to the relativistic domain was not correct, but it made the conceptual coupling of relativity and quantum statistical mechanics and indicated that Stoner's treatment implied a maximum value for the white dwarf mass. 

Stoner's response to Anderson arrived in a paper submitted in December 1929 \cite{Stoner1930}, where he worked out with more rigor the effect of the relativistic change of mass, still for the idealized case for a sphere of uniform density and formulating  the correct relativistic equation of state \cite{Nauenberg2008,Thomas2011}.
 Stoner calculated that ``For spheres of increasing mass the limiting density varies at first as the square of the mass, and then more rapidly, there being a limiting mass ($2.19 \times 10^{33}$ grams) [i.e. of the order of the sun's mass] above which the gravitational kinetic equilibrium considered will not occur'' \cite[p. 963]{Stoner1930}, thus confirming Anderson's unexpected result of a critical mass for white dwarfs. On p. 952 of Stoner's article a figure with curves showing variation of limiting electron concentration with mass appears, comparing Anderson's and Stoner's results with the straight line which is obtained when special relativity is neglected. This figure is clearly showing  that a limiting mass is obtained when the crucial role of special relativity is considered. In the following page he commented that ``The number of stars known to be of the white dwarf type is small, but this does not necessarily indicate that stars of very high density are uncommon. Dense stars of ordinary mass will have a small radius, and so will be faint objects [\dots] `Black dwarfs' (to use Fowler's term) would not be observed''. 
 
 Neither Stoner, nor Anderson speculated in these papers about what might be the fate of more massive stars. Stoner simply noted that ``gravitational kinetic equilibrium will not occur'' \cite[p. 963]{Stoner1930}. In 1936 Anderson then published his habilitation thesis: \textit{Existiert eine obere Grenze f\"ur die Dichte der Materie und Energie?} [Does it exist an upper limit for the density of matter and energy?] \cite{Anderson1936}.

\begin{figure}[h]
\centering

\resizebox{0.70\columnwidth}{!}{%
  \includegraphics{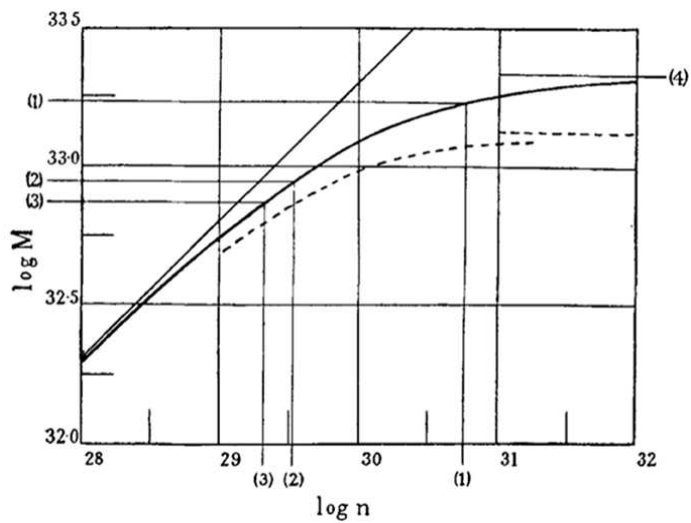} }
\caption{Variation of limiting electron concentration ($n$) with mass (M) in a sphere of uniform density. The points (1), (2), (3), (4) correspond  to Sirius B, $o_2$ Eridani B, Procyon B, and the limiting mass M ($2.19 \times 10^{33}$) \cite[p. 952]{Stoner1930}.}
\label{Stoner1930}     
\end{figure}

\section{Edward Milne and the idea of a condensed core in stars}

The Stoner-Anderson debate on the structure of white dwarfs was someway embedded in a hot controversy between Milne and  Eddington about the problem of stellar structure and the source of energy in stars, in which several other people became involved. At the end of 1929, Milne  presented his investigations on the relation between the masses, luminosities, and effective temperatures of the stars from a standpoint which was ``philosophically different'' from that adopted by Eddington \cite[p. 17]{Milne1929}. Milne criticized Eddington's theory on the ground that stars with a point source of energy, and a point concentration of mass, at the center (or a reasonable physical approximation to this arrangement) would be more stable than Eddington's models; in such stars the central temperatures would be very high indeed.\footnote{For an excellent detailed discussion on the Milne-Eddington controversy and for a comprehensive history of theories of stellar structure see \cite{Shaviv2009}.}
 
On the other hand, underscored Milne \cite[p. 16]{Milne1930a}, ``The theory of Sir Arthur Eddington does not claim to account for the observed division of stars into dense stars and stars of ordinary density, nor does it establish the division of ordinary stars into giants and dwarfs. On the other hand, it claims to establish what is known as the mass-luminosity law from considerations of equilibrium only, that is, without introducing anything connected with the physics of the generation of energy''. Inspired by dense matter in white dwarfs --- and by the Grant-Anderson-Stoner discussions on a dense core --- Milne's investigations materialized in attempts to construct stellar models by using the properties of degenerate matter and by introducing the idea of collapsed models, too massive for their gas pressure to support their mass if perfect-gas conditions prevailed, and of centrally-condensed models, whose density and temperature rose to enormous values at the centre \cite{Milne1930a} \cite{Milne1930b}.

Previous more vague ideas of condensed cores within stars  already mentioned were now acquiring a definite status in Milne's theory and were meant to play a role in the constitution of \textit{all} stars.

 In the opening lines of his `Analysis of stellar structure' \cite{Milne1930a}, Milne stressed how, according to the then current theory (by which he meant that of which the researches of Eddington were the basis), the ordinary stars (giants and dwarfs) were considered masses of perfect gas with central temperatures of the order of $10^7$ degrees and with central densities of the order of 50 times the mean density. However,  stressed Milne, the current theory  failed entirely to account for `white dwarfs', so that he wanted to show in the paper that ``a perfect-gas star in a steady state is in nature an impossibility, and that actual stars must either possess a small but massive core of exceedingly high density and temperature, or else must be almost wholly (that is, save for a gaseous fringe) at a very high density'' \cite[p. 4]{Milne1930a}. According to Milne, this appeared to be a fundamental property of steady-state configurations, and it corresponded to the observed division of the stars into ``ordinary stars'' (giants and dwarfs) and ``white dwarfs''. ``This result, added Milne, was not a consequence of any special hypothesis, but was flowing naturally from the method of analysis: ``The division of `ordinary stars'  into `giants'  and `dwarfs' would appear to be less fundamental and not to indicate any special difference of structure''. 
 
 According to their luminosity they must be either `centrally-condensed' or `collapsed'. In such centrally condensed models, with super high temperatures and density, the perfect-gas law could not hold. Actually, for the first time, the term `collapse' in the context of `collapsed configurations' that would become of common usage in the future, was first introduced in astrophysics by Milne in this theory. 
 
Milne's `Analysis' paper had been preceded by another on the origin of stellar energy and the mechanism of its evolution \cite{Milne1930c} in which he located the source of stellar energy ``In the intensely hot, intensely dense nucleus where the temperatures and densities are high enough for the transformation of matter into radiation to take place with ease'', very provocatively stressing how ``The consequences [of his investigations] amount to a complete revolution in our picture of the internal constitution of the stars''. His ideas about the source of energy in stars were mixing on one side the hypothesis of matter annihilation, and on the other the possibility of a ``synthesis or radioactive elements'' thanks to the extremely high temperatures ``of the order of $10^{10}$ degrees or higher'' that were ``to be assigned to the central condensation''.\footnote{Milne went back to this problem in a new article with a quite similar title \cite{Milne1931a}.}

  Milne's papers on the analysis of stellar structure aroused a large and vigorous debate. These issues, that had for long already been a subject of disagreement between James Jeans and Arthur Eddington, entered  a new phase through the work of Milne, whose views aroused a great interest because of the novelty underlying his proposal. The whole of the meeting of the Royal Astronomical Society of January 9, 1931, was dedicated to a general discussion on such fundamental topics, but Milne's ideas on the composite model for stars were not generally accepted. Milne's model was challenged especially by Norris Russell, Thomas Cowling and Bengt Str\"omgren. The latter reported the main critics in  \cite{Stromgren1931}.

 ``The irony of all this'', as emphasized in \cite[p. 238]{Shaviv2009},  ``is that Eddington thought at the beginning that his theory explained the gaseous giant stars and not the dwarf (main sequence) stars. As it turned out, Eddington's theory explains the dwarfs and Milne's theory explains the giant stars''.

However, at the threshold of the nuclear era, these animated discussions among astrophysicists contributed to the spreading of ideas about the ``core'' as a brand new physical object made of `nuclear' dense matter within stars, ready for more specific ``physical'' investigations. 
  
  \section{Chandrasekhar enters the lions' den} 

 While Anderson and Stoner were publishing the results of their work about the relativistic effects on electron degeneracy in white dwarfs, the 19-years-old Subrahmanyan Chandrasekhar,  known throughout his life as Chandra in the scientific world, was traveling on a ship from Bombay to Europe, determined to study and carry on research under Fowler at Cambridge. 
   As a student at Presidency College from 1925, Chandra found a growing liking for physics and mathematics and an ongoing attraction to English literature.\footnote{For biographical portraits of Chandra see \cite{Wali1990} and \cite{Parker1997}.} 
    In the autumn of 1928, during his trip around the world, Sommerfeld visited India and lectured at Presidency College in Madras. Chandra made it a point to meet Sommerfeld, from whose book \textit{Atombau und Spektrallinien} [\textit{Atomic Structure and Spectral Lines}] he had learned quantum theory:  ``I saw in the newspapers that he was going to be there, and so I went and saw him in the hotel [\dots] He gave me the copy of his papers on the electron theory of metals which were then in press, and his papers were clear enough for me to understand the Fermi statistics''. Chandra was taken aback to learn that the old Bohr quantum theory, on which Sommerfeld's book was based, was superseded ``by the discovery of wave mechanics by Schr\"odinger, and the new developments due to  Heisenberg, Dirac, Pauli and others'', and that the Pauli exclusion principle replaced Boltzmann statistics with Fermi-Dirac statistics \cite[p. 62]{Wali1990}.  The young student was someway shocked by such revolutionary news, but Sommerfeld offered him the galley proofs of his as-yet-unpublished paper on the new Fermi-Dirac quantum statistics and its application to the electron theory of metals.
   
   ``At about the same time, added Chandra, I read Eddington's \textit{Internal constitution of the stars}. It's quite readable. And it was the simultaneous knowledge of Eddington's \textit{Internal constitution of the stars}, together with modern statistics, at least modern as of then, through Sommerfeld, that turned my interest into the theory of white dwarfs and related matters''. \cite{Chandra1977}.  
  Glancing through the \textit{Monthly Notices of the Royal Astronomical Society} he found Fowler's paper containing still another application of the Fermi-Dirac statistics to the dense stellar matter in the form of degenerate electrons in white dwarfs: ``So it seemed to me that there was an area in which one could go right in. I could understand the Fermi statistics; I knew the theory of polytropes; I had read Fowler's paper; I could understand it. Right there, there was something which I could do. So that is  how I started'' \cite[p. 62]{Wali1990}. Within a few months he had enough mathematical preparation to understand the new statistics and was able to apply his knowledge to the problem of Compton scattering \cite{Chandra1929}. In January 1929 Chandra sent it to Ralph Fowler, whose monumental book \textit{Statistical Mechanics} had just come out. So Chandra's scientific career began with a series of papers on `the new statistics' published between 1929 and 1930, when he was only an eighteen-year-old undergraduate student.
During Chandra's final year at Presidency College, Werner Heisenberg went on a lecture tour and Chandra was in charge of Heisenberg's visit to the college and was entrusted with the responsibility of showing him around Madras. He had a wonderful occasion of discussing with him his papers and of increasing his expectations towards the idea of perfecting his studies in Europe. Even before completing his final examinations, Chandra was awarded the Government of India scholarship and on 31 July 1930, when he was only 19-years-old, he left Bombay on the steamer Pilsna, a liner of Lloyd Triestino travelling from Bombay to
Venice across the Arabic Sea, the Channel of Suez and the Mediterranean.
  
During his  travel towards Venice, from where he would travel by rail to London, he continued to work on a paper he had completed just before his departure. In it he had developed Fowler's theory of white dwarfs further, combining it with Eddington's mathematical model for an isolated mass of gaseous stellar material in equilibrium under its own gravitational forces, the so-called polytropic gas sphere, a crude approximation to more realistic stellar models with a simple relationship between pressure $p$ and density $\rho$: $p=K \rho^{1+1/n}$, where $n$, the polytropic index, and $K$ are constants that depend on the properties of the particles making up the gas. In working out the statistical mechanics of the degenerate high density electron gas at the center of the white dwarf, he realized, as Fowler had not, that the upper levels of the degenerate electron gas (which are those affected by changes in density and temperature) are relativistic. This meant that the pressure supporting the star against gravity grows no faster than the increasing gravitational force as the star contracts, in contrast with the familiar nonrelativistic situation where the pressure increases more rapidly than the gravitational forces ultimately providing a sufficient pressure to block further contraction. Chandra thus found that this limiting form of the equation of state had a dramatic effect on the predicted mass-radius relation: instead of predicting a finite radius for all masses, the theory was now predicting that the radius must tend to zero as a certain limiting mass is reached, above which the internal pressure of the white dwarf cannot support the star against collapse. 

The value of the limiting mass found by Chandra was $5.76\mu_e^{-2} M_\odot$ where $\mu_e$ denotes the mean molecular weight per electron. For the expected value $\mu_e=2$, the limit is 1.44 solar masses.\footnote{Both Stoner and Chandra assumed that the mean molecular weight of the gas is 2.5. This value that would soon have to be substantially reduced in the light of the evidence presented by H. N. Russell and B. Str\"omgren that stars contain large amounts of hydrogen, even if in the case of the white dwarf stage the hydrogen has been largely consumed.}
The existence of this limiting mass meant that a white-dwarf state does not exist for stars that are more massive.

This paper on the limiting mass was rather puzzling also for Fowler, and for this reason its publication was rather delayed, as recalled by Chandra \cite{Chandra1977}: ``I had written this paper in July; and I gave it to Fowler in September and he never did anything with it, whereas he sent my other paper to the \textit{Philosophical Magazine}. And fundamentally it is because neither Milne nor Fowler wanted to accept the fact that there was a maximum mass\dots''. It eventually appeared in July 1931 in the \textit{Astrophysical Journal},  published by the University of Chicago,  at the time much less important than the \textit{Monthly Notices of the Royal Astronomical Society} where Eddington and Milne generally published their work.

When he was writing his first paper, during the summer 1930, Chandra did not understand what the mass limit meant: ``I didn't know how it would end, and how it related to the 3/2 low-mass polytropes [\dots] I knew it must be significant, because Milne was working on the 3/2 polytropes at that time. He thought that every star must have a white dwarf core. And I couldn't see how that could be true''. But Chandra also recalled: ``I would say that I fully understood its implications by the end of 1930''. That is after having worked at his second and especially at his third paper on the theory of white dwarfs communicated by Milne himself at the Royal Astronomical Society \cite{Chandra1931c}.\footnote{His second paper on the subject, `The density of white dwarf stars', actually appeared in the Supplement of February 1931 to the \textit{Philosophical Magazine} \cite{Chandra1931b}. Now Chandra, who had become aware of Stoner's work \cite{Stoner1929} \cite{Stoner1930}, reconsiders the problem of the density from the point of view of the theory of polytropic gas spheres, deriving a formula for the mean density ``on considerations which are a much nearer approximation to the conditions \textit{actually existent} in a white dwarf''. He thus avoided  Stoner's assumption that the density is uniform throughout the star, but recognized that ``At any rate\dots the \textit{order of magnitude} of the density which one can on purely theoretical considerations attribute to a white dwarf is the same'' \cite[p. 595]{Chandra1931b}.}
  
   Chandra entered the field of stellar structure, the central thread of theoretical astrophysics, at a time when the Eddington-Milne controversy was on the verge of exploding. In this doing, he had to confront towering figures like the so-called `triumvirate' represented by Jeans, Eddington and Milne, that dominated the scene in the British area, and not only. Especially Eddington and Milne were invariably cited by anybody entering the field of the internal constitution of stars. 
  
On November 14, 1930, Fowler invited Chandra  to attend the meeting of the Royal Astronomical Society, because he wanted to introduce him to Milne, who that same day was presenting his paper `destroying' Eddington's view of the interior of stars: ``The paper is supposed to cause a lot of sensation,'' Chandra wrote to his father \cite[p. 85]{Wali1990}. He also added: ``Milne's results are in a sense a generalization of my own on the density of dwarf stars [\dots] Mine is one of the limiting cases of Milne's formulae. I think he will refer to my papers''. But this was not the case, Milne did not mention Chandra's contribution, even if he followed his work and encouraged him,  later even suggesting a collaboration. However he did not really understand the importance of Stoner's and Chandra's results, which were clearly disturbing his theory according to which \textit{all} stars must have a white dwarf core. At the same time, his interest in Chandra's work was justified.  He hoped that Chandra's work might support him in his rivalry with Eddington. One of the main consequences of Milne's analysis was the explanation not only of the existence of white dwarfs --- his collapsed configurations --- but also of the principal characteristics of these configurations. In his third contribution to the subject of white dwarfs \cite{Chandra1931c},  following Milne's analysis of stellar structure \cite{Milne1930a,Milne1930b}, Chandra wanted to develop the theory of collapsed configurations a stage further and with this aim he performed a very detailed analysis introducing the relativistically degenerate core. 

 In the meantime, the same idea of using the Lane-Emden equations for polytropes \cite[pp. 84-182]{Chandra1939}, taking into account the special relativistic effects in the equilibrium of stellar matter for a degenerate system of fermions, came independently to Lev Davidovitch Landau, who, like Chandra, was quite explicit in pointing out the existence of the critical mass. His paper \cite{Landau1932}, appearing in 1932, had the roots in Milne's proposal of a composite model for stars, and was meant as a critical contribution to the Eddington-Milne debate. But it went much beyond and had a key role in exporting Milne's idea of dense cores into the realm of physics. This migration to a different cultural context determined the starting of a new career for superdense cores within stars, which in Landau's theoretical investigations transformed into a well defined and promising physical object, whose potentialities would later be fully revealed with the advent of the neutron era. But at the moment Landau was not able to grasp the deep implications of his work, since he wrote his paper a year before Chadwick announced the discovery of the neutron.

\section{Landau 1932: a transition paper} 

Since the mid-nineteenth century, spectral analysis as applied to the study of stars had established a new relationship between physics and astronomy. When quantum theory decoded the enigma of spectral lines, astrophysics became to grow as a branch of physics. The lively debate about the structure of stars and their source of energy which saw dense stars at the crossroad of different discussions implying the quantum behavior at microscopic level, gradually began to shift towards the nuclear realm. Already in 1920, in investigating models for the nucleus, Rutherford had put forward the idea of a neutral particle formed by a bound state of proton and electron that he named `neutron'. This idea resurfaced in particular towards the end of the 1920s, when attention of physicists was more and more shifting from the outer electron layers of the atom (whose theory had been definitely settled by quantum mechanics) to the nucleus, that was now considered as the new frontier. Within this general trend, dense stars were being once more rediscovered as physical laboratories for speculating on nuclear processes.

In May 1931, well in advance with Chadwick's breakthrough short note to \textit{Nature} \cite{Chadwick1932} announcing the `Possible existence of the neutron', the particle postulated by Rutherford in his Bakerian Lecture to the Royal Society in 1920 \cite{Rutherford1920}, R. M. Langer and Nathan Rosen of MIT proposed that the combination of an electron and a proton ``would be very useful in explaining a number of atomic and cosmic phenomena'' \cite{LangerRosen1931}. Their main aim was to ``offer a way of describing the process of building up of the heavier elements''  but they also proposed that a part of the packing energy released in the formation of neutrons from hydrogen could be ``radiated in a single quantum''  thus explaining the production of cosmic radiation as observed by Millikan and others.\footnote{In 1931, the theory of cosmic rays as charged particles had not yet been established, so they referred to the spread theory of cosmic radiation as very high energy gamma rays.}  As a third application, in the section `High density matter in stars', they proposed the neutron to be at the origin of the formation of very dense cores in stars:  ``The usual explanation of the white dwarfs involving a high degree of ionization of the atoms is not the only one. There are in fact great advantages from this point of view in favor of our neutron. Being small it has a great mean free path and is comparatively insensitive to light pressure. It therefore goes easily to the center of a gravitating mass. Being neutral and having an extremely small external field, it permits high densities to build up before it deviates appreciably from perfect gas behavior''.

A growing interest in the nuclear dimension as a realm of relevant processes in stars is also testified to by  an article by Seitar\^o Suzuki, `Constitution of the white dwarf stars', in which he mentions white dwarf degenerate matter, assuming that all atoms of various elements are stripped of their extranuclear electrons and that all kinds of nuclei of atoms are formed entirely from protons and electrons. He then concluded that the heavy radio-elements exist abundantly in the white dwarfs \cite{Suzuki1931}.

Again white dwarfs' dense matter is used to reason at a nuclear level, where still protons and electrons are the protagonists of nuclear processes. It is in this context that a much celebrated paper by Landau,  `On the theory of stars', dated February 1931, but appearing only in February 1932, was conceived \cite{Landau1932}. Landau had begun his scientific activity in 1926 at the Leningrad Physical Technical Institute (now the Ioffe Physical Technical Institute, St. Petersburg) and graduated from Leningrad State University in January 1927, at the age of 19, having as supervisor Yakov I. Frenkel, head of the Theoretical Physics Department \cite{KapitzaLifshitz1969}. Landau's early entourage included George Gamow and Dimitri Iwanenko. Their group was  known as the Three Musketeers at the University of Leningrad. Between 1926 and 1928 Landau worked with Iwanenko on quantum theory, but during this period he  became particularly close to the brilliant Matvey Bronstein, who had in the meantime joined the so called jazz-band group. Gorelik and Frenkel \cite[pp. 23-24]{GorelikFrenkel1994} outline how Bronstein, fascinated by astronomy, introduced his physicist friends to astronomers like Viktor Ambartsumyan, who would later become a first class astrophysicist.\footnote{Still during his student years, Bronstein  wrote his first  astrophysical papers, which contained an important contribution to the theory of stellar atmospheres in the form of the so-called Hopf-Bronstein relation \cite{Bronstein1929,Bronstein1930}. Milne himself, one of the founders of the field, recommended Bronstein's second paper for publication in the \textit{Monthly Notices of the Royal Astronomical Society}. In 1931, \textit{Uspekhi Fizicheskikh Nauk} published a detailed survey by Bronstein entitled `The Modern State of Relativistic Cosmology' \cite{Bronstein1931}. It was the first review of cosmology in the USSR. Landau and Bronstein continued to collaborate even after the former had moved to Kharkov in 1932.}

In 1929, on an assignment from the People's Commissariat of Education, and  later  thanks to a  Rockefeller grant, Landau travelled abroad and for one and a half years worked in Denmark, Great Britain and Switzerland. 
The most important experience was his stay in Copenhagen where, at the Institute of Theoretical Physics, theoreticians from all Europe were attracted by Niels Bohr. Like many others, and in particular like his friend Gamow, Landau was strongly influenced by Bohr, whom he always considered his only teacher.

Rudolf Peierls met Landau in Zurich for the first time during the autumn of 1929: ``[\dots] we discussed things a lot [\dots] I cannot remember all the things, we discussed, but certainly he was then already very interested in astrophysics'' \cite{Peierls1977}. In Zurich, at that time, young physicists around Pauli, like Peierls, Bloch, Leon Rosenfeld, were struggling mostly on problems of metals and, of course, quantum electrodynamics,  the most debated subject at that time \cite{Rosenfeld1963}.
According to Arkadii Migdal \cite{Migdal1977}, ``Landau's idea was that theoreticians should not be devoted to one special part of physics''. 

In the early spring of 1931, Landau shared his time between Copenhagen and Zurich.\footnote{In particular, he visited Bohr from 8 April to 3 May 1930, 20 September - 20 November 1930, and 25 February - 19 March 1931 \cite[p. 359]{Pais1993}.}
 Gamow, too, was there from September 1930 to May 1931, and during that last month he completed his book on the constitution of atomic nuclei and radioactivity, the first one ever on theoretical nuclear physics \cite{Gamow1931}\cite{Gamow1968}: ``I remember that Landau was helping me with the mathematics, with calculating the perturbation and so on. And these formulas were all derived by Landau''.\footnote{As remarked by Pais, \cite[p. 325]{Pais1993}: ``Most of the people Gamow thanked for valuable advice belonged to the Copenhagen circle: Bohr himself and also Gamow's friends and contemporary fellows at the institute, Hendrik Casimir from Leiden, Lev Davidovich Landau from Leningrad, and Nevill Francis Mott from Cambridge''.}

It happened that, in that same period, on August 19, 1930, Fowler wrote to Bohr that he had ``very exciting news from Milne. He is convinced now that he has found exactly, where Eddington is wrong in his astrophysical theories'' and that he would tell more in detail on his arrival in Copenhagen in September. Milne had just published in \textit{Nature} his article on stellar structure and the origin of stellar energy \cite{Milne1930c}, where he put forward the idea that a core of very dense material would form within stars, ``a kind of `white-dwarf' at its centre, surrounded by a gaseous distribution of more familiar type; the star is like a yolk in an egg [\dots] It is to this nucleus that we must look for the origin of stellar energy, a nucleus the existence of which has previously been unsuspected''. His detailed analysis of stellar structure outlining his theory of a composite model for stars would appear in the following months \cite{Milne1930a}, but it was clearly the involved question of the stellar energy problem that definitely triggered the physicists' interest.
 On August 26 Bohr answered Fowler's letter: ``It shall be a great pleasure  to discuss  the many actual problems and not least the interpretation of the astrophysical evidence in which I am very interested. He added that they were expecting in September various visits, in particular Rosenfeld and Gamow. He did not mention Landau, who actually had written him on August 23 from Cambridge about his intention of visiting Copenhagen some time in the middle of September.\footnote{Archives for the History of Quantum Physics, Bd. AHQP/BSC 19: Niels Bohr. Scientific correspondence, 1930--1945.}  

Fowler arrived in Copenhagen on September 11. All this meant that already during the summer and early fall of 1930 Milne's theory was largely discussed in Copenhagen.\footnote{Milne's long paper appeared in the November number of the \textit{Monthly Notices of the Royal Astronomical Society}  \cite{Milne1930a}, together with related papers by R. H. Fowler, N. Fairclough, and T. G. Cowling, presented during the meeting held at the Royal Astronomical Society on January 9 1931, completely devoted to a debate on the subject.}

Landau's background  easily explains the motivations behind his interest in these questions. Under the spell of Bronstein's passion for astrophysics, and  having studied the unusual magnetic properties of the degenerate electron gas in a metal, like others at that time, Landau extended his investigations on the behaviour of a relativistic degenerate electron gas in a more extreme and challenging realm: the interior of stars.\footnote{In 1930, Landau collaborated with Frenkel in an article on the quantization of free electrons in a magnetic field. During his stay in Cambridge he tried to explain Pyotr Kapitza's results concerning the dependence of the electrical conductivity of metals on an external magnetic field applying the methods of quantum mechanics to the problem of the anomalous properties of the electric conductivity of bismuth in strong magnetic field. This work resulted in his theory of diamagnetism, that became a basis of research in solid-state physics.} 
 As  will be clarified in the following, Bohr himself was especially interested in the problem of stellar energy, so that the whole matter certainly became a hot topic during Landau's stay in Copenhagen. The first relevant paper applying quantum mechanics to stellar element synthesis by nuclear reactions --- that can be actually regarded ``as one of the pioneering contributions to nuclear astrophysics'' \cite[p. 85]{Kragh1996} --- had been written in 1929 by Fritz Houtermans and Robert d'Escourt Atkinson based on Gamow's theory of $\alpha$-decay \cite{AtkinsonHoutermans1929b}. As a first step they had actually sent a short note to \textit{Nature} `Transmutation of the lighter elements in stars' \cite{AtkinsonHoutermans1929a}. Theoretical nuclear physics was entering stellar interiors: solving the problem of energy generation in stars might also account for the abundances of the various chemical elements. As acknowledged by the authors, Gamow had been deeply involved in ``numerous discussions'', and it is easy to imagine that in turn he must have abundantly discussed these topics with Landau going back to Russia that same 1929, after having travelled through Europe. 

Landau's paper is dated February 1931, but soon after he returned to Leningrad  and only on January 7, 1932, he  submitted it to the brand new \textit{Physikalische Zeitschrift der Sowjetunion} --- the first Soviet physical journal published in languages other than Russian --- that is more than one month before Chadwick's announcement of having detected the neutron  \cite{Landau1932}.  That same 1932 Landau moved to Kharkov, where he became head of the Theoretical Division of the newly organized Ukrainian Physicotechnical Institute, an offshoot of the Leningrad Institute.\footnote{The establishment of the \textit{Physikalische Zeitschrift der Sowjetunion} had been promoted by Iwanenko, who was in the Kharkov Institute of Physics from 1929 to 1931 as first director of its theoretical division \cite{Sardanashvily2014}.}

Landau's paper has generally attracted  attention as one of the milestone's --- actually as a starting point --- in the path towards the idea of compact collapsed objects \cite{Yakovlev2013}.
 What is especially relevant here, is its genesis in the pre-neutron era, notably in the period preceding Fermi's solution to the problem of beta decay, that definitely banished electrons from the nuclear realm. In this sense Landau's paper is acting like a prism refracting different controversies both in the physical and the astrophysical realms. Landau's interests were extremely wide, moreover, in the mid-1930s, as he  himself explained, ``theoretical physics, unlike experimental physics, is a small science open to perception in its entirety by any theorist''  \cite{Gorelik2005}.   Landau's excursion in the astrophysical realm of stellar theory, was triggered by the debate about the Milne-Eddington controversy. Moreover, during his stay in Copenhagen, Landau interacted with Bengt Str\"omgren, who acknowledged Landau's assistance in the first paper he wrote for the \textit{Zeitschrift f\"ur Astrophysik} \cite{Stromgren1931} where he discussed Milne's ideas of a stellar nucleus of extreme density and temperature along lines differing from those followed by Eddington and examined the question of the existence of stellar configurations with a nucleus of this character. Str\"omgren, who had studied physics in Bohr's Institute for theoretical physics, was working at the Copenhagen Observatory, but he frequently attended conferences there, thus having  the occasion to become familiar with foreign visitors. During his studies for the Master's degree, he became much impressed with the latest developments: ``I had the idea that the time was ripe for applications of the new quantum mechanics to astrophysical situations''. He also remembered that \cite{Stromgren1978} ``Landau was a frequent visitor, and he was deeply interested in these questions, but had his own views that differed radically from those of other people [\dots] And his idea was that you have a very high density core, in the sun, and that release of gravitational energy therefore plays a role. For this reason he was inclined to disregard all of Eddington's work''. Str\"omgren eventually became one of the leading theoretical astrophysicists in the world. 
Landau's motivation thus aroused within the hot topic of the stellar equilibrium in gravity, particularly the maximum mass of white dwarfs, where electron degeneracy pressure stands against gravity. Milne's theory aspired to explain the existence of all types of stars including the white dwarfs, explain the energy generation and eliminate the problem of the stellar absorption coefficient, by supposing that the mass, the luminosity, and the absorption coefficient were completely independent. Landau started criticizing Milne's arguments against Eddington's mass-luminosity relation, also annoyed by his excessive reliance on what he considered `mathematical eccentricities', far from physics: ``The astrophysical methods usually applied in attacking the problems of stellar structure are characterised by making physical assumptions chosen only for the sake of mathematical convenience'' like, for instance, Mr. Milne's proof of the impossibility of a star consisting throughout of classical ideal gas''. This proof, added Landau, rested on the assertion that, ``for arbitrary $L$ and $M$, the fundamental equations of a star consisting of classical ideal gas admit, in general, no regular solution''. Landau then stressed that Milne seemed ``to have overlooked the fact, that this assertion results only from the assumption of opacity being constant throughout the star, which assumption {\it is made only for mathematical purposes and has nothing to do with reality} [emphasis added]''. Only in the case of this assumption, recalled Landau, the radius $R$ disappears from the relation between $L$, $M$ and $R$, ``which relation would be quite exempt from the physical criticisms put forward against Eddington's mass-luminosity-relation''. Once clarified his position towards astrophysicists' way of tackling such problems, Landau declared: `It seems reasonable to try to attack the problem of stellar structure by methods of theoretical physics, i.e. to investigate the physical nature of stellar equilibrium'' \cite[p. 285]{Landau1932}. As Cowling commented much later, ``Both observing astronomers and physicists tend to wax critical of the mathematician, and sometimes with reason. Mathematicians try to construct models of stars: I remember Milne saying here in 1930 that he would no longer speak of stars, but only of spherical masses of gas'' \cite[p. 121]{Cowling1966}.

Landau criticized Milne's introduction of ``a condensed inner part of the system'' as an \textit{ad hoc} hypothesis, without explaining the reason why such condensations could appear at all, so that the connection between the condensed state and the normal state remained ``rather mysterious''. Then, independently, followed Chandra's in using the Emden-Lane equation investigating the statistical equilibrium of a given mass without generation of energy, and showing that in the case of classical ideal gas, there is no equilibrium at all: ``Every part of the system would tend to a point''. However,  ``the state of affairs becomes quite different when we consider the quantum effects''. He then discussed the extreme-relativistic case finding that a star of fixed mass would have to either expand or collapse to a point to attain a minimum of the energy and reach an equilibrium state. In order to find the criterion separating the two cases he solved the n=3 polytropic equation of Emden finding that an equilibrium state is reached only for masses smaller than a critical mass of about 1.5 solar masses, again of the order of values found by Stoner and Chandra. However, for masses greater than the critical mass, Landau remarked that ``there exists in the whole quantum theory no cause preventing the system from collapsing to a point''. 

However, he continued, ``As in reality such masses exist quietly as stars and do not show any such ridiculous tendencies we must conclude that \textit{all stars heavier than 1.5 $M_\odot$ certainly possess regions in which the laws of quantum mechanics} (and therefore of quantum statistics) \textit{are violated}''. Landau stressed that there was no reason to believe that stars could be divided into two physically different classes according to the condition of having a mass greater or smaller than the critical mass, so that he supposed that all stars should possess those ``pathological regions'' avoiding the necessity of such division, and even that ``just the presence of these regions makes stars stars''.

But if this is the case, reasoned Landau, there was no need to suppose that the radiation of stars might be due to ``some mysterious process of mutual annihilation of protons and electrons'' (he is here referring to Jean's old ideas on annihilation as a possible source of stellar energy, also mentioned by Milne) because protons and electrons in atomic nuclei are very close together and ``they do not annihilate themselves'', even being \textit{both} constituents of the nucleus, according to current ideas about nuclear matter. In dismissing astrophysicists' vague ideas on the sources of stellar energy, recently tackled from a physical point of view by Atkinson and Houtermans, Landau is mentioning ``a beautiful idea of Prof. Niels Bohr's'' according to which one could be able to believe that ``the stellar radiation is due simply to a violation of the law of energy, which law, as Bohr has first pointed out, is no longer valid in the relativistic quantum theory, when the laws of quantum mechanics break down''.\footnote{In this regard, see Gamow's interview recalling Bohr's unpublished theory \cite{Gamow1968}, Bohr's manuscript in his {\it Collected Works} \cite[vol. 9, p. 88]{Bohr1986} and \cite[pp. 63-82]{Gorelik2005}.}
 At that time there was the big problem of apparent non conservation of energy in $\beta$-decay.  Bohr speculated on the idea that perhaps energy conservation is not strictly valid in microscopic processes related to such nuclear transformations and that this might \textit{also} even explain mechanisms related to the production of stellar energy. Such a problem, in turn, posed another completely unsolved mystery, together with nuclear structure, $\beta$-spectra and, last but not least,  the famous Klein paradox, according to which an electron could not be confined within nuclei, a problem much debated since 1929. All this led Bohr to conclude  that, ``As soon as we inquire [\dots] into the constitution of even the simplest nuclei the present formulation of quantum mechanics fails completely''. As emphasized by Pais, in anticipating ``such drastic revisions of physics'', Bohr was  looking for  a comprehensive point of view that would all at once explain these four puzzles \cite[p. 367]{Pais1993}. 
Pauli, who definitely disagreed with Bohr, was reflecting on the possibility that there would be agreement with experiments if a new neutral particle took part in the beta-disintegration process carrying away the excess of energy and angular momentum. To Bohr's proposal about energy in stars Pauli thus answered: ``let the stars radiate in peace!''\footnote{Pauli to Bohr 17 July 1929, reprinted in \cite[vol. 6, p. 447]{Bohr1986}.} 
Rutherford, on his side, decided to wait and see before expressing an opinion, feeling that ``there are more things in Heaven and Earth than are dreamt of in our philosophy'', as he wrote to Bohr in November 1929.\footnote{It is well known how Fermi took Pauli's idea so seriously, to incorporate `Pauli's neutron', in the meantime renamed neutrino, in his ground-breaking theory of beta-decay, in which a new interaction was introduced using the language of quantum field theory. Based on the proton-neutron model of the nucleus, the mechanism of  particle creation --- the electron-neutrino pair ---  solved the problem both of the pathological `nuclear electrons' and of the missing energy in the decay process. But immediate reactions were not exactly enthusiastic and only gradually the theory was generally accepted.}

In following Bohr, Landau thought he was killing  two birds with one stone: not only was he avoiding catastrophic collapse to a point invoking non-conservation of energy, but was also obtaining a source of stellar energy. In January 1931, Landau, then in Zurich, had written with Peierls an article where they had already based on Bohr's idea in arguing that Pauli principle and thus ordinary quantum theory, did not apply in the nucleus, where special relativistic effects become relevant \cite{LandauPeierls1931}.\footnote{However, according to Gamow, later it was shown by Landau himself that ``the rejection of the conservation law for energy will be connected with very serious difficulties in the general gravitational theory, according to which the mass present inside a certain closed surface is entirely defined by the gravitational field on this surface'' \cite[p. 747]{Gamow1934}. Gorelik has mentioned that at that time Bronstein realized the need for `a relativistic quantum theory + the theory of gravitation in astrophysics' explaining it in a very simple way: ``If the sun were compressed to nuclear density, its radius would be comparable with the gravitational radius'' \cite[p. 1042]{Gorelik2005}. By that time cosmology was becoming for Bronstein the real great challenge: ``a solution to the cosmological problem requires first to create a unified theory of electromagnetism, gravity, and quanta''. Such a brilliant mind became one of the many victims of Stalin's Great Purges and was executed in 1938.}

Landau expected that the breakdown of quantum mechanics would occur ``when the density of matter becomes so great that atomic nuclei come in close contact, forming \textit{one gigantic nucleus} [emphasis added]''.  Landau did not specify what particles were involved, even if he must have clearly referred to nuclei as built out of protons and electrons, as they were still generally considered at that time. We have here a definite transition from Fowler's dense matter of a white dwarf, described as ``analogous to a giant molecule'' to a core of highly condensed matter forming ``a single giant nucleus'' surrounded by matter in ordinary state within the central region of the star. In the end, Landau supported Milne's idea about the central region of the star consisting of a core of highly condensed matter. However, Milne's theoretical `collapsed configuration' was transformed in Landau's hands in a full-fledged physical system on which physicists could  theorize. The price to be paid was to reject the possibility that stars' evolution might depend on their mass. 
 
To summarize: in this rather short note, Landau is pursuing very ambitious aims: finding conditions for the equilibrium of a star, establishing the existence of a limit mass, finding a source of energy for stellar radiation and trying to develop a theory of stellar structure. Analyzing it in hindsight, many critics and comments could be put forward, that should be discussed within the state of physics at the time. It is however to be remarked that, even being aware that the gravitational collapse was a consequence of his calculations, Landau rejected this possibility, heavily contributing with his influence to block acceptance of this catastrophic phenomenon. 
 Without any doubt, this article was appealing to physicists, because it spoke their language, and for this reason it was widely cited during the years and opened the way to fruitful theoretical developments for reasoning on superdense matter in stars. The first clue to a fundamental difference in the evolution and final stages of low and high mass stars had been provided, but at the same time it had become clear that the analysis has to be shifted to the still basically unknown realm of nuclear matter.

  \section{Interlude: Dense matter and the early universe. Georges Lema\^itre and the primeval super-atom}

 If a dense plasma of nuclei and electrons could exist within white dwarfs, ``like a gigantic molecule in its lowest quantum state'', forming a ``Star-Atom'', according to Eddington's colourful expression \cite[p. 127]{Eddington1927},  a super-compact atomic nucleus having a weight equal to the entire mass of the universe could well be at the origin of the whole universe itself, according to a proposal put forward by the Belgian physicists and cosmologist Georges Lema\^itre \cite{Lemaitre1931a}. During his university studies Lema\^itre had already tackled the general theory of relativity, and for this reason he decided  to use a grant he had received in the summer of 1923 to go to Cambridge and study under Eddington, whose influential personality as a scientist and especially as an expert in relativity, inspired him to address his research interests to what appeared to him as a most fascinating field. During his later stay in the United States, in the period 1924-1925, Lema\^itre prepared for a Ph.D in astronomy at MIT and being attached to Harvard Observatory he was also introduced to the latest developments in astronomy and in particular experienced the impact of Hubble's observations of the early 1920s according to which the spiral nebulae are galaxies outside the Milky Way. Being convinced of the relevance of this new perspective and of the redshift-distance relation for relativistic cosmology, Lema\^itre visited both Vesto Slipher at the Lowell Observatory in Arizona, who had been the first to discover in 1917 that most spiral galaxies have considerable redshifts, and Hubble himself at Mount Wilson Observatory \cite{Kragh2013}. 
 
Unaware of Alexander Friedmann's work of 1922 \cite{Friedmann1922}, showing that Einstein's equations have dynamical solutions, Lema\^itre's formulated the same cosmological differential equations. He proposed a dynamical cosmological model in his ``Un univers homog\`ene de masse constante et de rayon variable rendant compte de la vitesse radiale des n\'ebuleuses extra-galactiques'' \cite{Lemaitre1927}. But his approach was quite different, because, contrarily to Friedmann, who did not compare the models with astronomical data, Lema\^itre addressed the cosmological implications of general relativity combining mathematical results with the physical reality, in particular, with astronomical observations of the recession of the nebulae, that he viewed as a ``cosmical effect of the expanding universe''  \cite[p. 489]{Lemaitre1931}. ``The expansion of the universe is a matter of astronomical facts interpreted by the theory of relativity'' stressed Lema\^itre in October 1931, during a meeting of the  British Association for the Advancement of Science dedicated to the evolution of the universe, to which de Sitter, Eddington, Millikan, and Milne participated. His 1927 theory went rather unnoticed and was `rediscovered' around 1930 when Eddington and De Sitter contributed to make it widely known  \cite[ch. 2]{Kragh1996}. At first, both Friedmann and Lema\^itre were ignored. Lema\^itre himself became aware of Friedmann's work when he attended the 1927 Solvay Conference, during discussions with Einstein. Einstein recognized the similarity between the two theories, and had no objection in this sense, but his conclusive comment was unfavorable: he considered it definitely ``abominable'' from the physical point of view \cite[p. 370]{Deprit1984} \cite[p. 125]{Kragh1987} \cite[p. 8]{Eisenstaedt1993}. But actually Lema\^itre, in telling Einstein about the recessional velocities of galaxies, had the impression that the latter was not really informed about astronomical facts. 

Lema\^itre's physical cosmology,  in connection with current views of dense matter in bulk subject to quantum laws, that most probably concurred to inspire him, led to a proposal which Lema\^itre presented in a short note to \textit{Nature}. In `The beginning of the world from the point of view of quantum theory' \cite{Lemaitre1931a} Lema\^itre answered to Eddington's contribution  `The end of the world: from the standpoint of mathematical physics' \cite{Eddington1931} published on the same journal, where the latter had clearly stated that ``the notion of a beginning of the present order of Nature is repugnant''. Lema\^itre proposed instead that he was ``inclined to think that the present state of quantum theory suggests a beginning of the world very different from the present order of Nature''. Thermodynamic principles, he said, require that ``(1) energy of constant total amount is distributed in distinct quanta''  and that ``(2) the number of distinct quanta is ever increasing. If we go back in the course of time we must find fewer and fewer quanta, until we find all the energy of the universe packed in a few \textit{or even in a unique quantum} [emphasis added]''. 

If an atomic nucleus could be counted as a unique quantum, ``the atomic number acting as a kind of quantum number,'' one could conceive the beginning of the universe in the form of a unique atom, the atomic weight of which is the total mass of the universe''. This highly unstable universe-atom ``would divide in smaller and smaller atoms by a kind of super-radioactive process''. He thus believed that the primeval atom hypothesis provided a physical beginning of the universe and that its subsequent evolution was the result of a disintegration \cite[pp. 113-114]{Lemaitre1931c}:  ``In the atomic realm, we know a spontaneous transformation that can give us some idea of the direction of the natural evolution; it is the transformation of radioactive bodies [\dots] an uranium atom is eventually transforming into a lead atom and seven or eight helium atoms. This is a transformation from a more condensed to a less condensed [\dots] The natural tendency of matter to break in more and more numerous particles, which shows itself in so striking a way in the radioactive transformations, can be observed also in the grains of light or photons that form the different forms of radiation''. 
 The `super-radioactive' processes he mentioned suggest a kind of matter very similar to nuclear matter, consisting of electrons and especially of alpha particles, which were considered a sort of building blocks of the nucleus, because of their recognized stability as entities deriving from the decay of radioactive elements, in particular very heavy elements such as uranium and thorium, whose half-lives were of the order of billion years. Some remnant of this process, recalled Lema\^itre following Jeans's idea, might still be fostering the heat of the stars. All this found an experimental base on what can be considered the nuclear physics of the time, which was on the verge of entering its modern era with the detection of the neutron, but whose knowledge still derived mainly from the study of radioactive decays, which on the other hand were connected to the formation of new chemical elements, that had been studied since the early years of radiochemistry. 

Lema\^itre's primeval matter appears to be quite similar to the stuff of which dense white dwarfs were supposed to be made. But actually, he did not specify the nature of the `primeval atom', a term that is probably to be interpreted as something similar to the basic primordial entity very common in ancient cosmogonies. A hint about the nature of the primordial super-atom is provided by his contribution to the mentioned discussion of October 1931 at the meeting of the British Association for the Advancement of Science, a  longer contribution in which he fully outlined his views about the physical universe --- ``The expansion of the universe is a matter of astronomical facts interpreted by the theory of relativity''  --- and its origin from the disintegration of the primeval atom: ``We want a `fireworks' theory of evolution. The last two thousand million years are slow evolution: they are ashes and smoke of bright but very rapid fireworks''. He suggested that big stars were remnants of the successive splittings of the primeval atom and that, with their fireworks of radiation, they were the source of cosmic rays of high energy \cite[p. 704-705]{Lemaitre1931b}. The key of the problem, according to Lema\^itre, was afforded by the discovery of cosmic rays: ``the energy of cosmic rays is comparable in amount to the whole energy of matter [\dots] If the cosmic rays originated chiefly before the actual expansion of space, their original energy was even bigger [\dots] The only energy we know which is comparable to the energy of the cosmic rays is the matter of the stars. \textit{Therefore it seems that the cosmic rays must have originated from the stars} [emphasis added]''. Inspired by Jeans' ideas admitting the possible existence of atoms of considerably higher atomic weight than the known end decay products of radioactive decays of the heavies atomic elements, Lema\^itre stated that ``Cosmogony is atomic physics on a large scale --- large scale of space and time --- why not large scale of atomic weight? Radioactive disintegration is a physical fact, cosmic rays are like the rays from radium. Have they not escaped from a big scale super-radioactive disintegration, the disintegration of an atomic star, the disintegration of an atom of weight comparable to the weight of a star''. Cosmic rays would be ``glimpses of the primeval fireworks of the formation of a star from an atom, coming to us after their long journey through free space''. 

He immediately suggested that ``a possible test of the theory is that, if I am right, cosmic rays cannot be formed uniquely of photons, but must contain, like the radioactive rays, fast beta rays and alpha particles, and even new rays of greater masses and charges''.\footnote{According to Millikan's opinion cosmic rays were ``the birth cries of the elements'', high-energy photons arising from the building-up of elements in the depths of space. 
His theory  had recently been challenged by the Bothe and Kolh\"orster's experiment published in 1929 \cite{BotheKolhorster1929}, showing that cosmic rays were charged particles and not `ultra-gamma rays'. But it cannot be excluded that the `cosmic birth' context summoned by Millikan's  theory, played someway a role in Lema\^itre's reflections leading to the primeval atom theory. In any case, during his stay at MIT, Lema\^itre collaborated with the Mexican physicist Manuel Vallarta in complicated calculations of the energies and trajectories of charged particles in the Earth's magnetic field, making use of MIT's differential-analyzer computer developed by Vannevar Bush. They concluded that both Arthur Compton's data deriving from his world campaign, that had verified the existence of the latitude effect (``showing that the cosmic rays contain charged particles'') and their own computer calculations, were providing ``some experimental support to the theory of super-radioactive origin of the cosmic radiation'' \cite[p. 91]{LemaitreVallarta1933}.}

Whether this was ``wild imagination or physical hypothesis'', it could not be said. In order to solve the problem two things were needed, according to Lema\^itre: ``First, a theory of nuclear structure sufficient to be applied to atoms of extreme weights  [\dots] The second thing we want is a better knowledge of the nature of the cosmic rays''. 

What is relevant in our context is that Lema\^itre's  ambitious theory was relating the mathematical universe of General Relativity to an evolutionary physical universe whose nature as a physical system was being discovered by astronomers: ``A really complete cosmogony should explain both atoms and suns'' \cite[p. 113]{Lemaitre1931c}. But he also showed how the theory of the expansion of the universe could be adapted to the idea of a primeval atom through three different phases: a first period of rapid expansion during which the universe-atom breaks in star-atoms, a period of slowdown, followed by a third phase of accelerated expansion, that we are living now, which is responsible of the separation of stars in extra-galactic nebulae \cite[p. 119]{Lemaitre1931c}.

In his ambition to explain the Universe at a macroscopic and microscopic level as a physical system in continuous evolution, Lema\^itre put quantum theory and thermodynamics in connection with a state of superdense concentration of matter, having such universal character to give origin to all the observed distribution of matter in the universe: all the atomic nuclei were produced by disintegrations of the primeval quantum.  Moreover, for the second time, after Eddington's observation that general relativity must be connected to the observed spectra of dense stars like white dwarfs, Einstein's theory was connected to a primeval dense concentration of matter giving origin to the whole physical universe. 

It is difficult to assess the overall impact of Lema\^itre's speculations related to his physical cosmogony.\footnote{Attention has been given by Kragh to responses to  Lema\^itre's theory of the expanding universe (see \cite{Kragh2013} and references therein).} 
 It is quite clear that he influenced further bold speculations put forward by physicists like Fritz Zwicky and especially Gamow, who had been Friedmann's student and had a knowledge of general relativity since the beginning of his research activity. 

His ``wild imagination'' was offering such cosmic fireworks  to physicists who had the same bold attitude and whose minds resonated on Lema\^itre's words \cite[p. 706]{Lemaitre1931a}: ``Our world is now understood to be a world where something really happens; the whole story of the world need not have been written down in the first quantum like a song on the disc of a phonograph. The whole matter of the world must have been present at the beginning, but the story it has to tell may be written step by step''.

\section{Sterne 1933: neutronization of superdense matter in stars}

When the neutron officially became a new constituent of the nucleus --- even if it was not immediately clear  whether it was or not a bound state of proton and electron --- it opened a new era in nuclear physics and in particular in its application to the astrophysical stage. The long-standing problem of the origin of elements and of stellar energy could be discussed on a new base. As it had happened in the case of  the new statistics in connection with metals and white dwarfs, now dense stellar matter became a testing ground for nuclear reactions. 

It appears that the first to propose a systematic discussion on the equation of state of nuclear matter, and to apply it to stars interiors, was Theodor E. Sterne, who received his PhD from Cambridge University in the summer of 1931 with Fowler as his supervisor (like in Stoner's case, Fowler is again acting behind the scene\dots).

Sterne started his investigations on what was considered ``One of the most important problems requiring solution'' at that time, that is the production of energy in stars \cite{Sterne1933a}.
 It was generally agreed that  the principal, if not almost the entire, source of this energy must be subatomic. In 1932, disintegrations produced by artificially produced fast protons had been observed at Cavendish Laboratory by Cockcroft and Walton with large production of energy, as well as transmutations produced by bombardment of fast alpha particles, resulting in the emission of neutrons capable in turn of further transmutations in striking other nuclei. The possibility of induced transmutations had thus been established beyond any reasonable doubt by strong experimental evidence, and considerable absorption or liberation of subatomic energy were expected in most cases. These energies, said Sterne, must be intimately related to the abundances of the elements in the stellar matter during the changes between the different states. In March 1933, he announced his program \cite[p. 585]{Sterne1933a}: ``It is possible to consider by statistical mechanics an assembly containing radiation, atomic nuclei, electrons, and neutrons; when all possible transmutations of the nuclei occur without the `annihilation' of any ultimate particles. One can calculate the abundances of the nuclei of the various sorts in such an assembly, when it is in equilibrium, in terms of the atomic masses and packing fractions''.\footnote{At that time physicists still discussed whether the neutron was a real elementary particle and whether the positron, that Sterne included in his discussion, was identical with Dirac's `holes'. In this regard Sterne, mentioned Carl D. Anderson's observation of the positive electron at Caltech, as well as cloud-chamber experiments performed at Cavendish Laboratory by Patrick Blackett and Giuseppe Occhialini, who had observed the phenomenon of electron-positron pair production producing a strong support to Dirac's theory.}

In three further papers appearing in cascade in MNRAS \cite{Sterne1933b} \cite{Sterne1933c} \cite{Sterne1933d} Sterne discussed the formation of the chemical elements by nuclear reactions in stars and the liberation of energy by transmutations, apparently being the first systematic investigation in this sense.\footnote{By that time Sterne was at Jefferson Physical Lab Cambridge, Mass. He thanked  Cecilia Payne and Ralph Fowler, who communicated the papers to the Royal Astronomical Society.} 
In  \cite[p. 748]{Sterne1933b}, he investigated the gradual contraction of a star, with equilibrium composition gradually shifting as the density and temperature increased. He pointed out that, as determined by Chadwick, neutrons had packing fractions which are considerably greater than the packing fractions of other kinds of nuclei. Applying the Darwin-Fowler method to the statistical equilibrium among nuclei, he arrived at the conclusion that ``At sufficient enormous densities [greater than approximately 2.3$\times 10^{10} g/cm^3$ when $T \ll 6\times10^7\rho^{1/3}$] [\dots] the assembly at low temperatures should contain a preponderance of neutrons [\dots] At these high densities, matter at low temperatures would be literally squeezed together into the form of neutrons''. \cite[p. 750]{Sterne1933b}.\footnote{And indeed, in a short note on {\it Nature} \cite{Sterne1933a} he had presented his preliminary investigations on the equilibrium property of an assembly containing radiation, atomic nuclei, electrons, and neutrons based on the ``hypothesis that nuclei ({\it and neutrons}) are made of electrons and protons [emphasis added]''. In  \cite{Sterne1933b}, instead, he also considered the possibility that the neutron could be ``an ultimate particle''.} 

He concluded the article expressing the hope that ``the statistical theory here developed may prove to be of assistance to astrophysicists''. \footnote{Sterne's pioneering article was cited by Gamow in 1939 \cite{Gamow1939a}, at a time when nuclear astrophysics had already developed into a research field attracting physicists with a competence in theoretical nuclear physics. Gamow acknowledged that: ``It was first indicated by Sterne that, at very high densities and not-too-high temperatures, the formation of a large number of neutrons must take place because the free electrons are, so to speak, squeezed into the nuclei by the high pressures''. Gamow is also suggesting to look at Hund's review article of 1936 \cite{Hund1936}, which will be discussed later.} 

In parallel with Sterne's theoretical work in which it was clarified that compression of cold matter to high densities would induce neutronization, the role of neutrons in the structure of stars was widely discussed in a PhD dissertation written under Max Born in G\"ottingen by Siegfried Fl\"ugge \cite{Flugge1933}. While Sterne was more  relying on the idea that after all a neutron was a bound state of a proton and an electron, Fl\"ugge specified that as during $\beta-$decay processes a neutron is transmuted in a proton + an electron, one could imagine that an evaluation of the number of neutrons in stars could be done through a ``thermodynamical equation according to the Synthesis Proton+Electron = Neutron + Energy'' \cite[p. 278]{Flugge1933}. He also examined,  ``as a curiosity'' what would be the characteristic of a star consisting only of neutrons (``\textit{ein Stern, der nur aus Neutronen best\"unde}'')  and speculated how neutron capture by heavy nuclei could explain the production of stellar energy \cite[p. 282]{Flugge1933}.
 
Neutrons were beginning to become the great protagonists of nuclear processes taking place in stars. It is thus not  surprising that speculations on the existence of exotic stars consisting only of neutrons, mentioned by  Fl\"ugge as a curiosity,  were quickly incorporated in a theory on the most catastrophic cosmic event known at the time: the explosion of a star.

 \section{A not so lonely sailor: Fritz Zwicky}
 
 As outlined in the previous sections, during the 1920s many physicists addressed astrophysical problems, exploring the properties of very dense stars in order to derive basic properties of matter in conditions that could not be obtained in any terrestrial laboratory. The growing relevance of the problem of stellar energy, and the related difficulties faced by physicists in their attempt to account for the actual production of such energy, went in parallel with the shifting of interest towards the nuclear realm during the 1930s, especially after the strong impact deriving from the confirmed existence of the neutron that opened the way to brand new theoretical and experimental investigations. 
 
 Theories about the stellar interiors included the new particle in discussions about the structure, equilibrium and generation of energy in stars. Papers on the phenomenon of neutronization of matter in stars with increasing density certainly did not escape the attention of Fritz Zwicky, a Swiss theoretical physicist working at the California Institute of Technology since the 1920s.\footnote{Born in Bulgaria in 1898, Zwicky grew up in Switzerland, and then studied in Zurich. He studied solid-state physics and worked in crystallography research before moving to California on an International Education Board post-doctoral fellowship in 1925.} He was familiar with quantum theory, as well as with dense matter in metals and crystals, a field in which he was still working during the early 1930s. 

At the same time, the Caltech campus is near the Mount Wilson Observatory, which had the world's largest telescope, and where Edwin Hubble was working since the end of the 1910s. In 1929, Zwicky was intrigued by Hubble's results \cite{Hubble1929} showing a roughly linear correlation between the apparent velocity of recession and the distance of galaxies \cite{Zwicky1929} and his interest in astrophysics grew with the arrival of the German astronomer Walter Baade from Hamburg in 1931. Baade was studying novae and together they came to the conclusion  that the population of novae consists of two types: the ordinary novae and the `supernovae', which are very rare but much more energetic. In December 1933, during the annual meeting of the American Physical Society at Stanford, they proposed that ``In the \textit{supernova} process \textit{mass in bulk is annihilated}. In addition the hypothesis suggests itself that \textit{cosmic rays are produced by supernovae}''. Basing on the assumption that ``in every nebula one supernova occurs every thousand years'' they accordingly evaluated the expected intensity of cosmic rays, comparing it with Millikan and Regener's observed flux. They concluded the abstract with a bold proposal: ``With all reserve we advance the view that  supernovae represent the transitions from ordinary stars into  \textit{neutron stars} which in their final stages consist of extremely closely packed neutrons'' \cite{BaadeZwicky1933}.\footnote{According to a review article by Zwicky \cite[p. 85]{Zwicky1940}, he and Baade introduced the term supernovae in seminars and an astrophysics course at Caltech in 1931 then used it publicly in 1933 during the just mentioned meeting of the American Physical Society held at Mount Wilson Observatory.}  
 
Such a star, they explained in a more detailed article, ``may possess a very small radius and an extremely high density. As neutrons can be packed much more closely than ordinary nuclei and electrons, the `gravitational packing' energy in a \textit{cold} neutron star may become very large, and, under certain circumstances, may far exceed the ordinary nuclear packing fractions. A neutron star would therefore represent the most stable configuration of matter as such'' \cite[p. 263]{BaadeZwicky1934a}. They were fully aware that their suggestion carried with it ``grave implications regarding the ordinary views about the constitution of stars'' and therefore would require ``further careful studies'' \cite[p. 77]{BaadeZwicky1934b}. 

Speculations on planetary nebulae,  as originating in novae, with their gaseous expanding shells as the remains of past outbursts, even suggesting an origin in outbursts of several stars, provided a well defined scenario --- on a large space-time scale --- of a phenomenon suggesting a process in which matter expanded after an explosion. Already in 1923, for example, J. H. Reynolds concluded an article on gaseous planetary nebulae with the following words: ``The old idea that the gaseous nebulae were the primitive forms of matter from which stars were evolved must, it seems, be given up for the exactly contrary hypothesis that they had their origin in stellar outbursts, where matter passed from complex to simpler forms by atomic disintegration under the stress of extreme temperature development'' \cite{Reynolds1923}. As already mentioned, the idea of stellar explosions  associated with collapse to a superdense configuration had been already suggested in connection with discussions on white dwarfs. In 1926, in comparing the nuclei of planetary nebulae to white dwarfs, Donald H. Menzel said in a section entitled `The physical state of the nuclear stars (white dwarfs)': ``Novae arise from giants and dwarfs, that is they are outbursts from dwarf stages of stars, that are probably experiencing these outbursts many hundred times during their history'' \cite[p. 307]{Menzel1926} However, the first very explicit description of the idea of stellar explosions associated with collapse to a dense configuration can be found in Milne's talk at  the meeting of the British association of October 1931 (Discussion on the Evolution of the Universe) \cite[p. 716]{Milne1931b}. Milne had recalled that during the contraction a star is losing gravitational energy, which is set free as heat and light, this shrinking must thus be ``the actual origin of the brightening [\dots] Since the rate of brightening is very rapid, we infer that the process of shrinkage is very rapid --- in fact cataclysmic. The process of shrinkage is a veritable collapse. In a nova outburst the star is seen to be collapsing on itself; and the suddenness of the collapse, and the resulting enormous amount of gravitational energy that must be got rid of in the short time available, conspire to produce the huge brightening of the star as observed. This sudden liberation of energy produces enormously increased radiation, which in turn expels the outer layers of gas. Such is the probable explanation of the origin of novae, or `new stars'''.  Milne also specified that ``the mass of the star, after the outbursts, is practically the same as before, yet it occupies a much smaller volume, hence its mean density must be much larger than before [\dots] The gases expelled from the star during the outburst are chips from the old block; but the star itself does not remain an old block; it becomes very much of a new block --- a very dense block''. Of course Milne immediately mentioned other dense stars, known as white dwarfs, and the nuclei of the planetary nebulae, both having probably undergone the process of collapse: ``It is reasonable to assume [\dots] that every white dwarf has been at one time a nova''. 
 
These speculations provided the astrophysical background, while the novelties derived by the new status of nuclear matter inspired Zwicky's further conjectures which resulted in an attempt to fill the collapse idea discussed by Milne and others with a more physical content. It is rather plausible that this part of their proposal came from Zwicky himself. His experience with dense  matter in crystals and metals most probably led him in a most natural way to reason on super dense neutronic matter in stars. The close packing of neutrons within dense stellar cores could explain the energy release in supernovae which he estimated to be equivalent to the annihilation of the order of several tenths of a solar mass. However, he could only guess at the scenario for forming neutron stars; all the physical mechanisms of the implosion, including the behaviour of matter in the core during the process and the actual emission of energy, remained completely unknown. What they estimated was the evaluation of energy involved in supernova explosions as if produced by particles or photons  that in turn was compared to the observations of the intensity of cosmic rays  made by Regener, and by Millikan and his collaborators. Lema\^itre's hypothesis of cosmic rays ``as remnants of some super radioactive process which took place a long time ago'' was mentioned by Zwicky exactly at that time.

 What has always been  duly termed a  `prescient' idea, was thus not coming out of the blue. It cannot be excluded that many of Zwicky's reflections about neutrons were inspired by the work of his colleague Langer, who was especially interested in the properties of neutrons, and also in the origin of cosmic rays, topics that he discussed at the same Stanford meeting of December 1933 in three different talks. The guiding concept in Baade and Zwicky's proposal appears in fact to be the problem of the origin of cosmic rays, seen as a mysterious radiation whose `cosmic' nature was still attracting the main attention, notably at Caltech, because of Millikan's presence. Millikan, the director of the Norman Bridge Laboratory of Physics at Caltech, had since the 1920s advocated that cosmic rays were high-energy gamma rays produced during the birth of elements in the universe, and had undertaken a major study of the radiation. Zwicky was thus definitely familiar with the problem. That same 1932,  a worldwide measurement campaign investigating a possible dependence of the rate on magnetic latitude was led by Arthur H. Compton and established beyond any doubt that a part of the primary radiation consists of charged particles. Moreover, parallel experiments also proved the existence of the east-west effect, hypothesized in 1930 by the Italian cosmic ray physicist Bruno Rossi. According to his prediction there should be an azimuthal asymmetry in the intensity of cosmic rays that would depend on the sign of the charge of the primary particles.   Both the charged nature of cosmic rays (also verified by the latitude effect) \textit{and} the sign of the charge, were determined by such experiments \cite{Bonolis2014}. Research on cosmic rays was already becoming strongly related to the emerging field of elementary particle physics, and the problem of their origin was gradually less investigated, at least up to the 1940s, when it was possible to establish the nature of the primary radiation. At that time the problem of their origin again became a hot subject, also in connection with other astrophysical developments. 

 Zwicky, Baade, and all other astronomers in Pasadena  were following Hubble's work and had witnessed Lema\^itre's lectures on the expanding universe and the primeval-atom hypothesis during his journey in the U.S. Already in early September 1932, during the Fourth General Assembly of the International Astronomical Union, which took place at Cambridge, Massachusetts. There, Eddington's public lecture on the expanding universe was a climax event and Lema\^itre's ``fireworks theory of the beginning of things'' was widely discussed \cite[p. 373-375]{Deprit1984}. Lema\^itre remained for some time working with Vallarta on his hypothesis for the origin of cosmic rays and both participated to the meeting of the American Physical Society that same November, where Arthur Compton presented the preliminary results of his survey of the intensity of cosmic radiation at a large number of stations scattered all over the world, widely confirming previous observations and ruling out the hypothesis that the radiation consisted of photons alone and that it was made up at least partly of charged particles. This question, according to Lema\^itre, was very likely bound up with general cosmogonical problems, even if the question as to their origin remained unanswered. Moreover, in November Lema\^itre was invited by Percy H. Robertson to give a seminar on his cosmology in Princeton, obviously attended by Einstein, and in December he moved to Caltech, where he also met Hubble. His  seminars in which he discussed his astounding theories on the  expanding universe and on the cosmic rays as the remains of the primordial universe, were widely spread by a long article on the \textit{New York Times} Magazine appearing in February1933. By that time Zwicky had already begun his speculations on the origin of cosmic rays, and the red-shift phenomenon of far away galaxies.
  
  In January 1933, Zwicky investigated the problem of the origin in an article entitled `How far do cosmic rays travel?' in which he tried ``to establish a relation between them and the red shift of extragalactic-nebulae'' examining two entirely different hypotheses: the one suggesting that cosmic rays must be of local origin (upper atmosphere, planetary system, etc.) and the second one, especially advanced by Robert A. Millikan, that they were produced throughout interstellar or intergalactic spaces \cite{Zwicky1933a}. Zwicky had in fact concluded from the results of observations on the red-shift of extragalactic nebulae, that the amount of dark matter in the Universe must be grater than that of luminous matter, and he thus tried to establish a connection between these two phenomena \cite{Zwicky1933b}.\footnote{Zwicky measured the velocity dispersion of the galaxies in the Coma cluster and found that there must be about 100 times more dark, or hidden, matter as compared with visible matter in the cluster. In this article Zwicky discussed redshift in connection with cosmological theories and explicitly mentioned: ``Another important proposal was made by Friedmann, Tolman, Lema\^itre and Eddington, whose work shows that according to the theory of relativity a static space is dynamically unstable and therefore tends to contract or expand. This result was interpreted by him to imply that the redshift would correspond to a factual expansion of space''. In his editorial note to the English translation of Zwicky's paper \cite{Zwicky2009}, J\"urgen Ehlers suggests that Zwicky did not specify which of the four names he meant, but that in reality this proposal was first made by Lema\^itre \cite{Ehlers2009}. Actually Tolman himself became really involved in cosmology around 1930-1931, in connection with Hubble's results about the red-shifts of the extragalactic nebulae being proportional to their distances \cite{Hubble1929} and when Lema\^itre's work became widely known also in the United States.}
  The connection between the origin of cosmic rays and the redshift phenomenon related to the expanding universe in Zwicky's research, is strongly suggesting that Lema\^itre's ideas on the expansion of the universe and especially about the primeval atom and its explosive nuclear processes provided a strong conceptual platform as a starting point for reflections on relativistic cosmology and in particular on the problem of cosmic rays, eventually leading to the theory of supernovas. Baade and Zwicky mentioned the possibility that either the cosmic rays ``originate in intergalactic space or that they are survivors from a time when physical conditions in the universe were entirely different from what they are now (Lema\^itre)'', but they considered both hypotheses to be very unsatisfactory and for this reason they made ``an entirely new proposal'' removing some of the major difficulties concerning the origin of cosmic rays \cite[p. 260]{BaadeZwicky1934a}.  In 1931 Regener, too, had  speculated on cosmic rays as a remain of an original explosion in connection with Einstein's closed universe \cite{Regener1931}.
   
 Lema\^itre's theory of a dense primeval state whose ``explosive'' expansion could gave origin also to cosmic rays, in connection with the growing role of neutrons in astrophysical realm, might well explain why a star consisting only of neutrons, that Fl\"ugge had considered a mere `curiosity', became a basic assumption in Baade and Zwicky's theory of neutron stars as remnants of supernova explosions, that in turn became the source for high energy cosmic rays. Milne himself had suggested \cite{Milne1930a} that novae resulted from the collapse of stellar cores, then becoming white dwarfs, that is very dense stars. In turn, the collapse to a superdense configuration had led to Sterne's and Fl\"ugge's suggestions that compressed matter in stars would result in neutronization. All this was part of Zwicky's conscious and unconscious imagination.

Baade and Zwicky did not mention Landau and Chandrasekhar, or any other work about the maximum mass of white dwarfs. Any connection would require a far deeper knowledge of nuclear theory and nuclear reactions. In any case, no relationship was established at the moment between these two compact objects: white dwarfs and the hypothetical neutron stars. However, as astronomers, they had  recognized the existence of a special class of stars, the supernovae, that during several weeks radiate as much energy as a whole galaxy of stars. This suggested that observation of these unique objects would furnish valuable information on fundamental problems such as the generation of energy in stars, the evolution of stars and stellar systems, the origin and characteristics of cosmic rays. Baade and Zwicky thus felt strongly motivated to start a systematic search of supernovae, that promised to be particular significant.

 \section{Chandrasekhar and the final fate of a white dwarf}

Towards the end of 1931, Chandra began to feel uneasy. His results on model stellar photospheres presented at the January 1932 meeting of the Royal Astronomical Society were much appreciated by both Milne and Eddington, who were following his work with great attention, apparently because they hoped that new results would confirm their own theories. However, he was still a PhD student, and in trying to measure up to such established and incredibly influential astrophysicists such as Eddington and Milne he was in reality an outsider within this small scientific community. Moreover, he felt that: ``Physics, was at the center, not astrophysics'' \cite[p. 98]{Wali1990}. Later Chandra recalled that Dirac told him \cite{Chandra1977}: ``Well, if I were you, I would be interested in relativity, rather than astrophysics''. Chandra then asked him: ``One time you did write a paper on astrophysics\dots'' and Dirac answered: ``Oh, that was before quantum mechanics''. 
 All this made Chandra feel afraid that astrophysics was considered inferior by most physicists. He felt alone and even thought of entering the field of theoretical physics. He greatly admired Dirac, with whom he had developed a friendly relationship, and told him how unhappy he was in Cambridge, so that Dirac suggested him to spend some time at Niels Bohr's Institute in Copenhagen, where Chandra went during his final year, before the end of his Government scholarship. He stayed there from August 1932 to May 1933, finding a friendly, informal and international atmosphere. During this period he established a strong relationship in particular with L\'eon Rosenfeld, who was much interested in Chandra's work, and at the same time could discuss common research issues with Bengt Str\"omgren, who very often visited Bohr's institute and had a strong physical background. Both Chandra and Str\"omgren represented, even if in different perspectives, a new figure of astrophysicist, strongly familiar with the physicists' community also because of university education.

At the time Bohr told Chandra that ``Well, I've always been interested in astrophysics, but the first question I should like to know about the sun is: where does the energy come from? And since I can't answer that question, I do not think a rational theory of the stellar structure is possible''. In recalling this conversation, Chandra added \cite{Chandra1977}:  ``Well, great as Bohr is, that remark of Bohr's is invalid. Later on, if one found the right nuclear reactions, it was because one had found out earlier the right temperatures and physical conditions by their ingenuity''. Here Chandra is certainly referring to what Bethe himself recognized about his theory on stellar energy and how it was inspired by the insight coming from Str\"omgren's work, that will be explained later.

 In a report written by Bohr in October 1933, concerning the work of Chandrasekhar during his stay in Copenhagen from August 1932 to May 1933, he declared: ``I am glad to take this opportunity for expressing my high appreciation of the scientific work which Mr. Chandrasekhar has performed in the course of his studies in this institute since September 1st 1932. During this time he has been successfully engaged in the theoretical treatment of a number of important astrophysical problems, and as well in the choice of these problems as in the methods used for their solution he has shown great ingenuity and ability. In my opinion he may be regarded as one of the most competent among the younger astrophysicists, as to whose future scientific activity great expectations are justified''.\footnote{Archives for the History of Quantum Physics, Bd. AHQP/BSC 19: Niels Bohr. Scientific correspondence, 1930--1945.}
 
By the end of 1932 Chandra had published four papers on rotating self-gravitating polytropes, which became his Ph.D. thesis. In  \cite{Chandra1932} he considered stars whose mass exceeds the critical mass and concluded that for these stars ``\textit{the perfect gas equation does not break down, however high the density may become, and the matter does not become degenerate. An appeal to Fermi-Dirac statistics to avoid the central singularity cannot be made}''.  The only way out of the singularity, added Chandra, ``is to assume that there exists a maximum density $\rho_{max}$ which matter is capable of''. However, at the very end of the article he wrote: ``We may conclude that great progress in the analysis of stellar structure is not possible before we can answer the following fundamental question: \textit{Given an enclosure containing electrons and atomic nuclei, (total charge zero) what happens if we go on compressing the material indefinitely?}''.\footnote{In this article, written in Copenhagen, Chandra cited \cite{Landau1932} and thanked Str\"omgren for advice. The latter most probably attracted his attention on Landau's paper.}

In October 1933 he was elected to a Trinity Fellowship, ``one of the most gratifying events that can happen to one'', as remarked by Milne in a letter he hastened to send him as soon as the news was announced \cite[p. 109]{Wali1990}. The Fellowship put him in contact with the Cambridge scientific society and he also got invitations from abroad. In particular from Boris P. Gerasimovi\v{c}, who had just become the director of the Pulkovo Observatory, near Saint Petersburg.\footnote{Later Stalinist purges in 1936-1937 devastated Russian astronomy and destroyed Pulkovo as an active research institute and the effect on Russian astronomy was to be felt for a very long time \cite{Eremeeva1995}.}

They had been in contact for some time and Chandra was eager to see Russia. During this four-week trip, he met Landau and Viktor A. Ambartsumyan and gave two lectures at  Pulkovo, one of which about his work on white dwarfs and the limiting mass. The brilliant Ambartsumyan, who was organizing the Soviet Union's first department of astrophysics, fully grasped the significance of Chandra's work on dwarf stars and suggested that he investigate the problem in greater detail working out the exact, complete theory of white dwarfs, (i.e., by direct radial integration of the equations, using the complete pressure-density relation), devoid of some simplifying assumptions, and to examine the entire range of densities, within the framework of relativistic quantum statistics and the improved knowledge of stellar interiors. Chandrasekhar felt again encouraged to tackle such immense problem.

Since the beginning,   Chandra's work had actually been related to fundamental issues involved in the Milne-Eddington controversy on the nature of the boundary conditions one should use in determining the equilibrium configurations of stars.  The existence of a limiting mass contradicted Milne's idea that \textit{all} stars had a degenerate core surrounded by outer layers of stellar material obeying the perfect gas equation of state. During the period 1932-1934, Chandra had been occupied with finishing his degree, moreover there had not been so much impact from his work. But now, Ambartsumyan's suggestion to explore again the problem represented a new challenge that might also settle the controversy. Eddington, who was personally interested in this new work, hoping that his ideas would prevail, even lent him  a Brunsviga hand calculator, that was a fundamental tool for solving numerically the differential equations related to the equations of hydrostatic equilibrium for each white-dwarf star of his sample. 

By the end of 1934 Chandra had completed a detailed analysis on the problem of the limiting mass, distinguishing between dense matter obeying the equation $p\sim\rho^{5/3}$ and ultradense matter which obeys the equation $p\sim\rho^{4/3}$. He reached a conclusion that a limiting mass is obtained only for the ultradense case, which he stated in the following terms \cite[pp. 373-377]{Chandra1934a}: ``\textit{The life-history of a star of small mass must be essentially different from the life-history of a star of large mass. For a star of small mass the natural white-dwarf stage is an initial step towards complete extinction. A star of large mass cannot pass into the white-dwarf stage and one is left speculating on other possibilities} [emphasis added]''.\footnote{See also `Stellar configurations with degenerate cores' \cite{Chandra1934b}.}

\begin{figure}[h]
\centering

\resizebox{0.75\columnwidth}{!}{%
  \includegraphics{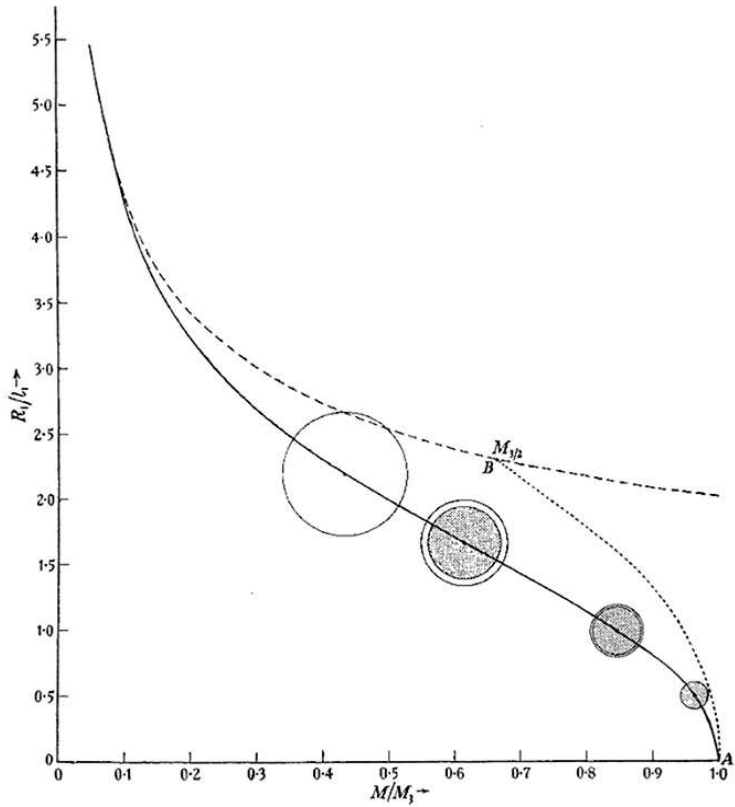} }
\caption{The mass of the white dwarf along the abscissa, is measured in units of the limiting mass (denoted by $M_3$) for a stable white dwarf, that is 5.728 divided by the average molecular weight squared, a ratio directly emerging from his theory. The full line curve represents Chandra's theory, showing the exact (mass-radius) relation for completely degenerate configurations, showing stars with highly collapsed configurations at different stages.  This curve tends asymptotically to the dotted line curve. As the mass of the white dwarf ($M$) approaches the maximum mass ($M_3$), the star shrinks while the radius $R$ becomes zero.  The dashed curve represents the relation $M=constant \times R^{-3}$ that follows from the non-relativistic equation of state $p=k_1(n_e)^{5/3}$ (low densities), thus representing Fowler's theory. The curve continues forever, thus showing that Fowler's theory does not predict a maximum mass; at the point B along this curve, the threshold momentum $p_0$ of the electrons at the centre of the configuration is exactly equal to $mc$. Along the exact curve, at the point where a full circle (with no inner circle) is drawn, $p_0$ (at the centre) is again equal to $mc$; the inner circles of the other circles represent the regions in these configurations where the electrons may be considered to be relativistic ($p_0\geq mc$). The dotted line shows the transition from the core in Fowler's theory to the one in Chandra's \cite[p. 219]{Chandra1935a}.}
\label{fig:Chandra1935}    
\end{figure}

On January 1, 1935, Chandra completed the paper ``The highly collapsed configurations of a stellar mass (Second paper)'' \cite{Chandra1935a}, a follow up of his \cite{Chandra1931c}, where he is clearly showing  that the existence of a limiting mass (that for a mean molecular weight per electron = 2 was 1.44 solar masses) meant that a white-dwarf state does not exist for stars that are more massive. This paper includes a figure [Fig. \ref{fig:Chandra1935}] exhibiting the mass-radius relation deduced on the basis of the exact equation of state allowing for the effects of special relativity of which equations $M=constant\times R^{-3}$ and $p=k_2(n_e)^{4/3}$ are the appropriate limiting forms, where $k_2$  is an atomic constant  and $n_e$ is the electron concentration. The effect of special relativity is to reduce the power of the pressure dependence on density from 5/3 to 4/3. This limiting form of the equation of state has a dramatic effect on the predicted mass-radius relation: the radius must tend to zero as a certain limiting mass is reached.

He remarked how one could notice clearly from these two curves ``how marked the deviations from the limiting curves become even for quite small masses,'' and how the relativistic effects are quite significant even for small masses. ``These completely collapsed configurations, continued Chandra, have a natural limit, and our exact treatment now shows how this limit is reached''.  He extended the discussion in a second paper dated January 4, and concluded that the developed methods and the results obtained ``would have to be extended for more general stellar models before any very definite conclusions could be drawn''. \cite{Chandra1935c}. 

Chandra gave an account of this work in the January 1935 meeting of the Royal Astronomical Society, of course showing Fig. \ref{fig:Chandra1935}, a clear definitive demonstration of what might happen to a white dwarf exceeding Chandra's maximum mass. Eddington attacked him frontally \cite[p. 38]{Eddington1935a}: ``Chandrasekhar shows that a star of mass greater than a certain limit remains a perfect gas and can never cool down. The star has to go on radiating and radiating and contracting and contracting until, I suppose, it gets down to a few kilometres radius when gravity becomes strong enough to hold the radiation and the star can at last find peace. Dr. Chandrasekhar had got this result before, but he has rubbed it in his latest paper; and, when discussing it with him, I felt driven to the conclusion that this was almost a \textit{reductio ad absurdum} of the relativistic degeneracy formula. Various accidents may intervene to save the star, but I want more protection than that. I think that there should be a law of nature to prevent the star from behaving in this absurd way''.   
Eddington recognized that Chandra had worked out correctly the astrophysical consequences of relativistic degeneracy, according to the current interpretation \cite[p. 195]{Eddington1935b}: ``I do not think that any flaw can be found in the usual mathematical derivation of the formula. But its physical foundation does not inspire confidence, since it is a combination of relativistic mechanics with non relativistic quantum theory''. In contending that the relativistic formula rested on a misconception (``It must at least rouse suspicion as to the soundness of its foundation''), Eddington examined this ``unholy alliance'' concluding that  the `relativistic' formula was ``erroneous'' and again correctly described the fate of a white dwarf with mass in excess of the critical value.\footnote{For a detailed discussion on Eddington and the controversy over relativistic degeneracy see \cite{Mestel2004}.} 

Having realized that relativistic degeneracy was incompatible with his theory, and yet having understood the alarming implications of Chandra's conclusions, Eddington paradoxically did not follow his own physical insight, accepting the physical reality deriving from relativistic degeneracy: in his eyes Chandra had actually revived the very same apparent difficulty solved by Fowler. Actually, already ten years before, Eddington had exactly described what would be the relativistic effects of a very powerful gravitational field exerted by a very big star with a mass between 10 and 100 times greater than the sun: ``It is rather interesting to notice that Einstein's theory of gravitation has something to say on this point. According to it a star of 250 million km. radius could not possibly have so high a density as the sun.  Firstly, the force of gravitation would be so great that light would be unable to escape from it, the rays falling back to the star like a stone to the earth. Secondly, the red-shift of the spectral lines would be so great that the spectrum would be shifted out of existence. Thirdly, the mass would produce so much curvature of the space-time metric that space would close up round the star, leaving us outside (i.e. nowhere)''. Eddington then added that the same argument could be found in the writing of Laplace (\textit{Syst\`eme du Monde}, Book 5, Cp. VI): ``A luminous star, of the same density as the earth, and whose diameter should be two hundred and fifty times larger than that of the sun, would not, in consequence of its attraction, allow any of its rays to arrive at us; it is therefore possible that the largest luminous bodies in the universe may, through this cause, be invisible'' \cite[p. 6]{Eddington1926}. Eddington of course perfectly knew the Schwarzschild solution, but the above arguments again show that he did not believe in its physical reality. 

In his paper `Stellar configurations with degenerate cores' \cite{Chandra1935c}, Chandra thanks McCrea, von Neumann,  Rosenfeld and  Str\"omgren ``for the encouraging interest they have taken in these studies and for many stimulating discussions''. All of them were his personal friends. However physicists did not want to enter openly the arena of such controversy, in part because Eddington was a most  influential scientist, but also because they did not take Eddington seriously any more and thought that it was not worthwhile losing time in sterile discussions of what they considered completely wrong ideas. Moreover, astrophysics was still a field  far away from the exciting new issues coming from theoretical and experimental physics of the early 1930s. On the other hand, Eddington was still admired as an authority by astronomers. So that on both sides, people chose not to be involved, or thought it was not worthwhile being involved, even if we know that physicists completely agreed with Chandra's work. As Chandra later recalled \cite{Chandra1977}: `` [\dots] all these people who supported me never came out publicly. It was all private''.
Actually, it was not completely like that. There was a solidarity from his young colleagues under the form of collaboration in articles. The more explicit one was one with Christian M\o ller  \cite{ChandraMoller1935}.  As Eddington had questioned the validity of the relativistic equation of state for degenerate matter, which by that time was generally accepted, they used Dirac's relativistic wave equation presenting arguments providing grounds ``for not abandoning the accepted form of the equation of state''. Eddington reacted to their article defending the relativistic degeneracy formula with a Note on `relativistic degeneracy' \cite[p. 20]{Eddington1935c}: ``In recent papers  I have contended that the  `relativistic' degeneracy formula is erroneous. This has led M\o ller and Chandrasekhar  to publish a note defending it. They give a derivation of the formula which is doubtless more up to date than those which I criticized. It therefore seems desirable that I should amplify my attack on the formula by showing why I am unable to accept M\o ller and Chandrasekhar's proof''.  

Chandra's relationships with young physicists is also testified by an investigation he carried on with L\'eon Rosenfeld on the deviation from perfect laws arising from causes other than degeneracy like the production of electron pairs and that resulted in a work published at that same time \cite{ChandraRosenfeld1935}.\footnote{In Chandra's biography \cite[pp. 129-131]{Wali1990} a correspondence with Rosenfeld, who was working with Bohr in Copenhagen, is mentioned in relationship to that period, January 1935. Bohr, too, expressed the opinion that there was nothing wrong in Chandra's formulation.  On January 29, 1935, Rosenfeld wrote Chandra, also on Bohr's behalf: ``Would you agree for us to forward confidentially Eddington's manuscript to Pauli, together with a statement of the circumstances and asking for an `authoritative reply'?'' About Eddington's manuscript, Rosenfeld remarked: ``After having courageously read Eddington's paper twice, I have nothing to change in my previous statements; it is the wildest nonsense''. Pauli declared that ``Eddington did not understand physics'', but, as Chandra wrote to Rosenfeld, ``astronomers continued to believe in Eddington''.}  
Later Rudolf Peierls recalled: ``I did not know any physicist to whom it was not obvious that Chandrasekhar was right in using relativistic Fermi-Dirac statistics, and who was not shocked by Eddington's denial of the obvious, particularly coming from the author of a well-known text on relativity. It was therefore not a question of studying the problem, but of countering Eddington. It was for this purpose that I wrote my paper in the \textit{Monthly Notices} \cite{Peierls1936} [\dots] I do not believe Eddington ever took any notice of my paper''  \cite[p. 135]{Wali1990}.\footnote{Peierls was referring to a controversy arisen as to whether the pressure-density relation of a degenerate relativistic gas enclosed in a certain volume would be independent of the shape of the volume \cite[p. 258]{Eddington1935d}. According to Peierls, ``This might seem sufficiently obvious to make a proof unnecessary'' but in view of the controversy it was worthwhile to give a proof\dots So that he assumed that ``the present form of quantum mechanics applies to the problem'', and only proves that from this theory one obtains the usual equation of state.} 

Many years later Chandra told  his biographer \cite[p. 143]{Wali1990}: ``Kamesh, suppose, just for a moment, Eddington had accepted my result. Suppose he had said, `Yes, clearly the limiting mass does occur in the Newtonian theory in which it is a point mass. However, general relativity does not permit a point mass. How does general relativity take care of that? If he had asked this question and worked on it, he would have realized that the first problem to solve in that connection is to study radial oscillations of the star in the framework of general relativity. It's a problem I did in 1964, but Eddington could have done it then in the mid-1930s! Not only because he was capable of doing it --- he certainly had mastered general relativity --- but also because his whole interest in astrophysics originated from studying pulsations of stars. And if he had done it, he would have found that the white dwarf configuration constructed on the Newtonian model became unstable before the limiting mass was reached. He would have found that there was no \textit{reductio ad absurdum}, no stellar buffonery! \textit{He would simply have found that stars became unstable before they reached the limit and that a black hole would ensue. Eddington could have done it}. When I say he could have done it, I am not just speculating. It was entirely within his ability, entirely within the philosophy which underlies his work on internal constitution of stars. And if Eddington had done that, he would stand today as \textit{the greatest theoretical astronomer of this century}, because he would have predicted and talked about collapsed stars in a completely and totally relativistic fashion. It had to wait thirty years'''. 

Such an exploration, commented Wali, ``was not outside Chandra's ability either.\footnote{See \cite[pp. 135-146]{Wali1990} for a description of Chandra's relationship with Eddington and the circumstances that led him to change his field of interest and go into something else: ``It was a personal decision I made at the time''. He definitely felt ``totally discredited by the astronomical community''.} 
 He reported some of his work on rotating white dwarfs at the 1939 Paris meeting, and the paper in the Synge volume published in 1972  contains an almost verbatim account of the work he had done in 1935''.\footnote{Wali is referring to `A limiting case of relativistic equilibrium' \cite{Chandra1972}. And actually, in 1962 Chandra decided to turn to general relativity --- a subject he was first introduced to during his first year as a graduate student in Cambridge. In 1964 he worked out the theory of pulsation of spherical stars in the framework of general relativity, proving their relativistic instability against gravitational collapse. This most cited work  marked Chandra's entry into the `seventh period' of his scientific life, which started around 1960, when he began to study general relativity thus being ready to work in relativistic astrophysics in coincidence with the discovery of quasars \cite{Friedman1996}.} 
 
In this regard, the most interesting of Chandra's friend of that time was John von Neumann. According to Chandra's later recollections, they became quite friendly during the period 1934-1935, when von Neumann was in Cambridge, on leave of absence from Princeton. This happened exactly at the time when Chandra had his controversy with Eddington. Chandra acknowledged that \cite{Chandra1977} ``Neumann was one of the people who privately supported me against Eddington [\dots] I got to know Neumann rather well. I was a fellow at Trinity at that time, Neumann used to visit me in my rooms in Trinity quite frequently. I think he was rather lonely in those days, so he would quite often come up to my rooms in the college and sit down and work in my rooms, and so I got to know him rather well [\dots] We used to go out for walks''.
 In the spring of 1935, they discussed Eddington's objections \cite[p. 143]{Wali1990}: ``John said, `If Eddington does not like stars to recede inside the Schwarzschild radius, one probably should try to see what happens if one uses the absolute, relativistic equation of state'. We started working on that together, but to go on we had to study equilibrium conditions within the framework of general relativity''. In 1934 von Neumann had discussed with Abraham H. Taub and Oswald Veblen the extension of the Dirac equation to general relativity \cite{NeumannTaubVeblen1934}, and was thus in the right position to recognize that Chandra's problem of the limiting mass almost naturally led to apply general relativity, as on the other hand Eddington's acrimonious comments were implying.
 As Chandra further recalled: ``at that time we started to work on some problems in relativistic gas spheres; it didn't go very far. I do remember our discussions of that year, and I did some work and published a paper in the late early seventies, on precisely the problem which Neumann and I discussed in 1934 --- the problem of isothermal gas spheres in general relativity. In a way, it shows Neumann's great insight. He said, `If objects are going to collapse, then they must collapse to smaller dimensions. We ought to look at it in the framework of general relativity\dots'. We were in the right direction. And in this instance I must say that it was Neumann who took the initiative''.\footnote{In `Stellar configurations with degenerate cores (second paper)' \cite{Chandra1935a}, Chandra cited an unpublished result of von Neumann, who ``has shown that the \textit{very} ultimate EOS [Equation of State] for matter should \textit{always} be $P=\frac{1}{3} c^2 \rho$''. And actually, in von Neumann manuscripts, there are notes written in 1935, which were published  by Abraham H. Taub in the 6th volume of his \textit{Collected Works} \cite{NeumannCW}. In the first note (p. 172), where he studied the nature of the `Static solutions of Einstein field equations for perfect fluid with $T^{\rho}_{\rho}=0$',  the space-time was assumed to be a static spherically symmetric one. The discussion of such solutions was reduced to the discussion of a differential equation in which pressure and density  satisfied $\rho=3p$ and the result was compared with that obtained in the classical theory. In the second note (p. 173), `On relativistic gas-degeneracy and the collapsed configurations of stars', von Neumann is approximating the equation of state of degenerate matter presumably occurring in white dwarf stars by different equations for various ranges of the density.} 
 
However, soon von Neumann left Cambridge and probably involved in different researches abandoned his work on the problem. Chandra on the other hand ``got sufficiently discouraged with the situation to leave the problem alone''. So, all this turned into a  lost occasion. 

In spite of his relationships, Chandra was still very young and moreover all the questions appeared to most physicists as a side problem respect the growing field of nuclear physics, the fundamental issue of the sources of stellar energy and other relevant theoretical developments like quantum electrodynamics and the emerging topic of particle physics. As Chandra told Wali  \cite[p. 145]{Wali1990}: ``I felt that astronomers without exception thought that I was wrong. They considered me as a sort of Don Quixote trying to kill Eddington''.   Wali, Chandra's biographer, immediately commented \cite[p. 144]{Wali1990}:  ``The moral is that a certain modesty of approach toward science always pays in the end. These people [Eddington, Jeans, Milne], terribly clever, of great intellectual ability, terribly perceptive in many ways, lost out because they did not have the modesty to say, `I am going to learn from what physics teaches me.' They wanted to dictate how physics should be''. As a matter of fact, Chandra's work had been ``completely and totally discredited by the astronomical community'', so that he decided ``to change the field of interest and go into something else''. In fall 1935 he received an offer from Harvard to lecture in `Cosmic Physics' and on November 30 he sailed from Liverpool bound to the New World, leaving behind his frustrating involvement in this clash of giants. 

Despite Chandra's feelings, theoretical astrophysics emerged during this period as a specialty dedicated to the physical interpretation of celestial phenomena. The strong connection established between the new generation of astrophysicists like himself and Str\"omgren with the physicists' community, was instrumental in their capacity of bringing new results from physics to bear on stellar problems. In turn, this interaction between the two communities, stimulated some theoretical physicists to tackle astrophysical problems from the point of view of nuclear physics, an exploding frontier field materialized by the new perspectives opened by the neutron. However, the extreme consequence of the limiting mass was still to be explored and this further fundamental step would be triggered by more systematic investigations on the pressing issue of generation of energy in stars, which during the 1930s evolved into a hot research topic within the physicists' community. An important premise in this sense were laid down by studies systematically analyzing the properties of neutronic matter in stars, a study inaugurated by Fl\"ugge  in his dissertation \cite{Flugge1933}.

\section{Hund and Kothari: neutronic matter in stars}

As early as 1936, an extensive review on the status of the theory of matter under high pressure and temperature was prepared by Friedrich Hund  \cite{Hund1936}. Hund's relevant work in the quantum theory of solids and in the electrons in crystal lattices, as well as his interest in the field of nuclear physics,  led him to analyze such physical aspects, using stars as cosmic laboratories providing information about the actual existence of such extreme states. At the same time, regularities in the observed properties of stars could provide support for the relevant laws of matter.  Basing on fundamental physical considerations, Hund systematically investigated the properties of a gas of electrons, nuclei, protons, and neutrons, when the temperature and density are extremely high: ``From what is known about the $\beta$ decay of nuclei, one can conclude that protons can transform into neutrons by absorbing electrons or emitting positrons.  Based on \cite{Sterne1933b,Sterne1933c,Sterne1933d}, Hund remarked that ``at high pressures it can prove to be favorable for the electrons and the nuclei together to transform into neutrons''. \cite[p. 230]{Hund1936}.

\begin{figure}[h]
\centering

\resizebox{0.90\columnwidth}{!}{%
  \includegraphics{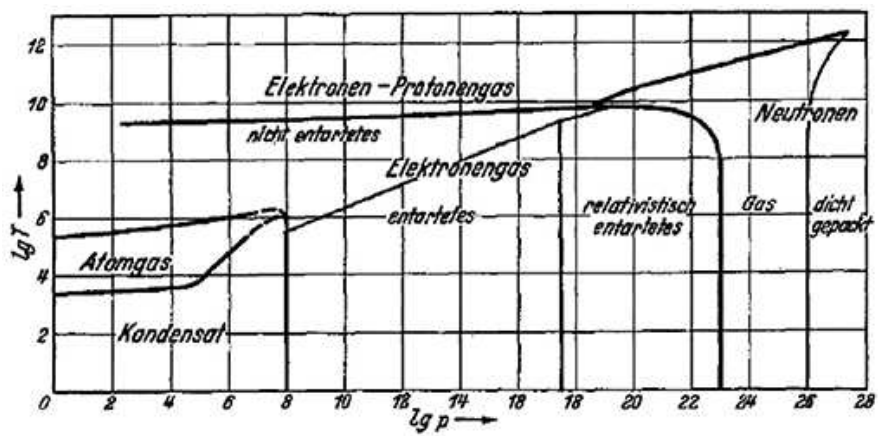} }
\caption{Phase diagram of matter from \cite[p. 232]{Hund1936}, showing the results of calculations for the following systems: a) nonrelativistic nondegenerate electrons and heavy particles; b) nondegenerate electrons and heavy particles, relativistic electrons, non-relativistic heavy particles; c) relativistic degenerate electrons, nonrelativistic nondegenerate heavy particles; d) relativistic degenerate electrons, nonrelativistic degenerate heavy particles.}
\label{Hund1936}       
\end{figure}

He then considered a gas consisting only of neutrons (see in particular the section `Das Neutronengas', p. 227) and the transformation processes occurring in regions of different equations of state of the particles. He was thus able to plot the boundaries between the different areas of electrons and nuclei, electrons and protons, and neutrons (Fig. \ref{Hund1936}). He found that beyond a certain value of the pressure the transformation of matter into neutrons occurred quite suddenly so that ``the nuclei and electrons rapidly disappear'' and ``matter behaves as a neutron gas''.  

After having systematically explored the properties of matter, Hund applied his investigations to make some order-of-magnitude predictions about the pressures and temperatures which occur within celestial bodies, such as planets, ordinary stars and, finally, `dense stars', as Hund  named white dwarfs, probably in one of the first explicit uses of this expression (see section `Die dichten Sterne', p. 253). He took for granted the existence of the limiting mass \cite[p. 254]{Hund1936}: ``Chandrasekhar   has calculated the structure of a star, for which the temperature is no longer important, with a more exact equation of state valid for the nonrelativistic and the relativistic electron gas. As the mass increases, the radius decreases; for the mass of the sun, the radius is approximately equal to that of the earth, and at even higher masses the radius tends rapidly to zero. The zero radius is reached for a finite mass only slightly larger than the mass of the sun. This last result should not be taken too literally, because, for calculating the equation of state, it was assumed to have unlimited validity for high pressures. This collapse to a zero radius (\textit{or to the corresponding value in the general relativity theory} [emphasis added]) stems from the high compressibility of matter in the state of the relativistic degenerate electron gas. If a star above the limiting mass were to have finite radius, the pressure would of course increase, but not fast enough to meet the corresponding increase in the weight of the above-lying layers''. Hund correctly remarked that the transformation of matter into neutrons would result in a greater limiting mass, but concluded that stars with sufficient mass could reach radii of the order of 10 km. How to avoid the `small radii problem'? A possible solution for stars with high masses would be to radiate large amounts of the gravitational energy set free in the process of contracting, reducing its mass significantly. ``As a \textit{possible final state in the evolution of stars}, concluded Hund, we are thus led to expect stars of moderate mass with very high densities''.
 
As a matter of fact, Hund was providing a physical base for the concept of neutron matter in stars. But, it was clear that at those pressures and densities the equation of state of nuclear matter was still far from being understood. And the generally spread hope was that something would intervene and save the star from a catastrophic collapse that was still a `black box', both  in term of the properties of matter in such extreme conditions and from the point of view of the collapse process itself, that nobody had still tackled. Hund, had someway briefly touched on the subject when he had mentioned the Schwarzschild radius, but only in brackets, as a side comment.

A step forward in this path was taken the  following year by Daulat Singh Kothari, a student of the renown astrophysicist Megh Nad Saha at Allahabad University and later of Ralph Fowler at Cambridge University \cite{Vardya1994}. He had written several papers on degeneracy and dense matter in celestial bodies and independently by Hund introduced the neutronization of matter in the interior of white dwarf stars by inverse beta decay process \cite{Kothari1937} calculating, for example, the value of the mass for which the electron concentration  would reach the maximum possible value beyond which all free electrons would combine with protons to form neutrons. But he was much more explicit in investigating  implications deriving from the transition to the neutron phase within superdense matter, thus setting the stage for a major role of neutrons within stellar cores.

  \section{The superdense core and the problem of stellar energy}

By the 1920s it had become clear that gravitational energy was insufficient as a source for powering stars. The radiation of the sun could not be maintained through a period of more than a billion years (the age of the earth at the time was estimated to be 3 billion years) solely through the release of gravitational energy. The release of nuclear energy through the transformation of hydrogen into helium was regarded as a likely mechanism. 

During the 1930s considerable progress was made in the field of nuclear physics, both through laboratory experiments and through further development of theory. The theory of stellar interiors had reached a point where the temperature, density and chemical composition of the central regions of main-sequence stars could be specified fairly accurately. Now the task was to compute, or estimate, which nuclear processes would be effective under such circumstances, what the reaction rates were, and how much nuclear energy would be produced per gram per second.  The physics of the nuclear processes in the sun naturally stood at the center of interest. Discussions on nuclear synthesis and stellar radiation were now based on neutrons, as units from which nuclei are built together with protons, and from which elements are formed in stellar interiors. A brief mention of the view ``that the stars contain central cores consisting largely of free neutrons'' since the early life of stars, where such large amount of free neutron would produce light and heavier elements by nuclear reactions, was made by Harold J. Walke in an article on nuclear synthesis and stellar radiation, \cite[p. 365]{Walke1935}.  He proposed ``a complete theory of nuclear synthesis by neutron capture and $\beta-$radioactivity'', regarding the neutron ``as a fundamental nuclear component, just as the electron is the fundamental extranuclear component''. On this theory therefore protons and $\alpha$-particles  would be formed mainly within nuclei as a result of the $\beta-$radioactivity. He also suggested that ``the initial condition of the universe'' consisted of a uniform distribution of neutrons and gamma-radiation. This primaeval gas, as previously suggested by Jeans, would be gravitationally unstable, and according to Walke it would condense ``to form huge non luminous nebulae''. As a result, hydrogen would be produced from the more frequent collisions between neutrons. Walke is also mentioning Baade and Zwicky \cite{BaadeZwicky1934b}: neutrons would accumulate at the centre of a star and thus, he concluded, element formation must take place in stellar interiors, where also  cosmic rays could originate \cite[p. 362]{Walke1935}.

Later, von Weizs\"acker, in remarking that at that time there wasn't very close contact between astronomy and physics, also added that, of course, in astronomy there was one great problem : ``every physicist who was working in fields like ours, like, for instance nuclear physics, knew that the problem of the interior of the stars was probably solved by Eddington, with the exception of the problem of the energy and that this was a problem of physics was clear, too. It was not clear how it was to be solved [\dots] we liked discussing this, of course [\dots]
I would say that people like say Nordheim, who at that time was also in G\"ottingen, or --- Placzek, Weisskopf, Bethe, the whole group, Bloch --- they all would have taken some academic interest. I mean, not an active interest, but some general interest in astronomical questions. But none of them, I think, had the idea that he would be working in astronomy''. While visiting Bohr's Institute in Copenhagen, von Weizs\"acker himself had discussed astrophysical issues with Str\"omgren, and had suggested in his monograph on atomic nuclei completed in the summer of 1936 that the quickly growing knowledge of nuclear reactions would suffice to resolve the stellar-energy problem \cite{Weizsacker1937a}. From these reflections arose his interest in seeking to explain how thermonuclear reactions could build elements up to their present abundances, thus opening the race to find a  solution of the stellar-energy problem \cite{Weizsacker1937b} \cite{Weizsacker1938}.\footnote{See \cite{Shaviv2009} for a detailed discussion of \cite{Weizsacker1937b} and \cite{Weizsacker1938}.}

\begin{figure}[h]
\centering
\resizebox{0.65\columnwidth}{!}{%
  \includegraphics{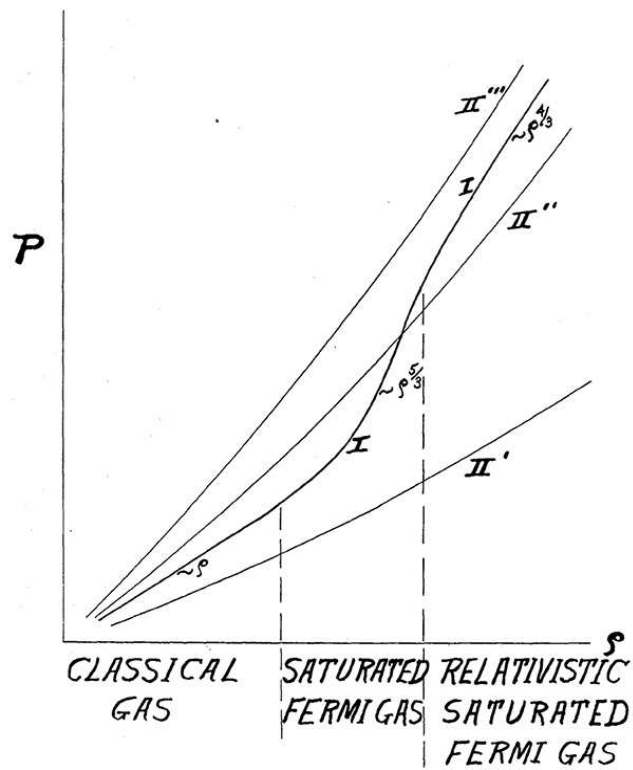} }
\caption{Diagrammatic representation of the outer pression-density curves corresponding to three different masses of the star (curves $II'$, $II''$ and $II'''$) compared with the pressure-density relation for an ideal electronic gas (curve I) \cite[p. 412]{Gamow1935}. 
}
\label{Gamow1935}       
\end{figure}

 By 1937-1938 it was a spread knowledge that energy-generation in stars is the conversion of hydrogen into helium. What was not established were the thermonuclear reactions involved in such process.  A turning point in these developments, was Gamow's growing interest for astrophysical issues, a new era in his scientific life. Already in 1933 he had written with Landau a paper investigating the process of thermal transformation of light elements in stars  \cite{GamowLandau1933} and was thus invited to give a talk in Paris on the evolution of stars. After participating to a meeting in London in 1934, he then emigrated to the United States. The issue of nuclear reactions powering stars, and the connected fundamental problem of the origin of chemical elements, was discussed by Gamow in a lecture at Ohio University, and later published in the \textit{Ohio Journal of Science}, a rather obscure journal \cite{Gamow1935}. After discussing nuclear transformations especially investigated by Fermi's group in Rome, Gamow shifted his attention from the experimental evidence obtained in the laboratory to the processes happening in the interior of stars. Apart from trying to outline the mechanisms for the building of elements, he also came ``to one of the most interesting questions concerning the physical state of the matter deep inside of stars [\dots] a mixture of two ideal gases: nuclear gas and electronic gas''. Basing on Landau's theory of 1932, according to  which most stars included a core of superdense `neutronic' matter of nuclear density, i.e. about $10^{12} g/cm^3$, Gamow gave a short account of Landau's calculations related to the equilibrium problem between the pressure of the electronic gas in the star's interior and the gravitational pressure of the outside layer and showed a diagrammatic representation for three different masses of the star (Fig. \ref{Gamow1935}).

Gamow observed that as far as the momentum is small compared with $mc$ the pressure $P$ of the ideal electronic gas is directly proportional to $\rho^{5/3}$. For larger densities velocities become relativistic and the pressure varies as $\rho^{4/3}$. The outside pressure $P'$ due to gravitation is proportional to $\rho^{4/3}$ and the coefficient of proportionality depends on the total mass $M$ of the star. If $M$ is small (curve $II'$) the inside pressure will be always larger; for somewhat larger mass (curve $II''$) there is a state of stable equilibrium between $P$ and $P'$ and the star will have inside a region filled up with non-relativistic saturated Fermi-gas. For still larger masses (curve $II'''$) the inside pressure would ``never be able to oppose the weight of stellar substance and the star would collapse into a mathematical point (!) unless, the further compression would be stopped by intranuclear repulsive forces between the particles of nuclear gas''. Here, the evolving knowledge on nuclear matter suggested Gamow a `nuclear argument'  to avoid the collapse that Landau had prevented by using Bohr's views about non conservation of energy in nuclear processes. Gamow then mentioned Landau's calculations on the limiting mass, and proposed that \textit{all stars possess such nuclei which evidently represent the sources of the stellar} energy radiated in such large amount into interstellar space''. Of course, he added that the question of the mechanism of energy-liberation was not yet quite clear. Moreover, proposed Gamow \cite[p. 413]{Gamow1935}, one could ``easily imagine that the stellar nucleus may not be considered as an inactive globe. The eruptive processes from the surface of the stellar nucleus will throw out the small pieces of nuclear substance which coming into the outside layer of the star will immediately disintegrate giving rise to the nuclei of different stable and radioactive elements''. Gawow expressed the hope that further investigations might clarify ``the relative importance of various processes and lead to a complete explanation of the relative abundance of different elements in the universe''.

In arriving to United States, Gamow was employed at George Washington University, where he always gave two regular courses, advanced courses \cite{Gamow1968}: ``relativity, quantum theory, nuclear physics''. But by this time he was more interested in applications of nuclear physics: ``Nuclear physics as such became boring for me, became too complicated, with all these complicated experiments and complicated theory, and I was doing nuclear astrophysics, so to speak --- the evolution of stars --- so I was mostly connected with astronomers, with people like Baade and Hubble [\dots] And whenever I went to California I was always going to talk to astronomers. I was in much closer intercourse with astronomers than with physicists at this time''.
At the same time, added Gamow, ``there was always this hostile feeling that astronomers, especially theoretical astronomers, didn't like me to invade their ground, because actually all these thermonuclear reactions in the stars were done by physicists --- me and Bethe and Houtermans and Weizs\"acker --- because astronomers didn't know about nuclear physics. They were sitting on their astronomical things [\dots] I started nuclear physics because in 1928 everybody was doing atomic and molecular structure, and van der Waals forces and doublets and triplets and spin and so on --- it was too
much. I didn't want to get mixed up with all this, so I decided to choose myself a corner where
nobody was doing anything, so I chose nuclear physics. And in time nuclear physics blew up into
a big thing, so I moved to nuclear astronomy, to nuclear astrophysics, cosmology''.

In the mid 1930s, Gamow thus fully focused on stars as a playground for his skills in the fast growing field of nuclear physics of which he had been a pioneer with his 1928 theory of quantum tunnelling. He had a great physical sense and great imagination. Focusing on nuclear processes, he  explored different stellar models in which problems of inner structure, energy sources and formation of elements in stars were all intermingled.\footnote{See \cite{Nadyozhin1995} and \cite{Cenadelli2010} for an analysis of Gamow's theorizing on stellar structure and evolution.} 
But apart from the fate of his models,  what is relevant for this narrative is that he continued to cultivate the idea that all stars might have a superdense core in their interiors. In his volume \textit{Atomic nuclei and nuclear transformations} \cite{Gamow1937} --- an upgrading of his \textit{Constitution of Atomic Nuclei And Radioactivity}, the very first textbook on nuclear physics \cite{Gamow1931} --- Gamow discussed the nuclear state of matter in the interior of a star in the preface, dated  May 1, 1936: ``For still higher densities [$>10^8g/cm^3$] electrons will probably be absorbed by the nuclei (an inverse $\beta-$decay process) and the mixture will tend to a state which can be described very roughly as a gas of neutrons''. For densities of the order of magnitude $\rho\sim 10^{12} g/cm^3$, average density of atomic nuclei, the conditions in the gas will become analogous to the conditions inside an atomic nucleus, pointed out Gamow, then citing Chandra and Landau in connection with the problem of pressure of degenerate matter in stars. He, too, mentioned the problem of `unlimited contraction' beyond a mass of about 1.5 solar masses, without any further comment.  

In showing that a gas of neutrons  could be compressed to a much higher density than a gas of nuclei and electrons, Gamow was calling such an extreme state of matter `the nuclear state' and  and the region of the star occupied by such nuclear matter the `stellar nucleus.'    

Although Gamow did not refer to it, because he completed the book in spring 1936, the microscopic descriptions of the equation of state of nuclear matter in beta equilibrium had also been independently given by Hund \cite{Hund1936}. In the chapter  `The new star model' of his \textit{Habilitationsschrift} published in 1936, Wilhelm Anderson, too, had talked of the formation in a few millions of years of a neutron core inside a star \cite[p. 72]{Anderson1936}.\footnote{The heavier neutrons would sink towards the center leaving behind a gas of electrons and protons. In this neutron-gas sphere, about half of the whole mass of the star would concentrate reaching enormous density and temperature unthinkable in the same condition for other kind of gases. In this way, so much contraction energy would be set free that it would be superfluous to look for other sources of energy for the sun. He then went on calculating on this model the contraction energy in the new model of star.}

In the final pages of the book, Gamow  then considers the conditions under which such stellar nuclei can really be formed. He then mentions \cite{Landau1932} and  \cite{Chandra1931a} arriving at discussing ``\textit{the final state of the star}'' up to the relativistic case, finding of course that for masses $M > M_0 \sim 1.5$ solar masses, ``equilibrium will never be possible for larger densities and the compression will proceed without limit'' \cite[p.  237]{Gamow1937}. He does not speculate further about the possibility ``for such unlimited contraction'', but he immediately remarks that ``unlimited contraction may start already for smaller masses than $M_0$, if we take into account the exchange attractive forces between particles [\dots] Thus we see that most of the stars, and possibly all stars, if the limiting mass $M_0$ is lowered by intranuclear forces, are subject to the formation of matter in the nuclear state in their interior at some period of their existence''.  Milne's dense collapsed cores, similar to white dwarfs tucked within stars, had now been transformed by Gamow into superdense neutron cores, possibly playing a role in fundamental nuclear processes within stars. ``The question whether most stars  actually possess such nuclei cannot, however, be answered definitely until the relevant astronomical evidence has been thoroughly examined, but there seems to be no reason why they should not'' concluded Gamow. 

In the very last lines of the volume \cite[pp. 234-238]{Gamow1937} he proposed the theory already exposed in the Ohio lecture, according to which ``eruptive processes of different types may go on continuously over the boundary between a large stellar nucleus and the surrounding matter in the ordinary gaseous state'' thus forming the nuclei of different elements. Moreover, one could easily see ``that pure gravitational energy liberated in the contraction to such immense densities will already be quite enough to secure the life of the star for a very long period of time''. This statement concluded the volume  that during the following years certainly contributed to the diffusion of views about the possible role of neutron cores in nuclear stellar processes and especially put a seal on the possibility of their existence. It represented a further important step towards the construction of a well founded physical model for Zwicky's speculative neutron stars.  
 
 In the meantime, the problem of stellar interiors and all the connected issues, especially the source of stellar energy, were being widely discussed within the physicists' community. On November 5, 1937, Landau sent to Bohr the English version of an article in which he proposed an upgraded version of his 1932 super dense core now transformed in a neutron core and asked him to send it to \textit{Nature}, if he would find that ``it contains some physical thoughts''.  And added that he would be very glad to hear his opinion on the article.

 On December 6 Bohr wrote to Landau, enclosing the proof of his letter to \textit{Nature} \cite{Landau1938}: ``As I think you know from my letter to Kapitsa, we were all in the Institute much impressed by the beauty of the idea and its promise. In the meantime we have, however, had a number of discussions on astrophysical problems, in which our attention has been directed to two reports in the Ergebnisse der Exakten Naturwissenschaften for 1936 and 1937, written by F. Hund \cite{Hund1936} and B. Str\"omgren respectively \cite{Stromgren1937}.\footnote{Str\"omgren throws some light on the connection between these two reviews recalling that ``one of those who came frequently to the Bohr Institute was Hund, and we discussed questions of stellar matters with him, and in the end it was agreed that he would write an article for the Ergebnisse on the physics of stellar-interior matter, and I would write the corresponding astrophysics review article [\dots] I found that, in the thirties, this is where they [physicists] got acquainted with stellar interiors, rather than through Eddington's book. For instance, a footnote by Tolman, shows how physicists got to know about the problem. There was also a limitation --- it was in German. But in those years, even in America, obviously German was studied [\dots] it was so necessary\dots German, in those years, when quantum mechanics was developing''. \cite{Stromgren1978}.} 

Landau answered on December 17 that he had added a citation to Hund's article, but on January 14, having received from Bohr the article of Str\"omgren, he stressed that after reading it he had not been able to find anything connected with his own work: ``Nur astrophysikalische Pathologie und etwas bekannte Kernumwandlungsphysik!'' [Only astrophysical pathology and some well known nuclear transformation physics]. On January 14 Landau was again writing to Bohr, after receiving a letter from M\o ller again discussing such topics on which he had reflected once more: ``The Str\"omgren's claims are unfortunately based on the wild Eddington's pathology which, as well known, is false not on one point but on all points. To unmask such pathology in a \textit{Nature} note is completely impossible, such unmasking would be longer and more complicated than the whole article''. As recalled by Peierls \cite[p. 163]{Peierls1996}: ``He was very critical, as was most of our generation of theoreticians, and the comment  `falsch oder trivial' about suspect papers, used often by Landau, was in common use. He was also fond of the term `pathologists' for people who wrote pathological papers, i.e. nonsense''. In spite of Landau's harsh comments, Str\"omgren's review was instrumental in introducing physicists to the problem of stellar interiors.

 In the starting lines of his article, Landau immediately stated that ``in bodies of very large mass'' the degenerate electron gas does not lead to extremely great densities, because of the `quantum pressure'. On the other hand, continued Landau citing \cite{Hund1936}, ``it is easy to see that matter can go into another state which is much more compressible --- the state where all the nuclei and electrons have combined to form neutrons [\dots] It is easy to compute the critical mass of the body for which the `neutronic' state begins to be more stable than the `electronic' state [\dots] When the mass of the body is greater than the critical mass, then in the formation of the `neutronic' phase an enormous amount of energy is liberated, and we see that the conception of a `neutronic' state of matter gives an immediate answer to the question of the sources of stellar energy. The sun during its probable time of radiation (about $2\times 10^{9}$ years according to general relativity theory) must have emitted something of the order of magnitude of $3\times 10^{50}$ ergs. The liberation of this amount of energy requires the transition of only about 2 per cent of the mass of the sun (with the assumption of constant density) or even only $3\times 10^{-3} M_\odot$ (with the Fermi gas model) to the `neutronic' phase [\dots] Thus we can regards a star as a body which has a neutronic core the steady growth of which liberates the energy which maintains the star at its high temperature'' Landau then expressed the hope that ``The detailed investigation of such a model should make possible the construction of a consistent theory of stars''. 
 
 As regards the question of how the initial core could be formed, Landau had already shown in 1932 that ``the formation of a core must certainly take place in a body with a mass greater than $1.5 \ M_{\odot}$''.  However, he now concluded the article with a challenging question regarding the stars with smaller mass, for which ``the conditions which make the formation of the initial core possible have yet to be made clear''.  

The last letter from Landau to Bohr is dated February 1, 1938 and he never replied to a letter by Bohr of July 5: ``As you know all here have been very interested in your most suggestive idea about stellar-constitution, and we have lately followed very closely the discussions about it, which have taken place among astrophysicists. We are all very eager to learn what progress you have made with it yourself''. He then continues with the exciting new perspectives ``about the origin of the nuclear forces opened by the discovery of the heavy electron [\dots] It would surely be most pleasant and instructive to all of us to discuss these various prospects with you and we hope very much indeed that you this year will be able once again to take part in our annual conference for the old and present collaborators of the Institute, which is planned to take place in the first week of October''.\footnote{Archives for the History of Quantum Physics, Bd. AHQP/BSC 19: Niels Bohr. Scientific correspondence, 1930--1945.}

In the meantime, Gamow, together with Merle Tuve and Edward Teller, was organizing the Fourth Annual Conference on Theoretical Physics sponsored by the George Washington University and the Carnegie Institution of Washington to discuss the burning problem of nuclear energy in stars.\footnote{When George Gamow  had been employed at George Washington University, the joint meetings organized by the Carnegie Institution and the University had come about as a condition for his employment in order to avoid the isolation from other theorists. Their style was obviously inspired by the conferences organized by Bohr in Copenhagen, having the same informal character and being very limited in size with no published proceedings. The first one, held in 1935, was devoted to a discussion of the latest problems of nuclear physics. The 1936 conference focused on molecular physics, and the third one on problems of the properties and interactions of elementary particles and the related questions of nuclear structure.}
 The fourth conference, devoted to `The problem of stellar energy and nuclear processes', was held in Washington, D.C., on March 21-23. It represented Gamow's official entering into the astrophysical realm, a circumstance well reflected by the mixed character of the invited scientists: astrophysicists studying the internal constitution of the stars (S. Chandrasekhar, B. Str\"omgren, T. E. Sterne, D. Menzel and others) and physicists working on different branches of nuclear physics (H. Bethe, G. Breit, G. Gamow, J. v. Neumann, E. Teller, M. Tuve, L. Hafstad, N. Heydenburg and others).  Chandra was at the time completing his book \textit{An introduction to the study of stellar structure}, and Str\"omgren had just written his review article on the theory of stellar interior and stellar evolution \cite{Stromgren1937}, both had recently moved to the University of Chicago. The stage was set for discussing astronomical observations, astrophysical theories and theoretical physics within a common perspective and establish a collaboration between astronomers, astrophysicists, and experimental nuclear physicists which led to the emergence of nuclear astrophysics as an established research field. 
 
It is not by chance that the first meeting was opened by Str\"omgren, who outlined in some detail the mathematical treatment and current status of the problem of temperature and density distribution and chemical composition in the interior of stars, with special reference to the critical features of the various particular stellar models used for these calculations. 
The bearing of current knowledge of nuclear reactions on the evaluation of the behavior of stars with nuclear sources of energy was reported by Gamow, while Bethe reported on the study of particular nuclear reactions which would lead to liberation of energy and to the building up of heavy elements. By that time Bethe's wide knowledge had just been displayed in his `trilogy' that later became known as the `Bethe Bible', presenting a complete coverage of nuclear physics published in the \textit{Reviews of Modern Physics} written between 1936 and 1937 in collaboration with his colleagues Bacher and Livingston.

Another question which brought about much discussion during the conference concerned the degree of central condensation of stars, together with the possible existence of a super-dense stellar nucleus, at least in some stars, as recently proposed by Landau. Chandra reported his investigations concerning the possibility of high central condensation in various known stars. His results lead to the conclusion that, whereas for giant stars the degree of central condensation is necessarily slight, there are stars for which as much as 90 per cent of the total mass is concentrated within less than half the radius from the center. Another aspect of the problem of central condensation was given by Sterne, who indicated the possibility of direct estimates of the density-distribution of double-star components from the observed characteristics of their orbits. The study of the stellar model having a highly condensed neutron-core of the Landau type was reported by Teller.  By direct integration of the equations of stellar equilibrium, one arrived for such models at extremely high temperatures ($\sim10^9$ C) and densities ($\sim10^9 g/cm^3$) near the surface of the core. Since under such conditions the already-known nuclear reactions would proceed with extremely high velocities, it was concluded that such a star model is inherently unstable.  Thus, as far as astrophysical evidence was concerned, the model of a star with a heavy stellar nucleus at the centre was \textit{not} confirmed, except possibly for supergiants, according to the report on the conference published on \textit{Nature} by Gamow, Chandra, and Tuve \cite{ChandraGamowTuve1938}.

 They made an interesting final remark:  ``Valuable contributions to the discussion of such superdense state of matter in a stellar interior from the point of view of general theory of gravitation was given by Neumann'' thus providing hints of a follow up of von Neumann's reflections on these issues going back to his discussions with Chandra at Cambridge in 1935.\footnote{At that time, Chandra included further unpublished results by von Neumann about the point-source model (in which it is assumed that the entire source of energy is liberated at the center of the star) in a dedicated section of his book \textit{An Introduction to the Study of Stellar Structure}, published in 1939. At page 332 he emphasized that ``Von Neumann's treatment of this problem is very powerful''. Two manuscript notes related to this issue, entitled `The point-source model' and `The point-source-solution, assuming a degeneracy of the semi-relativistic type, $p=K\rho^{4/3}$ over the entire star', were published in von Neumann's \textit{Collected Works}  \cite[pp. 175-176]{NeumannCW}.} 
 
  At this conference Hans Bethe was inspired, especially by Str\"omgren's new estimates of the solar interior temperature, to investigate those processes that produce energy in massive stars.\footnote{See what Bethe says at p. 2 of his autobiography `My life in astrophysics'  \cite{Bethe2003}: ``Str\"omgren, a well-known Scandinavian astrophysicist, reported that the central temperature of the sun was now estimated as 15 million degrees, not Eddington's 40. This is still the estimate. This change came as a result of assuming that the sun was predominantly hydrogen with approximately 25\% helium, rather than assuming it had about the same chemical composition as the earth''. However, Bethe is opening this contribution stating that his first involvement with astrophysics ``came as a result of Carl Friedrich von Weizs\"acker's suggestion to investigate the fusion of two protons to form a deuteron, namely $H+H\rightarrow D+e^{+}+\nu$''. Actually Gamow had suggested to one of his graduate students, Charles Critchfield, to calculate the proton-proton reaction and in early 1938 the work was submitted  to Bethe. The latter found the calculations to be correct and they wrote a joint paper \cite{BetheCritchfield1938} paving the way to Bethe's celebrated paper of March 1939 \cite{Bethe1939}.}
   His paper published in March 1939 \cite{Bethe1939}, in which he showed that the most important source of energy in ordinary stars are the reactions of carbon and nitrogen with protons forming a cycle in which the original nucleus is reproduced, was a landmark paper that formed the basis of much work in astrophysics for decades.

 These results demonstrated how farsighted had been the organizers of the conference in gathering together nuclear physicists and astrophysicists for the first time. 
 The researches of the previous two decades into the constitution of the stars had resulted in considerable advance in the understanding of the physical processes in stellar interiors. The chief success of the investigations was the establishing of a mass-luminosity relation. This relation had been obtained without reference to the actual nuclear reactions that are the source of stellar energy, merely from consideration of the mechanical and thermodynamical equilibrium of the star. The problem of stellar energy had to be tackled by nuclear physicists who had devoted all their time to the field to sort it out. At the same time, they needed convincing results communicated by astrophysicists: the temperature, the density, the composition\dots without using at all the input from energy production. The chief success of the investigations is the establishing of a mass-luminosity relation.  As   Str\"omgren recalled \cite{Stromgren1978}: ``It was simply due to this situation that, whatever the mechanism, it must be one that gives a high degree on concentration of energy production in the central region. Then there's no doubt about the model and that fixes the temperature. Once this was understood by the physicists who were ready to accept this, in spite of what, shall we say, Landau said, then that communication was easier than the other. There were so many things that were very difficult for one who wasn't a nuclear physicist to appreciate''. 
   This turned the study of stellar structure from one containing a substantial degree of arbitrariness to one in which definitive models could be derived for any given star in any given state of evolution. 

A conclusion had been thus reached during the conference that stellar models with a concentrated nuclear core could not represent standard stars.\footnote{Within one month, the question was also discussed by Gamow and Teller during the  APS Meeting of  April 28-30 \cite{GamowTeller1938}.}
 However, shortly after the conference, Gamow relaunched the subject \cite{Gamow1938a} remarking that, as the stars of the giant class are distributed in the Hertzsprung-Russell diagram ``in a very peculiar way, very different from the main sequence''  the energy source in giants must be entirely different, probably due for example ``to the beginning of the formation of a dense neutron core in the centers of these stars'' since all giants ``have the masses larger than the critical mass of Chandrasekhar and Landau''. In this case, suggested Gamow in his tentative theory of novae \cite{Gamow1938b},  the formation and growth of a neutron core, ``representing a practically unlimited source of energy, should be expected. The growth of such a core [\dots] may bring the star into the Giant branch of the H-R diagram''.\footnote{On the other hand, a star deprived of any source of nuclear energy would progressively contract and eventually become a  white dwarf, for masses smaller than the critical mass.}
At this point, an explosion of these massive stars would occur, leading to extremely bright novae, that Gamow identified ``with the so-called super-novae of Baade and Zwicky''.
 
Gamow was in fact reacting to a paper by Zwicky concluding a wide search for super-novae that the latter had carried out during the last two years using an improved telescope \cite{Zwicky1938a}. The most important conclusion which he drew from these new observational results was that ``the existence of \textit{two classes of temporary stars}, super-novae and common novae, has been established beyond doubt''. The idea that a certain stage of contraction one might expect that the formation of a large amount of free neutrons would lead to a rapid collapse of the whole star and  to the liberation of tremendous amounts of gravitational energy was again related by Gamow to the Baade and Zwicky supernova mechanism the following year \cite{Gamow1939b}.
 
  Gamow's  pioneering role in connecting nuclear physics and astrophysics, is testified by a long report about the nuclear transformations as energy sources in stars submitted on May 25 to the \textit{Zeitschrift f\"ur Astrophysik}, where he also discussed very dense stars and the accretion of neutrons into an extraordinarily dense and well delimited core (``\textit{Landauschen Kern}''), which could not have been formed from ``usual contraction processes'' of a less dense material \cite[p. 155]{Gamow1938c}. One must thus suppose that they are produced by some external forces when the star was born. In particular he put forward the idea that if these cores existed in all star, one might speculate that, ``according to the theory of the \textit{expanding universe}, the whole space in the past must be of a quite small dimension and filled with matter of exceedingly high density. During the expansion process, this `seed of the world' [`Weltkern'] would disintegrate in smaller pieces, that now, embedded in less dense atmospheres are observed as shining stars''.  
  
  This was an early hint of Gamow's commitment in the cosmological problem connected with the building of elements that would become a major topic immediately after the war, and shows the influence of Lema\^itre's physical cosmology, which had been again recently discussed during a conference on the physics of the universe and the nature of primordial particles [cosmic rays] organized at the University of Notre Dame, Indiana, on 2--3 May by Arthur Haas, most probably in collaboration with Lema\^itre himself, who was at the time a visiting professor there \cite{DepPhys1938}. It gathered about a hundred scientists and was one of the first in which cosmology was a main focus. As a student Gamow had been especially fascinated by special and general theory of relativity, and for this reason he had followed Friedmann's course entitled `Mathematical foundations of the theory of relativity', and ``at first hand, directly from him'', he had learned the theory of the expanding universe. However, Friedmann prematurely died: ``This ruined my plans to continue my work on relativistic cosmology'' recalled Gamow in his autobiography \cite[p. 45]{Gamow1970}. Although he was only 21 years old when Friedmann died, he continued to consider himself a pupil of Friedmann.
   All this was now resurfacing and shows how he could well be inspired by Lema\^itre's theory of a `colossal explosion' of the primeval atom. Gamow's concluding lines of his report explicitly expressed the hope that ``the close collaboration between astronomers and physicists'' would soon lead to an answer to the question of the evolution of stars \cite[p. 160]{Gamow1938c}. 
   
  During that same 1938, von Weizs\"acker had published a new article on the problem of  energy production in stars in which he also proposed as origin of the universe  the result of a cosmic explosion from a superdense compressed nuclear state. His physical cosmology is very similar to Lema\^itre's primeval atom, however, as emphasized by Kragh \cite[p. 99]{Kragh1996}, like Gamow would do later, von Weizs\"acker ``did not refer to general-relativistic models and did not try to combine his nuclear-historical sketch with the geometrical history of the universe as given by the Friedmann-Lema\^itre equations. In this sense, it was only half a big-bang hypothesis''. However, its strong nuclear physics content would later provide an inspiring key for Gamow's later big-bang cosmology.

\section{Oppenheimer and Serber: the stability of a neutron core}

During that `hot' summer of 1938, the stage was set for Oppenheimer's entrance into the still open problem of stellar energy, at which Bethe had begun to work after the Washington conference,  and for which he would soon provide a solution.

The road to Oppenheimer's work on the problem of massive stars has been beautifully reconstructed by Hufbauer \cite{Hufbauer2006} with plenty of interesting details.

According to this reconstruction, before 1938 Oppenheimer came in contact in several occasions with problems belonging to theoretical astrophysics, starting of course from the already mentioned circumstance of his sojourn as a postgraduate student in 1925-1926 at Cambridge University, where he had Fowler as chief mentor. Apart from a series of interesting occasions described by Hufbauer which are forming a convincing background motivating Oppenheimer's interest in stellar theorizing, it is to be further emphasized  that during the early 1930s, nuclear physics, cosmic ray physics, and the emerging field of particle physics were still very much part of the same scenario, in which Oppenheimer was actively working. 

With the increasing knowledge on the nuclear realm Oppenheimer and many others  continued to keep an open eye on the problem of reactions in stars, because of the possibility of understanding how a whole series of nuclear processes that could not be reproduced in terrestrial laboratories, took place in stellar interiors. In a similar manner, the first accelerators used to bombard the nucleus helped in having experimental beams of particles, but could not compete with the high energies typical of the cosmic rays. These were thought to provide information of nuclear processes, in particular after the detection of the mesotron in cosmic rays, which for several years was identified with Yukawa's meson, the carrier of nuclear forces,  only during the late 1940s becoming the weak interacting muon. Since 1933 Oppenheimer had been deeply involved in work on the positron, on collisions processes generated by the interaction of high-energy particles or radiation with matter, as well as on the mechanisms underlying the formation of showers and `nuclear stars' following interactions generated by  cosmic ray particles of very high energy and in general in problems related to the interaction between neutrons and the nucleus. 
In 1937-1938  several of his contributions written  with his collaborator Robert Serber, focused on mesons and cosmic rays.  A couple of articles appearing between August 1937 and April 1938 on nuclear reactions involving transmutations of light nuclei tackled problems which were not very far from the processes which were at the time being discussed as possible sources of stellar energy. As emphasized in \cite[p. 187]{Thorne1994} Robert Oppenheimer was in the the habit of reading with care every scientific article published by Landau. Thus, Landau's article on neutron cores in the 19 February 1938 issue of \textit{Nature} caught his immediate attention.

In the meantime, the 4th Washington Conference focusing on the problem of stellar energy had taken place and things were becoming ripe for its solution. During the summer, following the Conference, Oppenheimer invited Bethe to lecture to his students and collaborators, and so there was plenty of time for talking about the interior of stars. Landau's model of the neutron core was widely discussed, especially in connection with its possible role in giant stars. Apart from this, neutron cores were in itself very appealing for physicists: It was nuclear matter, after all, resembling a giant nucleus made up mainly by neutrons, so that Oppenheimer thought that it was worthwhile exploring the physics involved. He was of course well aware of Gamow's arguments about a superdense core in stars discussed in his \textit{Structure of atomic nuclei and nuclear transformations} (pp. 232-238) that had been published the previous year, as well as of his latest articles published between spring and early summer.  Landau's article must have been widely discussed with Serber and with Oppenheimer's brilliant student George Michael Volkoff, who appears to have been involved  because of his longstanding interest in astronomy \cite[p. 38]{Hufbauer2006}.\footnote{Volkoff became a graduate student of Oppenheimer in 1936 and between 1938 and 1939 he was completing his thesis on `The equilibrium of massive neutron cores'.   Because of his Russian origin, and his proficiency in his native language,  he became an important bridge between the scientific communities of East and West during the cold war years \cite{Volkoff1990}.} 

 In that same June 1938, Oppenheimer had moved from Berkley's Physics department  to Caltech, as he used to do every year, and by that time his interest in the problem of stellar energy had ripened up to the point that together with William A. Fowler, working at Caltech, and Rudolf Minkowski (Carnegie Institution of Washington, Mount Wilson Observatory), he had organized a symposium dedicated to nuclear transformations and their astrophysical significance within the annual summer meeting of the Astronomical Society of the Pacific that was held in San Diego, California, on June 22-23, as a joint session with the American Physical Society \cite{PASP1938}.

 Oppenheimer  was scheduled to give a talk on `The physical problem of stellar energy', while Minkowski discussed `The Composition of Stellar Atmospheres'.\footnote{Minkowski, whose  uncle Hermann had been the famous mathematician at Zurich and G\"ottingen, had done his doctoral studies in Breslau on spectroscopic problems. After a year in G\"ottingen with James Frank and Max Born, he moved to the University of Hamburg, as an associate professor of physics and where he became Baade's friend and collaborator. He remained there until 1935 when he was dismissed by the nazi racial laws. In 1936 Baade invited him to work at Mount Wilson Observatory, where his competence as spectroscopist was especially appreciated leading to a close and fruitful collaboration.}
  William Fowler, who was working with Charles C. Lauritsen at Caltech's Kellogg Radiation Laboratory, endowed with a large high voltage X-ray tube accelerating charged particles, gave a talk on `Nuclear Reactions as a Source of Energy.'\footnote{William Fowler had got his Ph.D. in 1936 studying nuclear reactions of protons with the isotopes of carbon and nitrogen in the laboratory, the very reactions in the CN-cycle, that with Weizs\"acker and Bethe's proposals were already revealing their key role within processes governing energy production in stars.}

 Working in a laboratory that was at the cutting edge of nuclear physics, Fowler thus found himself as one of the founders of the emerging science of nuclear astrophysics.\footnote{For his theoretical and experimental studies of the nuclear reactions of importance in the formation of the chemical elements in the universe, Fowler would be  awarded the 1983 Nobel Prize in physics, jointly with Chandrasekhar.} 

  Fowler had actually followed Oppenheimer's course on theoretical nuclear physics at Caltech: ``that was really one of the highlights, because Robert was an excellent teacher, and he knew what was going on in nuclear physics''. Oppenheimer was also deeply involved in the research activities of the Kellogg's laboratory since 1933, when Lauritsen had been able to produce artificial neutrons with accelerators and played an enormous role in teaching them the    theoretical implications of their results: ``he understood so much more completely than either Charlie or I, or even Tolman, the meaning of what we were doing [\dots] he understood all the quantum mechanics and special and general relativity in a very deep way [\dots]  He was able to translate what we were finding in the laboratory into useful contributions to physics [\dots] If it hadn't been for Oppenheimer, I think we would have missed [laughter] practically all of the significance of what we were doing [\dots] Robert almost certainly was the first one to tell us that Bethe had pointed out the importance of these reactions in the sun and other stars''  \cite{Fowler1983a,Fowler1984/86}.\footnote{See \cite{GoodsteinGreenberg1983} on the beginning of nuclear astrophysics at Caltech.}

This last sentence is suggesting that the idea of the symposium was most probably triggered by Oppenheimer himself. What is to be emphasized is that the title of the symposium, `Nuclear transformations and their astrophysical significance' --- for the first time explicitly connecting nuclear physics and astrophysics --- represented a further step along the road of an integration of the two scientific communities, someway inaugurated by the fourth Washington meeting. 

In the meantime, Bethe, together with Gamow and Teller's  student Charles L. Critchfield, published in August the already mentioned paper  addressing the proton-proton reactions into deuterons and developing a quantitative scheme of a theory for stellar energy production \cite{BetheCritchfield1938}. This paper was clearly discussed with Oppenheimer, who was mentioned in a footnote. The same ``interesting discussion of these questions'' were also acknowledged  in a short note submitted by Oppenheimer and Serber on September 1 \cite{OppenheimerSerber1938}. They started acknowledging Bethe and Critchfield's recent work which ``could be made to account successfully for the main sequence stars'', but not for the enormous energy output of very massive stars such as the red giant Capella, that had a much lower density and temperature than the Sun. In his talk at the San Diego Meeting, Oppenheimer had already  presented  ``the details of the theory of the possible nuclear changes in the lighter elements and the possibility of their application to the interior of stars'' suggesting ``a new model with a high central concentration of neutrons'' \cite[p. 210]{PASP1938}. But now Oppenheimer and Serber left aside the problem of energy generation for standard stars, tackled by Bethe and Critchfield: it was the neutron core, investigated by Landau and Gamow, that represented an intriguing and challenging physical system. A condensed core appeared to be still an interesting hypothesis in connection with the giant stars, where it would form after all the thermonuclear sources of energy, at least for the central material of the star, had been exhausted. And in fact, they immediately clarified that, in this regard, it still seemed of some interest Landau's ``suggestion of a condensed neutron core, which would make essential deviations from the Eddington model possible even for stars so light that without a core of highly degenerate central zone could not be stable''. The last provocative lines of Landau's paper had hit the bull's eye\dots Thus Oppenheimer decided to play the game, and started it with Serber, tackling a fundamental problem, which was essential for a discussion of the role of such a core, that is ``the estimate of the minimum mass for which it will be stable''. At that time Landau was like many others a victim of the Great Terror and languished in Stalin's prisons since April of that year.\footnote{Landau's dear friend Matvey Bronstein had already been killed. Only the following year, after a courageous intervention of Pyotr Kapitsa, who wrote a letter directly to Molotov and Stalin threatening to quit the Institute for Physical Problems, Landau was released. He would have barely survived even a short period of further imprisonment. As promised by Kapitza, he was the right person to solve the mystery of superfluidity, an achievement for which Landau was awarded the Nobel Prize for Physics in 1962.}

According to Oppenheimer, Landau's evaluation appeared ``to be wrong''. Landau had not properly taken account of the attractive forces of gravity and had not considered the role of nuclear forces between neutrons, a force that actually was not fully understood, but on which some guess could be made, basing on the phenomenology of experimental work, of which Oppenheimer was deeply aware. Both his expertise as a theoretical physicist and his constant involvement in the interpretation of experimental results were crucial in what will be now discussed. In investigating the stability of such a core, Oppenheimer and Serber were reasoning on a large assembly on nuclear particles, confined by gravitational forces, comparing this system with an actual nucleus: ``The question of the actual stability of core models thus involves a consideration of the contribution of nuclear forces to the core-binding. The forces which must be known are those acting between a pair of neutrons; and no existing nuclear experiment or theory gives a complete answer to this question''. Based on different assumptions on the nuclear forces also derived from investigations by Critchfield and Teller \cite{CritchfieldTeller1938},\footnote{The authors thanked Bethe, Fermi, Gamow and Oppenheimer ``for helpful discussions''.} they concluded that ``even in the heaviest stars no core will be formed until practically all sources of nuclear energy have been, at least for the central material of the star, exhausted''. The arguments given did show that the nuclear forces considered precluded the existence of a core for stars with masses comparable to that of the sun. It was thus clear that Landau's idea --- originally inspired by the old Milne's proposal --- that a large neutron core could be tucked away in stars like the sun keeping it hot --- was definitely wrong. But this did not rule out the possibility of neutron cores in larger stars. Oppenheimer's investigations definitely shifted the attention to the possibility that such a core could form only when ``practically all sources of nuclear energy have been, at least in the central material of the star,  exhausted''.

Oppenheimer and Serber had thus showed that one cannot build a viable model of the Sun with its energy coming from a neutron core, and later Bethe definitely showed that  reactions of carbon and nitrogen with protons are ``the most important source of energy in ordinary stars'', so that interest in such stellar model declined. However, Bethe himself, in his 1939 paper \cite{Bethe1939}  left the problem of energy production in giant stars open. In that case it seemed rather difficult to account for the large energy production by nuclear reactions \cite[p. 450-451]{Bethe1939}: ``The only other source of energy known is gravitation, which would require a core model for giants''. In this regard he cited  \cite{Landau1938}, but according to Bethe, he got this  suggestion also by Gamow in a letter \cite[footnote 41]{Bethe1939}.\footnote{It appears that Fermi, too, who always had a great interest in astrophysical issues, speculated on the problem suggesting that normal stars with neutron cores would have the luminosity and spectral characteristics of red supergiants. The circumstance was mentioned by Kip Thorne \cite{Thorne1989} who added that  ``nobody seems to have built detailed models of such stars and verified Fermi's suggestion until the work of Thorne and $\dot{Z}$ytkow''. Starting from 1976, Kip Thorne and Anna $\dot{Z}$ytkow wrote in fact a series of several papers devoted to the question: What are the possible equilibrium states for a star consisting of a massive nondegenerate envelope surrounding a degenerate neutron core?}

\section{General relativity officially enters the stage of compact stars: Tolman and Zwicky}

In the meantime, during that same summer 1938, Zwicky had again entered the scene with a follow up of his proposal about the existence of neutron stars \cite{BaadeZwicky1934a} \cite{BaadeZwicky1934b}. 
At that time very little was known definitely about supernovae and it seemed certainly premature to discuss in any detail the formation of neutron stars as a possible cause for supernovae. However, since 1934 Zwicky had initiated with Baade the first systematic sky survey, and confirmed that a number of historical novae were indeed supernovae. 
Through the discovery with the 18-inch Schmidt telescope on Palomar mountain of eight supernovae, the existence of supernovae as a new special class of temporary stars might ``be regarded as established beyond reasonable doubt ''  \cite[p. 727]{Zwicky1938a}.\footnote{See list of published articles on these discoveries in footnote 3 of this article.}
These new observations, that unquestionably confirmed the existence of violent events in the universe, urged Zwicky to pick up again the  topic. Now that neutron cores were being discussed quantitatively by  influential scientist like Gamow, Landau and  also by Oppenheimer, based on nuclear physics, he felt that such theoretical investigations might well apply to his old idea of a collapsed neutron stars as remnants of supernovae explosions. It is to be remarked that Oppenheimer himself did not cite \cite{BaadeZwicky1934b}. Probably,  he wanted  to take a distance from such issues that on one side had not affected astronomers' interests being too speculative and had also been completely ignored by physicists because they definitely lacked a physical base. On the other hand, as we have seen, Oppenheimer's interest into the matter was aroused within a completely different context, much more related to the development of the neutron core idea as a support to the problem of stellar energy, with no apparent relationship to Zwicky's remnants of catastrophic explosions. Only vague hints connecting the two scenarios were actually existing up to that moment. 

 In any case, Zwicky made a brand new very bold attempt along this path employing general relativity. Already during the June San Diego meeting, he presented a talk entitled ``On neutron stars'' \cite{Zwicky1938b}, where he focused on collapsed neutron stars as representing ``\textit{states} of \textit{lowest energy} that matter may assume without being completely transformed into radiation''. The very rapid ``transformation of stellar matter into the neutron state'' might provide an explanation to the ``stupendous rate at which energy was liberated in some of the recently observed super-novae''.
 The abstract also mentioned the old Eddington's argument related to gravitational red-shift originating on the surface of dense stars as white dwarfs that he now presumed might be observable also in super-novae and in their remains, neutron stars, where the phenomenon would be much more pronounced. Last but not least, Zwicky made a second stronger connection to general relativity, according to which ``the mass of a star of given density cannot surpass a certain critical value (Schwarzschild limit)''. He then mentioned the energy that might be liberated at this limit ``because of gravitational packing'' as being $0.58 \ Mc^2$, where M is the ``\textit{proper mass} of the star''. The derivation of this result, explained Zwicky, had been obtained in discussion with Richard C. Tolman, and would be communicated elsewhere. From this remark, it appears that Tolman and Zwicky had been discussing the necessity of taking into account general relativity in the case of Zwicky's `neutron stars' at least in late spring - early summer 1938.

Zwicky discussed more in detail  the big issues mentioned in the previous abstract in an article dated August 8, 1938, entitled `On Collapsed Neutron Stars' \cite{Zwicky1938c}.
Here the divide between Zwicky's and Oppenheimer's approach --- and Landau's and Gamow's as well --- can be measured by the distance Zwicky took in this article from the  idea of neutron stars ``regarded as a giant nucleus composed of separate neutrons of precisely the same character as free neutrons''.  Instead Zwicky  specified that he used the term neutron star ``simply to designate a highly collapsed star, the average density of which is of the order of the density of matter existing inside of ordinary atomic nuclei''. Therefore the neutron composition of such a star should be rather taken ``as a short designation for an extended state of matter of nuclear density in which every region whose linear dimensions $d$ are larger than about $\delta=e^2/m_{e}c^2=2.8\times 10^{-13}$ cm is essentially electrically neutral''. The paper was divided in two parts. In the first part he discussed the Schwarzschild  solution and the second part dealt with the possibility of actually observing the formation of collapsed neutron stars. But Zwicky did not know enough nuclear physics and general relativity to tackle the problem in detail and rigorously, basing on such theoretical tools. Even if he had a longstanding familiarity with general relativity topic since when he studied in Zurich, also with Hermann Weyl.

In his views there were some properties of neutron stars, that appeared to support his hypothesis: First of all, ``a) Cold neutron stars, according to present knowledge, represent the states of lowest energy that matter may assume without being completely transformed into radiation. b) According to the general theory of relativity, a \textit{limiting mass} of stars exists for every given average density (Schwarszschild limit)\dots'' (here he cited \cite{Tolman1934}). 
Zwicky provided without proof the ``energy liberated because of gravitational packing'': $E=(1-4/3\pi)Mc^{2}=0.58 \ Mc^{2}$ (where M is the proper mass of the star), and mentioned as well the existence of a limiting mass (`$M_L=6.4\times10^{34} g$')  for an average density $\rho=10^{14} g/cm ^3$, announcing that ``The derivation of these results which was obtained in discussion with Professor R. C. Tolman  will be communicated in a joint paper with Professor Tolman''.  However, there is no trace in the published literature of such joint paper, but as a matter of fact Zwicky had discussed the problem with Tolman, his colleague at Caltech, at least before  June, according to the announcement made in the abstract of his talk at San Diego \cite{Zwicky1938b}. 

These circumstances, on the other hand, are clearly confirmed in a later paper by Tolman (`Static Solutions of Einstein's Field Equations for Spheres of Fluid') where the latter, in an unusually long footnote \cite[footnote 2, p. 365]{Tolman1939a} acknowledged that: ``My own present interest in solutions of Einstein's field equations for static spheres of fluid is specially due to conversations with Professor Zwicky of this Institute, and with Professor Oppenheimer and Mr. Volkoff of the University of California, who have been more directly concerned with the possibility of applying such solutions to problems of stellar structure''. We will soon come back to the crucial connection between Tolman and Oppenheimer. Tolman then continued with his description about Zwicky's recent attempt to introduce general relativity in the treatment of his neutron stars: ``Professor Zwicky [\dots] has suggested the use of Schwarzschild's interior solution for a sphere of fluid of constant density as providing a model for a `collapsed neutron star'''. Tolman then expressed the hope that ``the considerations given in this article may be of assistance in throwing light on the questions that concern him''.

Applying to his case an argument which had already been used within Newtonian gravity by John Michell  and by Pierre-Simon Laplace during the 1780s \cite{Israel1987} (and which was later recalled by Eddington both in 1926 \cite[p. 6]{Eddington1926} and during his controversy with Chandrasekhar), Zwicky concluded his paper remarking that  \cite[p. 523-525]{Zwicky1938c}: ``A star which has reached the Schwarzschild limiting configuration must be regarded as an object between which and the rest of the universe practically no physical communication is possible [\dots] It is, therefore, impossible to observe physical conditions in stellar bodies which have reached the Schwarzschild limit. It should, however, be possible to observe stellar bodies in stages intermediate between the ordinary configurations and the collapsed configurations of limiting mass just described, provided that such are accessible''. As a consequence, pointed out Zwicky: ``If supernovae are transitions from ordinary stars into neutron stars, the observation of light-curves and spectra of supernovae should furnish us with direct evidence of the neutron-star hypothesis [\dots] the surface of the central star of a supernova should be exceedingly hot, the acceleration of gravity very high, the light coming from this surface should be subject to enormous gravitational redshifts''.\footnote{In this regard, Zwicky is mentioning Minkowski's forthcoming huge contribution on the spectra of supernovae \cite{Minkowski1939}.}

The redshift effect had been mentioned by Ernest J. \"Opik in his theory of giant stars \cite[p. 3]{Opik1938}, as a phenomenon which ``may asymptotically tend to reduce the luminosity of a superdense contracting star to zero'' because in such cases stars ``should possess a superdense core containing the major fraction of the mass'' and the red-shift effect might be considerable. As white dwarfs had provided in 1915 a new test of Einstein's theory of general relativity well outside the solar system,  the neutron-star hypothesis, in conjunction with observations on supernovae might now lead to a further and far-reaching test of the general theory of relativity, in two different astrophysical situations: Zwicky's neutron stars and neutron cores in giant stars. A connection between these still theoretical astrophysical entities was thus established, and Zwicky's old speculations were beginning to transform into a more tangible reality.

\section{The Tolman-Oppenheimer-Volkoff equation}

A specially relevant question for this narrative is related to the strong personal and scientific friendship between Oppenheimer and Tolman, going back to the period 1929-1930, when Tolman was  embarking on his project on the connection between general relativity and thermodynamics following the cosmological issues arising from Hubble's discovery of the distance-redshift relation for galaxies and especially connected to Lema\^itre's proposal of an expanding universe.  
Being a physical chemist, but with a strong interest in astronomy and relativity, as early as 1922 he had also investigated the possibility of explaining the relative abundances of hydrogen and helium through chemical equilibrium reactions in what Helge Kragh \cite[p. 43]{Kragh2013} has duly termed ``a pioneering contribution to nuclear astrophysics''.

That same 1922 he had moved to Caltech. His early interest in the theory of relativity, later led him to tackle the cosmological implications of general relativity that culminated in the publication of his seminal book \textit{Relativity, Thermodynamics, and Cosmology}  a most cited text up to post-war years.\footnote{During the mid 1930s Robert Marshak and other students of Columbia College at Columbia University, New York, including Julian Schwinger, Herbert Anderson, Norman Ramsey and Henry Primakoff, all unsatisfied by the too formal approach they were taught, ``wanted to learn the physics of relativity'' and thus formed the Undergraduate Physics Club lecturing to one another \cite{Marshak1970}. They discovered Tolman's book ``which was very physical'' and Marshak remembered that  ``they went through that book very thoroughly''. }
 Tolman had always kept contacts with astronomers working at Mount Wilson Observatory near Pasadena, in particular with Hubble, with whom  he published  a paper on the nature of `nebular redshifts' \cite{HubbleTolman1935}. 
Tolman was thus a very peculiar figure, whose wide interests integrated  experimental, observational and theoretical aspects. Working at the boundary of different fields, he became crucial during the summer-fall 1938 in reorienting Oppenheimer's interests towards a new perspective for investigations on the neutron core, from which most of the nuclear stellar energy mechanisms had been stripped out, apart from some special situations. In this regard, the open questions that could be tackled were related to the evolution and in particular to the final fate of stars. In discussions with Tolman, the nature of the neutron core as a compact physical object definitely emerged: it was clear that it had to be tackled using general relativity. In the case of white dwarfs it was not so compelling, but now to really explore the behaviour and eventually ultimate fate of a superdense assembly of neutrons, Einstein's theory could not be avoided.

In the above mentioned footnote of his paper `Static Solutions of Einstein's Field Equations for Spheres of Fluid' \cite{Tolman1939a} in which he acknowledged discussions with Zwicky, Tolman outlined the respective research paths: ``Professor Oppenheimer and Mr. Volkoff have undertaken the specific problem of obtaining numerical quadratures for Einstein's field equations applied to spheres of fluid obeying the equation of state for a degenerate Fermi gas, with special reference to the particular case of neutron gas. Their results appear elsewhere in this same issue. My own solutions of the field equations, as given in the immediately following, can make only an indirect contribution to the physically important case of a Fermi gas, since it will be seen that they correspond to equations of state which cannot be chosen arbitrarily. My thinking on these matters has, however, been largely influenced by discussions with Professor Oppenheimer and Mr. Volkoff, and it is hoped that the explicit solutions obtained will at least assist in the general problem of developing a sound intuition for the kind of results that are to be expected from the application of Einstein's field equations to static spheres of fluid''.\footnote{See correspondence between Oppenheimer and Tolman, courtesy of Caltech archives: Tolman to Oppenheimer, October 19, 1938 (discussing the ``paradoxical character of the Schwarzschild interior solution'');  Tolman to Oppenheimer, November 9, 1938 (about having found ``two more possible solutions for the gravitational field of a static sphere of perfect fluid''); Oppenheimer to Tolman, from Berkeley, no date. See also \cite{Tolman1939b} \cite{Tolman1939c}, presenting a series of analytical solutions of Einstein's equations allowing to better understand the origin of the limiting mass, that Tolman discussed in two subsequent papers  on the \textit{Astrophysical Journal}.}

Tolman's paper, as well as the new contribution written by Oppenheimer  in collaboration with his student Volkoff, `On Massive Neutron Cores' \cite{OppenheimerVolkoff1939}, had been actually received in the \textit{Physical Review} the same day, January 3, 1939.  Both papers appeared in the same February 15 issue and contained the derivation of the equation of hydrostatic equilibrium for a spherically symmetric star in the framework of general relativity, since then  called the Tolman-Oppenheimer-Volkoff equation. 

Oppenheimer and Volkoff took up the problem where Chandrasekhar and von Neumann had left it: they studied ``the gravitational equilibrium of masses of neutrons, using the equation of state for a cold Fermi gas, and general relativity''.  In the view, which seemed plausible by that time, that the principal sources of stellar energy are thermonuclear reactions, at least in main sequence stars, then the limiting case considered by Landau in 1932 again became of interest in the discussion of what would eventually happen to a normal main sequence star \textit{after all the elements available for thermonuclear reactions are used up}.  Landau had showed that for a model consisting of a cold degenerate Fermi gas, a mixture of electrons and nuclei, there exist no stable equilibrium configurations for masses greater than 1.5 solar masses, all larger masses tending to collapse \cite{Landau1932}.  
Chandra had clearly highlighted the importance of such issue, when he remarked that the life history of a star of small mass must be essentially different from that of a star of large mass \cite[p. 377]{Chandra1934b}: the latter cannot pass into a white dwarf stage and ``one is left speculating on other possibilities''. 

When gravity becomes the sole and key governing force, a sufficiently massive star collapses under its own gravity, but at that time this was not felt as a fundamental key problem in astronomy and astrophysics, even if the collapsing process had been an ingredient of astrophysical theorizing on the structure of stars. But the growing role of nuclear processes had in the meantime completely transformed the issue of the limiting mass into a problem related to nuclear matter at densities beyond those found inside a nucleus. 

 Both Landau and Gamow had recently suggested that in sufficiently massive stars after all the thermonuclear sources of energy, at least for the central material of the star, have been exhausted a condensed neutron core would be formed \cite [p. 234]{Gamow1937}   \cite{Landau1938}.  Oppenheimer and Serber, taking into account some effects of nuclear forces had made a reasonable estimation of the minimum mass for which such a core would be stable (approximately 0.1 solar masses) \cite{OppenheimerSerber1938}. A neutron core with a mass less than about 0.1 solar masses would disintegrate into nuclei and electrons.
The gradual growth of such a core, with the accompanying liberation of gravitational energy, had been suggested by Landau, and in this connection it seemed now interesting ``to investigate whether there is an upper limit to the possible size of such a neutron core''. 
 
Landau had found un upper limit of about 6 solar masses, beyond which the core would not be stable but would tend to collapse. Two objections might be raised against this result: ``\textit{One is that it was obtained on the basis of Newtonian gravitational theory while for such high masses and densities general relativistic effects must be considered} [emphasis added]''. The second one was related to the assumptions used by Landau for the Fermi gas, now that the theory had  to be applied to the case of a neutron gas. They thus wanted to establish ``\textit{what differences are introduced into the result if general relativistic gravitational theory is used instead of Newtonian  and if a more exact equation of state is used} [emphasis added]''.\footnote{Tolman's \textit{Relativity, Thermodynamics and Cosmology}, of which especially cited were pp. 239--247, provided the theoretical basis for the discussion of the general relativistic treatment of the equilibrium of spherically symmetric distributions of matter, and the subsequent treatment of the special ideal case of a cold neutron gas. \cite{Chandra1935a} is cited, too.} 
Chandra had used a Braunschweiger calculator to compute the white-dwarf structure,  now Volkoff used a Marchant for the numerical integrations of some equations that could not be carried out analytically \cite[p. 377-378]{OppenheimerVolkoff1939}, as Tolman was trying to do for some specific cases. It was the beginning of what became known as computational relativity.

They found that for a cold neutron core \cite[p. 380]{OppenheimerVolkoff1939} ``there are no static solutions, and thus no equilibrium, for core masses greater than $m\sim 0.7 M_\odot$ [\dots] Since neutron cores can hardly be stable (with respect to formation of electrons and nuclei) for masses less than $\sim 0.1 M_\odot$, and since, even after thermonuclear sources of energy are exhausted, they will not tend to form by collapse of ordinary matter for masses under $1.5 M_\odot$ (Landau's limit), it seems unlikely that static neutron cores can play any great part in stellar evolution''. As this limit was lower than the Chandrasekhar mass limit of white dwarfs, 1.44 $M_\odot$, their limit appeared to create difficulties with the formation of neutron cores in ordinary stars. 

Moreover, they added that ``the question of what happens, after energy sources are exhausted, to stars of mass greater than $1.5 M_\odot$ still remains unanswered''.\footnote{These startling results  were  already announced by Volkoff at the annual meeting of the APS, held at UCLA on December 19, 1938 (`On the equilibrium of massive neutron cores')  \cite{Volkoff1939a}: ``No physically plausible modifications of the equation of state seem essentially to alter this conclusion, or to change radically the order of magnitude of $M_2$ [$M_2 \sim\ 0.75 M_\odot$]''.}

The conclusion was that there would then seem to be only two answers possible to the question of the final behavior of very massive stars: either the equation of  state they had used failed to describe the behavior of matter at density higher than nuclear density (so that their extrapolation of the Fermi equation of state could ``hardly rest on a very sure basis''), or the star would ``continue to contract indefinitely, never reaching equilibrium''. Both alternatives, concluded the authors required ``serious consideration''.
  
  They were beginning to lay a theoretical framework for investigating the fate of collapsing stars, even if many doubts persisted about the equation of state of highly compressed nuclear matter, that is, the extent to which  matter at supranuclear density might successfully resist further compression. 

A series of theoretical consideration about  nuclear forces, even in the case of $\rho>10^{15} g/cm^3$ having the extreme effect of making $p=\frac{1}{3}\rho$ in such a `critical' core,
 led them to conclude that it seemed likely that their limit of $\sim 0.7 M_\odot$ was ``near the truth''. This limit would be modified by future developments after the war, but conceptually it confirmed the existence of the mass limit within their theoretical frame.\footnote{The equation of state $p=\frac{1}{3}\rho$ had been used by von Neumann's in his 1935 notes ``Static solutions of Einstein field equations for perfect fluid with $T^{\rho}_{\rho}=0$'' and `On relativistic gas-degeneracy and the collapsed configurations of stars' \cite[pp. 172-173]{NeumannCW}. It appears that information about von Neumann's results  continued to circulate within the scientific community even after the war. At the beginning of chapter 3 of Wheeler and collaborators' contribution to the first Texas meeting \cite{WheelerEtAl1965} (`Hydrostatic equilibrium and extremal mass-energy'),  in citing  \cite{Tolman1934} as well as Oppenheimer and Volkoff's article, they  wrote: ``We have been told that in unpublished work at Cambridge in 1935 John von Neumann integrated the general relativity equation of hydrostatic equilibrium for the special case $p^*=\rho^*/3$''.}
 
  However, it appeared that ``for an understanding of the long time behavior of actual heavy stars a consideration of non-static solutions must be essential''. Among all spherical non-static solutions one would hope ``to find some for which the rate of contraction, and in general the time variation, become slower and slower, so that these solutions might be regarded, not as equilibrium solutions, but as quasi-static''. But as a final conclusion to the discussion they stated that ``For high enough central densities it is no longer justified to neglect even a very slow time variation; and the singular solutions which presumably represent very massive neutron cores cannot be obtained unless this is taken into account. \textit{These solutions are now being investigated} [emphasis added]''.
 
This new chapter of stellar structure differed from the preceding ones because, contrarily to what had happened with white dwarfs, all these models were derived as purely theoretical constructs, without any observed astronomical objects known at that time to which they might actually apply.   These `stellar neutron cores' were developed within a completely different research strand and far from any connection with Zwicky's `neutron stars'. 
  
  An attempt had been made by Zwicky to provide a  theoretical basis to the idea that neutron stars should be born in supernovae explosions, and to relate it to observational astronomy through the redshift of supernova spectra investigated by Minkowski. Interest in these issues was beginning to connect the two aspects, but a large divide still existed between the different cultures. 
Zwicky's new contribution `On  the theory and observation of highly collapsed stars' \cite{Zwicky1939}, of April 1939, was pursuing a new strategy investigating ``the general relativistic solution given by Schwarzschild of the problem of a homogeneous sphere of constant density'' and  ``the possibility of actually observing the formation of collapsed neutron stars''.  In stating that the hypothesis of the formation of a neutron star ``would run into serious difficulties if one should attempt to retain the classical theory of gravitation'', Zwicky specified that ``in the theory of neutron stars it is necessary to introduce general relativistic effects'', according to which he interpreted the redshift in the spectrum of the supernova IC 4182 as a general relativistic gravitational redshift \cite[p. 727]{Zwicky1939}, and estimated some of the physical characteristics of the central star of a supernova one year after maximum brightness: radius (100 km), average density ($10^{12} g/cm^3$) and temperature (greater than $5\times 10^6$ degrees). In making  a distinction  between the rest mass and the gravitational mass he was able to estimate the binding energy of a neutron star of mass M, and thus evaluate how much energy could be released during the core-collapse of massive stars.
  
Zwicky, was basing his theoretical investigations on new spectral studies of two bright supernovae (IC 4182 and NGC 1003) performed by his colleague Rudolf Minkowski at Mount Wilson Observatory  \cite{Minkowski1939}, that in his eyes fully justified a more detailed examination of the neutron star proposal.\footnote{Minkowski made a very detailed discussion of his observation basing on  Zwicky's assumption that the observed red shift might be caused ``by the increase of gravitational potential at the surface of a collapsing star'' \cite[p. 208] {Minkowski1939} and concluded that two different explanations of the red shift, as either a gravitational effect or as  Doppler effect, appeared possible. If a more detailed study of the radiative equilibrium did not lead to a rejection of one of these conceptions, a decision might be brought about by a theory of supernovae which could explain the similarity of the red shift in different supernovae.}
   
 But it was again Tolman who inspired investigations towards the application of the general theory of gravitation.  In the concluding lines Zwicky thanks him for discussions during which ``many of the results given in the first part of this paper were derived'', but no mention of Oppenheimer's papers with his collaborators can be found, apart from an article by Volkoff.\footnote{Volkoff  is investigating the difference in behavior between solutions of Einstein's field equations with infinite central pressure (that Schwarzschild had  dismissed as physically inadmissible because of this singularity) and the Oppenheimer-Volkoff cold neutron gas model leading to an upper limit on the size of a static sphere \cite{Volkoff1939b}.}
   
Among the special reasons for which the study of supernovae might ``eventually prove to be of considerable interest'', stressed Zwicky in the concluding lines of his new paper, the following was to be singled out: ``If the neutron star hypothesis of the origin of supernovae can be proved, it will be possible to subject the general theory of relativity to tests which according to the considerations presented in this paper deal with effects which in order of magnitude are large compared with the  tests so far available''. Apart from mentioning the possibility that cosmic rays originate in supernovae as an added incentive for pursuing such investigations \cite[p. 743]{Zwicky1939}, Zwicky made a further startling statement, that clarifies how deeply aware he had become of the possible implications of his original fascinating idea:  ``\textit{The general theory of relativity, although profound and exceedingly satisfactory in its epistemological aspects, has so far practically not lent itself to any very obvious and generally impressive applications. This unfortunate discrepancy between the formal beauty of the general theory of relativity and the meagerness of its practical applications makes it particularly desirable to search for phenomena which cannot be understood without the help of the general theory of relativity} [emphasis added]''.  

This statement is the best possible coeval comment to the Oppenheimer-Volkoff paper, in which for the first time general relativity was deliberately applied to tackle the problem of a compact astrophysical object, and that in Zwicky's words probably acquired a meaning going well beyond Oppenheimer's own intentions and at the same time represents the best introduction for the final phase of Oppenheimer's efforts in this direction in which he explored with his student Snyder the final fate of a collapsing stellar neutron core. 

Tolman later had an important role within the Manhattan Project, and, as revealed by Serber himself, it is remarkable that he was the first to put forward the idea of implosion as a way of compressing matter and triggering the explosion process of nuclear weapons \cite[p. xxxii]{Serber1992}. The similarity between stellar implosion-explosion problems and the building of nuclear and thermonuclear weapons would in turn attract the attention of a new generation of physicists deeply involved in these activities during World War II --- notably John A. Wheeler and Ya B. Zeldovich --- towards the connections between general relativity and the interior of a compact star. Such similarity also suggested the adaptation of bomb design codes to simulate stellar implosions  \cite{ColgateWhite1966}.

\section{From the neutron-core to the neutron star. The last chapter of Oppenheimer's trilogy.}

Officially  Zwicky was completely ignored, not being cited in the Oppenheimer-Volkoff paper, that deliberately took a distance from `neutron stars', considered a  fruit of Zwicky's speculations, without any clear physical content.\footnote{It is to be remarked that, still in 1964, three years before the discovery of pulsars, when  it was ``accepted with a reasonable assurance'' that the supernova explosion of a star is first triggered by the collapse of its core, Hong-Yee Chiu stated in a review on `Supernovae, neutrinos, and neutron stars' that the possibility that ``neutron stars may be the remnants of supernovae has so far been accepted only with skepticism'' \cite[p. 368]{Chiu1964}.} On the other hand, Chandrasekhar's classic white-dwarf work, too, was scarcely credited by Oppenheimer --- as well as by Landau --- thus favoring  an interpretation in the direction of a divide between fields and scientific styles.  
From Oppenheimer's point of view, all that was restricted to the theory of relativistic electron degeneracy as needed for a full investigation of the white-dwarf problem, a very specific astrophysical problem also having an interest for astronomical observations. In his articles he explicitly mentions \cite{Landau1932} whose investigations on the physical nature of the equilibrium of a given mass of material had been performed using ``a model consisting of a cold degenerate Fermi gas''. Following Landau, but with improved knowledge on nuclear matter, Oppenheimer and collaborators used astrophysics, as a realm providing a  {\it physical system} at extremely high density for their investigations about its stability and the existence of an upper limit to its possible size \cite{OppenheimerSerber1938,OppenheimerVolkoff1939}.

In any case, because of his  commitment to subnuclear processes generated by cosmic-rays and their relationship with the emerging modern particle physics, Oppenheimer could not ignore Zwicky's plausible speculations on supernovae explosions as sources of high energy particles, even if such investigations were in turn connected with further speculations on a collapsed compact astrophysical object made up of neutrons.

From the outlined arguments presented up to now, it is quite evident that Oppenheimer's work with his collaborators was being carried on within a wider related context, and especially within the ongoing and spreading interest towards superdense neutron cores in stars. However, Zwicky's idea was gaining momentum as one can see from Gamow's article `Physical possibilities of stellar evolution' \cite{Gamow1939a}, submitted in November of 1938, where he outlined a picture of stellar evolution on the basis of the Bethe-Weizs\"acker theory. Apart from attempts to explain the energy production in red giants, he discussed the ``contractive stage where the energy liberation is purely gravitational'' and ``the possibility of neutron-core formation in heavier stars, in application to the explosion phenomena observed in supernovae''.\footnote{Apart from Baade and Zwicky and Chandrasekhar, Gamow cited Sterne and Hund, so that all the implications contained in these pioneering works were beginning to be fully appreciated by this time.}
 Gamow recognized that for stars with large masses,  no stable finite state does exist and so they ``must undergo continuous unlimited contraction''. However, he saved the situation proposing that ``such a process will never continue indefinitely because, since all stars possess an angular momentum, the centrifugal forces will soon become large and will, most probably, cause the breaking of such a massive star into several smaller pieces with the masses below the critical value. These pieces will then continue to exist indefinitely in the form of white dwarfs''. Thus, according to Gamow, the existing white dwarfs did not represent a finite stage of evolution of a single star, but must be considered the fragments resulting from the explosion of heavy stars.

In extending his astrophysical investigations, Gamow, in collaboration with Teller \cite{GamowTeller1939}, also addressed the problem of the origin of great nebulae within the framework of an expanding universe: ``The type of expansion necessary for the formation of nebulae indicates that space is infinite and unlimitedly expanding''.   In the last section, they considered the cosmological consequences which they discussed from the form of the fundamental (Friedmann--Lema\^itre) equation for the expanding universe as given by Richard Tolman in his textbook of 1934 \cite{Tolman1934}. They deduced that the nebulae are the largest assemblies of matter which can be kept together by gravitation against the dispersing effect of the random velocities of the stars. Moreover, they pointed out that the mutual velocities of neighbouring nebulae are of the same order as the random velocities of stars in a nebula. This supported the theory that all nebulae originated from the same very limited region of space: ``It seems much more likely that such an odd occurrence as our planetary system might be formed in the original highly condensed state of the universe than in the present dilute one''. Since 1937 Gamow had actually shifted his interests from nuclear physics proper and had decided to give a graduate course at George Washington University on general relativity and its connections with cosmology \cite[p. 21]{Hufbauer2009}.
This work is an early hint of Gamow's developing research interest in cosmology, which, would be officially inaugurated by the Eighth Washington Conference on theoretical physics devoted to `Stellar Evolution and Cosmology' held in 1942 \cite{GamowFleming1942}, where he spoke about his new ideas on cosmological nucleosynthesis.  Immediately after the war, Gamow would fully merge cosmology with his wide competence as a nuclear astrophysicist, formulating what became successfully known as the big-bang theory of the universe further developed with his collaborators Ralph A. Alpher and Robert Herman who predicted in a later paper that the cooled remnant of the hot early phases should be present in the Universe today and estimated that the temperature of this thermal background should be about 5 K \cite{Alpher2012}.

During that spring-summer 1939, Tolman himself did not resist the temptation of examining the connection between the stability of stellar models and the origin of novae \cite{Tolman1939c}, in an article in which he acknowledged discussions with Oppenheimer. As he remarked in the introduction: ``With the help of such studies one might ultimately hope to understand not only the existence of the great majority of stars in steady states and of a limited classes of stars in pulsating states producing variations in luminosity, but also ``the existence of some --- perhaps nearly all --- stars in states that can lead to the occasional formation of novae, or to the related case of supernovae''.

 In his March 1939 paper, Bethe, too, had tackled the problem of the last stages of stellar evolution: ``It is very interesting to ask what will happen to a star when its hydrogen is almost exhausted. Then, obviously, the energy production can no  longer keep pace with the requirements of equilibrium so that the star will begin to contract [\dots] In the white dwarf state, the necessary energy production is extremely small so that such a star will have an almost unlimited life [\dots] For heavy stars, it seems that the contraction can only stop when a neutron core is formed [\dots] However, these questions obviously require much further investigation'' \cite[p. 456]{Bethe1939}.\footnote{In coming back from the 1938 Washington conference organized by Gamow and Teller, Bethe was very excited, and thus triggered his PhD student Robert Marshak's interest in astrophysics and especially in white dwarfs and in their energy source.  In his PhD thesis (`Contributions to the Theory of the Internal Constitution of Stars') Marshak investigated in detail the state of matter in the interior of a white dwarf star and concluded that no hydrogen could be present. Under these circumstances the radius of the star is uniquely determined by its mass, according to the theory of degenerate configurations. However, in his calculations Marshak found a serious discrepancy between the theoretical radius of Sirius B (``only $5.7 \times 10^8$ cm, as compared with the observed radius of $13.6 \times 10^8$ cm): ``The present investigation has at least established almost beyond question that the claim of astrophysics is in direct conflict with the claim of nuclear physics and that there really seems to be no simple explanation of the radius discrepancy for Sirius B'' \cite{Marshak1940}. Marshak's work later served as a fundamental reference for the understanding of this type of stars \cite{Marshak1970}. Marshak's attention was then diverted from astrophysics and he started working on other problems which were more directly connected with particle physics broadly interpreted.}

Novae (and thus supernovae) and white dwarfs, were definitely an issue at stake during 1939, and in fact  a specific conference was organized  in July in Paris, at the Coll\`ege de France,  in order to study these two categories of stars, which were in the foreground of the current research. On that occasion, Chandra \cite{Chandra1941} stated again his conclusions regarding the limiting mass based on relativistic degeneracy and connected his theory to the supernova phenomenon suggesting that a star which has exhausted its nuclear fuel and whose mass was exceeding such an upper limit would collapse with a huge release of gravitational energy. Such energy would in turn fuel the explosion, leaving a very compact neutron core.

That same July, F\'elix Cernuschi, working at MIT with Sandoval Vallarta, discussed in three articles the problem of supernovae, `neutron-core stars' and the origin of cosmic rays, in the new perspective of the discovery of fission \cite{Cernuschi1939a,Cernuschi1939b,Cernuschi1939c}. The titles of the three articles (`Super-Novae and the Neutron-Core Stars', `A Tentative Theory of the Origin of Cosmic Rays' and `On the Behavior of Matter at extremely high temperatures and pressures') are a clear indication of the constant interest for the dense neutron cores from the point of view of the quick development of nuclear physics, but also in connection with a growing interest for Zwicky's theory of supernovae as sources of cosmic rays and as stellar objects representing the transition of an ordinary star into a neutron star. But his first objection to Zwicky's theory was that ``an ordinary star is a gaseous star without neutron core'' and thus it appeared difficult to imagine how a supernova could result from such a transformation  \cite[p. 120]{Cernuschi1939a}. He assumed instead that white dwarfs are stars with an unstable neutron core and that such instability  derived from fission processes of very heavy nuclei such as uranium and thorium whose existence in the neutron core he had postulated. In this way it was possible to imagine that the same process of fission of single a giant nucleus of atomic number 10000 would produce energies of the order of $10^{12}$ eV, so that , ``a super-nova would not be a transition of an ordinary star into a neutron star, but would result from the explosion of the neutron core of a white dwarf'' in a cascade of successive fissions \cite[p. 121]{Cernuschi1939b}. Under these assumptions it seemed also possible ``to imagine a concrete physical mechanism which might underlie the production of cosmic rays''. One can see here once more the pervasive influence of Lema\^itre's primaeval atom.

Neutrinos, too, are beginning to populate the interior of superdense cores: ``if the neutrino does exist, it will be of great importance in the internal constitution of the stars, due to the fact that this particle should have an extremely high penetrability and, therefore, under certain conditions the transport of heat resulting from the neutrinos might not be negligible beside the flow of radiation''. In this sense neutrinos are beginning to play a role in supernovae explosions.  Cernuschi is also trying to reconcile the divide between astronomers and physicists investigating whether Landau's theory might support Zwicky's proposal.

 Cernuschi's article on the \textit{Physical Review} is followed back to back by the third paper of the Oppenheimer and collaborators' trilogy, submitted in early July 1939: `On continued gravitational contraction.' With his student Hartland Snyder, Oppenheimer took general relativity far beyond Zwicky's possibilities and focused on how the neutron core would evolve once it became unstable \cite{OppenheimerSnyder1939}.
  
  Volkoff and Oppenheimer  had already made clear that assemblies of neutrons  are so compact that general relativity is no longer a small correction and can no longer be neglected because it is central to the stability of such astrophysical objects. They had been able to show that the general relativistic field equations do not possess any static solution  for a spherical distribution of cold neutrons, if the total mass of the neutrons is greater than $\sim 0.7 M_\odot$, and had established that a star under these circumstances would collapse under the influence of its gravitational field. In the meantime, with Snyder, Oppenheimer had explored  the process of gravitational collapse itself, where the full consequences of Einstein's theory of gravitation could be seen at work.\footnote{Already during his presentation of December 1938 at UCLA meeting, Volkoff had  mentioned that nonstatic solutions for the cases of masses beyond the critical mass were being investigated \cite{Volkoff1939a}.} 
  Oppenheimer and Snyder were now definitely stripping `neutron cores' (the Oppenheimer-Volkoff ``spherical distribution of cold neutrons'') of any outer envelope, openly studying what were actually `neutron stars'. Even if they did not call them as such, and only referred to `heavy stars', made of course mainly of neutrons. These investigations were now waiting to officially enter the field of theoretical astrophysics, but had already begun to re-write the chapter of `compact stars', up to that time only containing theorizing on white dwarfs. 
  
  The very first sentence of the abstract itself, once excluding other possible situations, left no hope for a star with a critical value of the mass:    ``When all thermonuclear sources of energy are exhausted a sufficiently heavy star will collapse. \textit{Unless} fission due to rotation, the radiation of mass, or the blowing off of mass by radiation, reduce the star's mass to the order of that to the sun, \textit{this contraction will continue indefinitely}'' [emphasis added]. The concluding lines further emphasized their expectations: ``this behaviour will be realized by all collapsing stars which cannot end in a stable stationary state''.

  Oppenheimer and Snyder found that, as seen by a distant observer, general relativity predicted that the star would asymptotically shrink to its Schwarzschild radius, light from the surface of the star would be progressively reddened, being able to escape over a progressively narrower range of angles. According to the scenario already outlined by Eddington more than a decade before \cite[p. 6]{Eddington1926}, the star would close itself off from the rest of the universe except for its intense gravitational field: ``The mass would produce so much curvature [\dots] that space would close up round the star, leaving us outside (i.e. nowhere)''. These inescapable general arguments were confirmed by the study of analytic solution of the field equations for the case that the pressure within the star could be neglected. It showed that, although the collapse would formally take an infinite time when viewed from large distance, the time measured by an observer comoving with the star would be finite as would also be the time until a distant observer would find the star to be undetectably faint as a consequence of the general relativistic effects. 
  
  That Oppenheimer and Snyder were venturing into  unknown territory is someway testified also from the fact that, apart from the obvious reference to the Oppenheimer-Volkoff paper,  the only citation is to an article by Tolman of 1934 \cite{Tolman1934a}. Tolman is also thanked for ``making a portion of development available''. 
  The simple scenario they had used to describe the collapse process was rather idealized and far from an actual physical model of a collapsing star. But in establishing the physical reality of a phenomenon deeply rooted in the theory of general relativity, the Oppenheimer-Snyder paper gave rise to a startling and unexpected consequence in the real world of astronomy.  
  For the first time, Schwarzschild's purely mathematical solution to the general theory of relativity was systematically discussed within a framework related to a specific physical object. 
  
  As stressed in \cite{Eisenstaedt1993}, these results were actually derived using the so called ``dust solution'', a general solution of the field equations for the case of spherical symmetry and no pressure \cite{Lemaitre1932}, and which were well known to Tolman, who had worked with him during Lema\^itre's stay for two months at Caltech in the early 1930s. In this remarkable contribution, whose 1933 version is cited in  \cite{Tolman1934a}, Lema\^itre also demonstrated that the Schwarzschild singularity is only an apparent singularity. Eisenstaedt emphasizes in his detailed discussion about Lema\^itre's pioneering results \cite[p. 11]{Eisenstaedt1993}, that he tackled this problem once he realized that \cite[p. 200]{Lemaitre1932}: ``The equations of the Friedman universe admit [\dots] solutions in which the radius of the universe goes to zero. This contradicts the generally accepted result that a given mass cannot have a radius smaller than [\dots] $2m$'' (in natural units: $G=c=1$). 
  
  It is to be emphasized that, in the last section, where Lema\^itre is discussing the physical interpretation of the ``zero value of the radius'', he is remarking that matter should find a way to avoid the vanishing of its volume and as matter is formed by stars, this would be `manifestly impossible''. Lema\^itre is comparing this situation to ``the interior of the companion of Sirius'': it appears that even for a degenerate gas nothing might oppose to such a condensed form of matter. At distances between atomic nuclei and electrons of the order of $10^{-12}$,  subatomic  forces opposing to penetration between particles would dominate and certainly be able to stop contraction: ``The universe would thus be comparable to a giant atomic nucleus''. He immediately added that once the contraction is blocked, the process should restart in the opposite verse: ``These solutions in which the universe is expanding and then contracting, periodically reducing to an atomic mass having the dimensions of the solar system, definitely have a  poetic charm and make us think to the phenix of the legend,'' concluded Lema\^itre in the last lines of the paper. 
  
  However, even appreciating that the Schwarzschild singularity could be locally eliminated by a coordinate change, Lema\^itre did not provide an overall picture of collapse to a black hole.   Interestingly, in 1934 Synge wrote a paper entitled `On the Expansion or Contraction of a Symmetrical Cloud under the Influence of Gravity', in which he studied the evolution of a small cloud of particles finding that a collapse beyond the Schwarzschild singularity is possible, at least in the pressure-free case. Eisenstaedt duly remarks \cite[p. 14]{Eisenstaedt1993} that this paper, which might be of great interest for the Oppenheimer-Snyder problem, was almost never cited, even by Synge himself, who actually later found the complete extension of the Schwarzschild solution \cite{Synge1950}.

  When the expansion of the universe was becoming an accepted phenomenon, general relativity had potentially revealed something new and quite unexpected on the universe. As Wheeler much later stressed, \cite[p. 6]{Wheeler1968}: ``No test of Einstein's theory is more dramatic than the expansion of the universe itself, and none has a closer bearing on the phenomenon of collapse'' well expressing that between the predicted and observed expansion and the gravitational collapse of a star ``there is not one significant difference of principle''.  However, this flurry of interests connecting general relativity and astrophysics was interrupted by the outbreak  of World War II, and the two aspects remained separated up to the end of the 1950s - early 1960s, when a novel closer alliance would be established between the general theory of relativity and the physical universe. By that time, the period of stagnation of the theory, the ``low-water-mark'' of general relativity \cite{Eisenstaedt1987,Eisenstaedt1987b} had given way first to the ``renaissance'' \cite{BlumLalliRenn2016} and then to the ``golden age'' of general relativity \cite[pp. 74-80]{Thorne2003}, during which relativistic astrophysics was established as a novel research field and consensus about the existence of extreme physical implications of the theory such as gravitational waves and black holes had formed. 
 
Both the theoretical demonstration of an inescapable process such as the gravitational collapse within Einstein's theory, and the established existence  of supernovae as a new class of astrophysical objects --- a striking evidence for the existence of violent events in the universe --- as well as the continuously evolving stage of an expanding universe, combined all together in marking the end of the `Aristotelian vision' that had dominated astronomy for about 2000 years: the heavens as the domain of an eternal perfect harmony, contrasted with Earth as the realm of conflict and change.

Most astronomers, however, paid little attention to such reality, generally believing that in the final stage of collapse sufficient material would always be ejected to bring the mass of the resulting body down to below the Chandrasekhar limit --- or to below the Oppenheimer-Volkoff limit which is the corresponding maximum mass of a neutron stellar core. The awkward character of the questions aroused from the problem is also testified by the lack of any mention of the collapse for masses beyond the limiting mass in Chandra's comprehensive textbook \textit{An introduction to the study of stellar structure} \cite{Chandra1939}.

However, within a year, Gamow and the young Brazilian physicist Mario Sch\"onberg, who had studied with Fermi and Pauli, investigated for the first time the physical process of  `catastrophic collapse' from a point of view of nuclear physics \cite{GamowSchonberg1940}, arguing that rapid cooling due to extensive neutrino losses by what they  called, for brevity, ``urca-processes'' in inverse beta-decay, would result in a catastrophic failure of pressure support near the core, unable to support the weight of the overlying collapsing layers  \cite[p. 540]{GamowSchonberg1941}.\footnote{They named it the the `urca-process'  because it results in a rapid disappearance of thermal energy from the interior of a star, similar to the rapid disappearance of money from the pockets of the gamblers in the Casino da Urca, in Rio de Janeiro, where they discussed the problem when they first met  \cite[p. 137]{Gamow1970}.}
They did not mention Cernuschi's articles, and at the same time rejected Zwicky's hypothesis of the collapse as being due to the formation of a large number of neutrons with subsequent closer ``packing'' in the central regions. What they wanted was the instantaneous removal of large amounts of gravitational energy produced by contraction, in order to have a collapse ``with a velocity comparable to that of `free fall' independent of the kind of particles existing in its interiors''. Processes of absorption and reemission of free electrons could lead to tremendous energy losses through neutrino emission and  cause the collapse of the entire stellar body. During the last ten years, neutrinos had gained a considerably important position in nuclear physics, in spite of the fact that all the attempts at their direct observation had failed. However, exactly their capability of passing through many thousands of kilometers of matter without suffering absorption \cite{BethePeierls1934}, made them the right agent to remove the surplus energy from the interior of a contracting star, whose body was completely transparent for neutrinos. They also emphasized that \cite[p. 541]{GamowSchonberg1941}, ``while the neutrinos are still considered as highly hypothetical particles because of the failure of all efforts made to detect them, the phenomena of which we are making use in our considerations are supported by the direct experimental evidence of nuclear physics''.
 In speculating that neutrinos might play an important role in stellar evolution, particularly in the collapse of evolved stars, they ushered in the advent of particle astrophysics. Such a hypothesis was quite bold for the time because neutrinos, which had been proposed by Pauli in 1930, were not directly detected until the mid 1950s. The intense neutrino flux emitted during the process was dramatically confirmed by Supernova 1987A whose observation in 1987 coincided with a burst of 11 neutrinos, detected by Super-Kamiokande in Japan, by 8 further neutrinos registered independently in Ohio, and by 5 events at the Baksan Neutrino Observatory on the Caucasus mountains.
The fast removal of energy due to the emission of neutrinos would induce ``the collapse of the entire stellar body with an almost free-fall velocity'' while rapid contraction would increase the central temperature. Stars possessing a mass larger than the critical mass would undergo a much more extensive collapse and their ever-increasing radiation would drive away more and more material from their surface: ``The process will probably not stop until the expelled material brings the mass of the remaining star below the critical value''. This process might be compared with the supernovae explosions, in which case the expelled gases would form extensive nebulosities such as the Crab-Nebula \cite[p. 546]{GamowSchonberg1941}, leaving behind a faint star, that according to its observed properties, was classified at the time as a very dense white dwarf \cite{Minkowski1942}. This view, already clearly expressed by Chandra \cite{Chandra1935a,Chandra1935c}, was destined to endure for a long time. The following year, Sch\"onberg became Chandra's post-doc student and with him he wrote a paper in which they discussed the problem of what is the maximum mass of a star's  hydrogen-exhausted core that can support the overlying layers against gravitational collapse \cite{ChandraSchonberg1942}. The result of their investigation was that the helium core reached the maximum mass it could attain without collapsing when just about 10\% of the hydrogen had been consumed. This is known now as the Chandrasekhar-Sch\"onberg limit. In the concluding lines they again stressed that ``the supernova phenomenon may result from the inability of a star of mass greater than $M_3$ [upper limit to the mass of degenerate configurations] to settle down to the final state of complete degeneracy without getting rid of the excess mass'' thus assuming that the final state would be that of a white dwarf, and without considering the possibility that the star might collapse to nothing.

By 1939, the problem of what happens to a compact star core made entirely of degenerate fermions (electrons and neutrons) had been studied by a handful of researchers. Oppenheimer, however, --- and Tolman as well --- did not recognize that what they had tackled was conceptually quite similar and had been already anticipated by  B. Datt's simple but more general model, published in May 1938, in the \textit{Zeitschrift f\"ur Physik} \cite{Datt1938}.\footnote{It was dated Kalkutta, Presidency College, 10 September 1937.}
 Working at Presidency College, where Chandrasekar had studied and where later Abdus Salam would move his first steps as a physicist, Datt used general relativity to examine the final fate of an idealized  homogeneous pressure-less spherically symmetric, massive cloud with no rotation and internal stresses, collapsing under its own gravitational attraction.\footnote{See comment to the English translation of Datt's article by Andrzej Krasi\'nski \cite{Krasinski1999}.}
This classic scenario became later  known as the Oppenheimer-Snyder-Datt model (OSD). 

But at the moment all this sounded like an exotic problem, and nobody realized, not even Oppenheimer himself, how innovative their contribution was: on one side nuclear matter --- and particle physics --- were becoming essential for the description of matter at such extreme densities. On the other hand, it had become clear that such superdense objects could be described \textit{only} within Einstein's theory of gravity: the door had been opened on the world of relativistic astrophysics.   There was at least one physicist who was deeply  aware of the relevance of these results. Landau, who was again free after a year of imprisonment and was working at his celebrated theory of superfluidity, added the Oppenheimer-Snyder paper to his `Golden List', according to what Evgeny Mikhailovich Lifschitz told Kip Thorne many years later \cite[p. 219]{Thorne1994}. It even appears that ``So great was Landau's influence that his view took hold among leading Soviet theoretical physicists from that day forward''.

Most probably it is not by chance that in May 1939, not long after the appearance of Datt's contribution on  the \textit{Zeitschrift f\"ur Physik} --- but nearly in parallel to the Oppenheimer-Snyder paper --- Einstein himself submitted a contribution in which he worked out how a swarm of particles would behave as they collapsed through gravity \cite{Einstein1939}.  As stated by Einstein, this investigation arose out of discussions conducted with Howard P. Robertson, Peter G. Bergmann and Valentine Bargmann  on the \textit{mathematical} and \textit{physical} significance of the Schwarzschild singularity, which had played a role in Zwicky's paper on the collapse of neutron stars, but especially in Tolman's article written in parallel with Oppenheimer and Volkoff's contribution, that should not have escaped the attention of Einstein and Robertson,  both having a longstanding personal relationship with Tolman. In particular, during the 1930s, Robertson had worked on the problem of the Schwarzschild space-time, but he did not publish it \cite[p. 328]{Eisenstaedt1987}.  However, according to Bergmann, Einstein was not aware of Oppenheimer's papers. In his \textit{Introduction to the Theory of Relativity}, whose first edition was printed in May 1942, with a foreword by Einstein himself mentioning the many hours spent in discussing the text, Bergmann summarized as follows Robertson's view \cite[p. 203-204]{Bergmann1942}: ``Robertson has shown that, if a Schwarzschild field could be realized, a test body which falls freely toward the center would take only a finite proper time to cross the `Schwarzschild' singularity, even though the coordinate time is infinite; and he has concluded that at least part of the singular character of the surface $r = 2m$ must be attributed to the choice of the coordinate system''.  At the end of the section dedicated to the Schwarzschild singularity, Bergmann introduced a short description of Einstein's article with the following clear-cut sentence: ``\textit{In nature, mass is never sufficiently concentrated to permit a Schwarzschild singularity to occur in empty space} [emphasis added]'' \cite[p. 204]{Bergmann1942}. 
In any case, no reference to Oppenheimer's works with his collaborators can be found in Bergmann's book. 
 
 Starting from the Schwarzschild's solution of the static gravitational field of spherical symmetry, and from the vanishing of the $g_{44}$ term of the equation, Einstein tackled the question whether it was possible ``to build up a field containing such singularities with the help of actual gravitating masses or whether such regions with vanishing $g_{44}$ do not exist in cases which have physical reality''. As a field-producing mass he chose a system formed by a great number of small gravitating particles moving freely under the influence of the field produced by them all together. The particles in Einstein's ``spherical star cluster'' all moved in circular orbits around a common center, and he calculated that material particle orbits could not have radii less than one and a half Schwarzschild radii, in Schwarzschild coordinates, so that the essential result of this investigation, concluded Einstein, was ``a clear understanding as to why the `Schwarzschild singularities' do not exist in physical reality. Although the theory given here treats only clusters whose particles move along circular paths it does not seem to be subject to reasonable doubt that more general cases will have analogous results. The `Schwarzschild singularity' does not appear for the reason that matter cannot be concentrated arbitrarily. And this is due to the fact that otherwise the constituting particles would reach the velocity of light''. In the concluding lines of the paper, Einstein drastically stated that it is not possible to attain the Schwarzschild radius in nature and thus the problem of the mathematical and physical significance of the Schwarzschild singularity, ``quite naturally  leads to the question, answered by this paper in the negative, as to whether physical models are capable of exhibiting such a singularity''. As underlined by Jean Eisenstaedt within a discussion on Robertson's work on the Schwarzschild singularity, Einstein's starting hypothesis of circular orbits for his system of gravitating particles, as a matter of fact excluded the possibility of reaching the Schwarzschild singularity, because the radius of a gas cloud described by strictly circular orbits must necessarily be greater than 3/2 the Schwarzschild radius. In this sense, Einstein's article is based on a `circular logic' \cite[p. 337]{Eisenstaedt1987}.

Oppenheimer and Snyder had emphasized that although the Schwarzschild singularity occurring at radius $r=2m$ is not actually a singularity, there is still a space-time singularity at the centre, where the density of the dust becomes infinite. 
According to Roger Penrose \cite{Penrose1996}, ``Since there is still a singularity in the Oppenheimer-Snyder collapse model (at $r = 0$), the Chandrasekhar dilemma [on the existence of a maximum mass for white dwarf stars] is not removed by their collapse picture.
However many people remained unconvinced that this description would necessarily
be the inevitable result of the collapse of a star too massive to be sustainable as
either a white dwarf or neutron star. There were a number of good reasons for some
scepticism. In the first place, the equations of state inside the matter were assumed
to be those appropriate for pressureless dust, which is certainly far from realistic for
the late stages of stellar collapse. Moreover, the density was assumed to be constant
throughout the body. With realistic material, there are many alternative evolutions to
that described by Oppenheimer and Snyder. For example, nuclear reactions set off
at the centre could lead to an explosion --- a supernova --- which might perhaps drive
off sufficient mass from the star that a stable equilibrium configuration becomes
possible''.

In January 1939, the Fifth Washington Conference on Theoretical physics had been held, having low temperatures as a focus for the discussion. However, as it is well known, Bohr, who had just arrived from Europe, brought with him news about the Frisch-Meitner explanation of fission as a physical process, immediately arousing an incredible excitement and putting in motion a series of events which would deeply affect the whole scientific community in connection with the dramatic developments on the world stage. On September 1, when Oppenheimer and Snyder's paper appeared in the \textit{Physical Review},  Nazi troops marched across the Polish border. In the  same issue Bohr and Wheeler, working together at Princeton, outlined an account of the mechanism of nuclear fission on the basis of the liquid drop model of  nuclei \cite{BohrWheeler1939}. By the end of 1939, actual neutrons in heavy nuclei had already become the protagonists in a completely  different realm, eventually leading to the building of the first nuclear reactors and the first nuclear weapons. Oppenheimer himself would be heavily involved in these efforts, heading the Manhattan Project's secret research laboratory at Los Alamos. The curtain apparently closed on the march leading to the first applications of general relativity to an astrophysical compact object, but behind the scenes new premises for a great renewal of interest in superdense matter and compact objects  were laid during the war period  which would eventually flourish  within the new conditions provided by post-war science.

\section{Acknowledgements}

I am deeply indebted to Gordon Baym, Wolf Beiglb\"ock, Werner Israel, Roberto Lalli, and an unknown referee, for a critical and constructive evaluation of the manuscript and for many most useful suggestions. Special thanks go to J\"urgen Renn for arousing my interest in the fascinating history of astrophysics.

I also gratefully acknowledge that this research has made a systematic use of NASA's Astrophysics Data System Bibliographic Services.

\end{document}